\documentclass[epj]{svjour}
\usepackage{graphics}
\usepackage{graphicx}
\usepackage{bm}
\usepackage{epic,eepic}
\usepackage{epsfig}
\usepackage{pifont}
\usepackage{subfigure}

\usepackage{citesort}

\usepackage{caption}
\usepackage{amsmath,amssymb,amsfonts,mathrsfs,dsfont,mathtools}
\usepackage{url}
\usepackage{color}
\usepackage{float}
\usepackage{mathptmx}
\usepackage{latexsym}

\usepackage[toc,page]{appendix}

\newcommand{\Id}{\mathds{1}}            % D x D identity matrix

\newcommand{\be}{\begin{equation}}
\newcommand{\ee}{\end{equation}}

\newcommand{\tshear}{\tau_{\mbox{\tiny {shear}}}}

\newcommand{\Ca}{\mbox{Ca}}

\newcommand{\Ren}{\mbox{Re}}
\newcommand{\De}{\mbox{De}}
\begin{document}
\sloppy
\title{A Lattice Boltzmann study of the effects of viscoelasticity on droplet formation in microfluidic cross-junctions}
\author{Anupam Gupta \and Mauro Sbragaglia% etc
}                     % Do not remove
\institute{Department of Physics and INFN, University of ``Tor Vergata'', Via della Ricerca Scientifica 1, 00133 Rome, Italy}
\authorrunning{A. Gupta, M. Sbragaglia}
\titlerunning{A Lattice Boltzmann study of the effects viscoelasticity in microfluidic cross-junctions}
\date{Received: date / Revised version: date}
\abstract{
Based on mesoscale lattice Boltzmann (LB) numerical simulations, we investigate the effects of viscoelasticity on the break-up of liquid threads in microfluidic cross-junctions, where droplets are formed by focusing a liquid thread of a dispersed (d) phase into another co-flowing continuous (c) immiscible phase. Working at small Capillary numbers, we investigate the effects of non-Newtonian phases in the transition from droplet formation at the cross-junction (DCJ) to droplet formation downstream of the cross-junction (DC) (Liu \& Zhang, {\it Phys. Fluids.} {\bf 23}, 082101 (2011)).  We will analyze cases with {\it Droplet Viscoelasticity} (DV), where viscoelastic properties are confined in the dispersed phase, as well as cases with {\it Matrix Viscoelasticity} (MV), where viscoelastic properties are confined in the continuous phase. Moderate flow-rate ratios $Q \approx {\cal O}(1)$ of the two phases are considered in the present study. Overall, we find that the effects are more pronounced with MV, where viscoelasticity is found to influence the break-up point of the threads, which moves closer to the cross-junction and stabilizes. This is attributed to an increase of the polymer feedback stress forming in the corner flows, where the side channels of the device meet the main channel. Quantitative predictions on the break-up point of the threads are provided as a function of the Deborah number, i.e., the dimensionless number measuring the importance of viscoelasticity with respect to Capillary forces.
 % end of PACS codes
} %end of abstract
\maketitle
\section{Introduction}
\label{intro}
Microfluidic devices are important in the studies that require control over droplet size in small scale hydrodynamics~\cite{Christopher07,Christopher08,Teh08,Baroud10,Seeman12,Glawdeletal}. Common droplet generators used are T-shaped geometries~\cite{Demenech07,Demenech06} and flow-focusing devices~\cite{LiuZhang09,LiuZhang11,Garstecki13}. In T-shaped geometries, a dispersed (d) phase is injected perpendicularly into a continuous  (c) phase~\cite{Demenech07}. Forces are created by the cross-flowing continuous phase which periodically break off droplets; the flow-focusing geometry, instead, creates droplets by focusing a fluid thread into another fluid~\cite{Gordillo,Link04,Anna,LiuZhang09,LiuZhang11}. The operational regime of these devices is characterized by the Capillary number $\Ca$, which quantifies the importance of the viscous forces with respect to the surface tension forces, and the flow-rate ratio $Q$ of the two fluids. Previous research has been mainly restricted to Newtonian fluids. However, there are many situations of practical interest which result in considering a non-Newtonian behaviour~\cite{Arratia08,Steinhaus,Husny06,Arratia09}. The formation of viscoelastic droplets in Newtonian continuous phases was investigated in various flow-focusing geometries by Steinhaus {\it et al.}~\cite{Steinhaus}. The effects of elasticity on filament thinning was studied by Arratia {\it et al.}~\cite{Arratia08,Arratia09} using dilute aqueous polymeric solutions with different molecular weights. In these studies it is observed that elasticity prolongs the processes of thinning of the filament and significantly increases the interval required for break-up to complete. In a recent paper, Derzsi {\it et al.}~\cite{Garstecki13} presented an experimental study of the effects of elasticity in microfluidic flow-focusing devices: the authors find that the elasticity of the focusing liquid facilitates the formation of smaller droplets and leads to transitions between various regimes at lower ratios of flow and at lower values of the Capillary numbers in comparison to the Newtonian case. Complementing these results with systematic investigations by varying deformation rates and fluid constitutive parameters would be of great interest~\cite{Demenech07,Demenech06,LiuZhang09,LiuZhang11,Wuetal}. In a recent paper, Liu \& Zhang~\cite{LiuZhang09,LiuZhang11} performed numerical simulations of three-dimensional microfluidic cross-junctions at low Capillary numbers. A regime map was created to describe droplet formation at the cross-junction (DCJ), downstream of the cross-junction (DC), and stable parallel flows (PF). The influence of flow-rate ratio, Capillary number, and channel geometry was then systematically studied in the squeezing-dominated DCJ regime.\\ 
In the present paper we study the impact of viscoelastic phases in the transition from DCJ to DC regimes. At fixed flow conditions, these effects will be shown to be particularly relevant for the case where viscoelastic properties are confined in the continuous phase, due to the fact that the action of the co-flowing liquid is supplemented with the polymer feedback stresses that are well pronounced in correspondence of the corners where the side channels meet the main channel. Quantitative details on how the polymer feedback stresses change the transition from DCJ to DC regimes are provided.\\
The paper is organized as follows: in Sec.~\ref{sec:model} we will present the necessary mathematical background for the problem studied, showing the relevant equations that we integrate in both the continuous and dispersed phases. In Sec.~\ref{sec:rheo} useful benchmarks for the shear and elongational rheology of the numerical model will be provided for the typical parameters used in our study. In Sec.~\ref{sec:results} we will present the numerical results and characterize the effects of viscoelasticity on the various regimes. Conclusions will follow in Sec.~\ref{sec:conclusion}.

\section{Macroscopic equations}\label{sec:model}

Numerical modeling of viscoelastic fluids often relies on coupling constitutive relations for the stress tensor, typically obtained via approximate representations of some underlying micro-mechanical model for the individual polymer molecules, with a Navier-Stokes (NS) description for the solvent. We will use a hybrid algorithm combining a multicomponent Lattice-Boltzmann (LB) model with Finite Differences schemes (FD): the solvent part of the model is obtained with LB models~\cite{Benzi92,Succi01,Zhang11,Aidun10}, which proved to be extremely valuable tools for the simulation of droplet deformation problems~\cite{Xi99,vandersman08,Komrakova13,Liuetal12}, droplet dynamics in open~\cite{Moradi,Thampi} and confined~\cite{Liuetal12,Gupta09,Gupta10} microfluidic geometries. LB is instrumental to solve the diffuse-interface hydrodynamic equations of a binary mixture of two components~\cite{Yue04,Yueetal05,Yueetal06a,Yueetal06b,Yueetal08,Yueetal12}: the resulting physical domain can be partitioned into different subdomains, each occupied by a ``pure'' fluid, with the interface between the two fluids described as a thin layer where the fluid properties change smoothly. The solute part of the model is based on a FENE-P constitutive model~\cite{Wagner05,Lindner03}. The FENE-P constitutive equations are solved with a FD scheme which is coupled with the solvent LB as described in~\cite{SbragagliaGuptaScagliarini,SbragagliaGupta}. Our numerical approach has been extensively validated in our previous works~\cite{SbragagliaGuptaScagliarini,SbragagliaGupta}, where we have provided evidence that the model is able to capture quantitatively rheological properties of dilute suspensions as well as deformation and orientation of single droplets in confined shear flows. The main essential features of the model are recalled in the Appendix.\\ 
For the geometry studied in this paper, we will provide quantitative details on how the viscoelastic model parameters affect the break-up properties. We will analyze separately cases with {\it Droplet Viscoelasticity} (DV), where the viscoelastic properties are confined in the dispersed (d) phase undergoing the break-up process, as well as cases with  {\it Matrix Viscoelasticity} (MV), where the viscoelastic properties are confined in the continuous (c) phase. In the MV case, the equations we solve in the continuous phase are the Navier-Stokes (NS) equations coupled to the FENE-P constitutive equations
\be\label{NSc}
\begin{split}
\rho_{c} & \left[ \partial_t \bm u_{c} + ({\bm u}_{c} \cdot {\bm \nabla}) \bm u_{c} \right]  =  - {\bm \nabla}P_{c}+\\ &  {\bm \nabla} \left(\eta_{c} ({\bm \nabla} {\bm u}_{c}+({\bm \nabla} {\bm u}_{c})^{T})\right)  +\frac{\eta_{P}}{\tau_{P}}{\bm \nabla} \cdot [f(r_{P}){\bm {\bm {C}}}].
\end{split}
\ee
\be\label{FENEP}
\begin{split}
\partial_t {\bm {C}} + (\bm u_{c} \cdot {\bm \nabla}) {\bm {C}}  =  {\bm {C}} \cdot ({\bm \nabla} {\bm u}_{c}) + & {({\bm \nabla} {\bm u}_{c})^{T}} \cdot {\bm {C}}  \\  -& \left(\frac{{f(r_{P}){\bm {C}} }- {{\bm I}}}{\tau_{P}}\right).
\end{split}
\ee
Here, ${\bm u}_c$ and $\eta_c$ are the velocity and the dynamic viscosity of the continuous phase, respectively. $\rho_c$ is the solvent density, $P_c$ the solvent bulk pressure, and $({\bm \nabla} {\bm u}_c)^T$ the transpose of $({\bm \nabla} {\bm u}_c)$. As for the polymer details, $\eta_{P}$ is the viscosity parameter for the FENE-P solute, $\tau_P$ the polymer relaxation time,  ${\bm {C}}$ the polymer-conformation tensor, ${\bm I}$ the identity tensor, $f(r_P)\equiv{(L^2 -3)/(L^2 - r_P^2)}$ the FENE-P potential that ensures finite extensibility, $ r_P \equiv \sqrt{Tr({\bm {C}})}$ and $L$ is the maximum possible extension of the polymers~\cite{bird,Herrchen}. In the dispersed phase we just consider the NS equations 
\begin{eqnarray}\label{NSd}
\rho_d \left[ \partial_t \bm u_{d} + ({\bm u}_{d} \cdot {\bm \nabla}) \bm u_{d} \right] 
&=&  - {\bm \nabla}P_{d} \nonumber \\ && + {\bm \nabla} \left(\eta_{d} ({\bm \nabla} {\bm u}_{d}+({\bm \nabla} {\bm u}_{d})^{T})\right)                                              
\end{eqnarray}
where the different fields have the same physical meaning but they refer to the dispersed phase. Immiscibility between the dispersed phase and the continuous phase is introduced using the so-called ``Shan-Chen'' model~\cite{SC93,SC94,SbragagliaGuptaScagliarini} which ensures phase separation with the formation of stable interfaces between the two phases characterized by a positive surface tension $\sigma$.\\
For the DV case, we consider the reversed case, where the FENE-P constitutive equations are integrated in the dispersed phase (i.e. \eqref{NSc}-\eqref{FENEP} with c $\rightarrow$ d), while only the NS equations are considered in the continuous phase (i.e. \eqref{NSd} with d $\rightarrow$ c).\\
As for the geometry used, we consider the simplest case where channels have a square cross-section $H \times H$.  The square cross-section is resolved with $H \times H = 50 \times 50$ grid points, while the main and side channels are resolved with $L_x = 1150$ and $L_y = 250$ grid points, respectively. Besides the geometrical parameters, the Newtonian problem is described by six parameters characterizing the flow and material properties of the fluids. These parameters are the mean speeds of the continuous and dispersed phases, $v_c$ and $v_d$, respectively; the viscosities of the two fluids $\eta_c$ and $\eta_d$ of Eqs. (\ref{NSc}) and (\ref{NSd}), the interfacial tension $\sigma$, and the total density $\rho_c=\rho_s=\rho$ (the same for the dispersed and continuous phases). We will assume perfect wetting for the continuous phase, while the dispersed fluid does not wet the walls. As usual with these kinds of systems~\cite{Demenech07,LiuZhang09,LiuZhang11}, we choose the following groups: the Capillary number calculated for the continuous phase,
\be\label{eq:Ca}
\Ca = \frac{(\eta_{\mbox{\tiny{TOT}},c}) v_c}{\sigma}
\ee the Reynolds number $\Ren =\rho v_c H/ (\eta_{\mbox{\tiny{TOT}},c})$, the viscosity ratio $\lambda$, and the flow-rate ratio
\be\label{eq:Q}
Q=\frac{v_d}{v_c} = \frac{Q_d}{Q_c}
\ee
where $Q_d=v_d H^2$ and $Q_c =v_c H^2$ are the flow-rates at the two inlets.  For the flow regimes under consideration, the Reynolds number is small ($\Ren \approx 0.01-0.1$), and does not influence the droplet size, which leaves us with three governing parameters: $\Ca$, $\lambda$ and $Q$. Notice that the total viscosity in the continuous phase $\eta_{\mbox{\tiny{TOT}},c}$ is either $\eta_{\mbox{\tiny{TOT}},c}=\eta_c+\eta_P$ (for MV) or $\eta_{\mbox{\tiny{TOT}},c}=\eta_c$ (for DV). In the outlet, we impose pressure boundary conditions and use Neumann boundary conditions for the velocity field. A Dirichlet boundary condition is imposed at the inlets by specifying the pressure gradient that is compatible with the analytical solution of a Stokes flow in a square duct~\cite{vandersman08}. As for the polymer boundary conditions, we impose a Dirichlet type boundary condition by linearly extrapolating the conformation tensor at the boundaries.\\

\section{Viscoelastic parameters and rheology of the numerical model}\label{sec:rheo}

%%%%%%%%%%%%%%%%%%%%%%%%%%%%%%%%%%%%%%%%%%%%%%%%%%%%%%%%%%%%%%%%%%%%%%%%%%%%%%%%
%%%%%%%%%%%%%%%%%%%%%%%%%%%%%%%%%%%%%%%%%FIG 1%%%%%%%%%%%%%%%%%%%%%%%%%%%%%%%%%%%%%%%%%%%%%%%%%%%%%%%%%%%%%%%%%%%%%%%%%%%%%%%%%%%%%%%%%%%%%%%%%%%%%%%%%%%%%%%%%%

\begin{figure*}[t!]
\subfigure[{\scriptsize Polymer shear stress}]
{
\includegraphics[width = 0.45\linewidth]{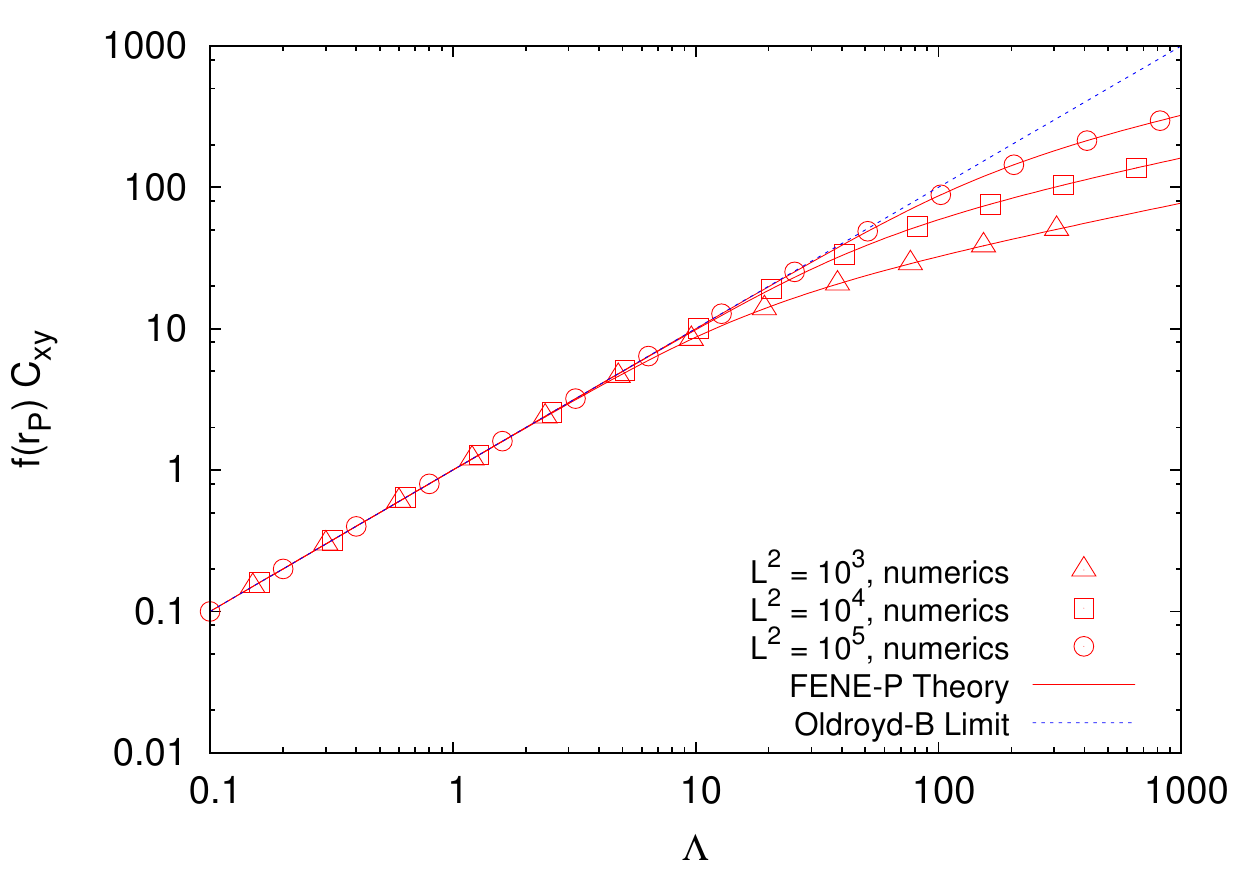}
}
\subfigure[{\scriptsize Polymer shear viscosity}]
{
\includegraphics[width = 0.45\linewidth]{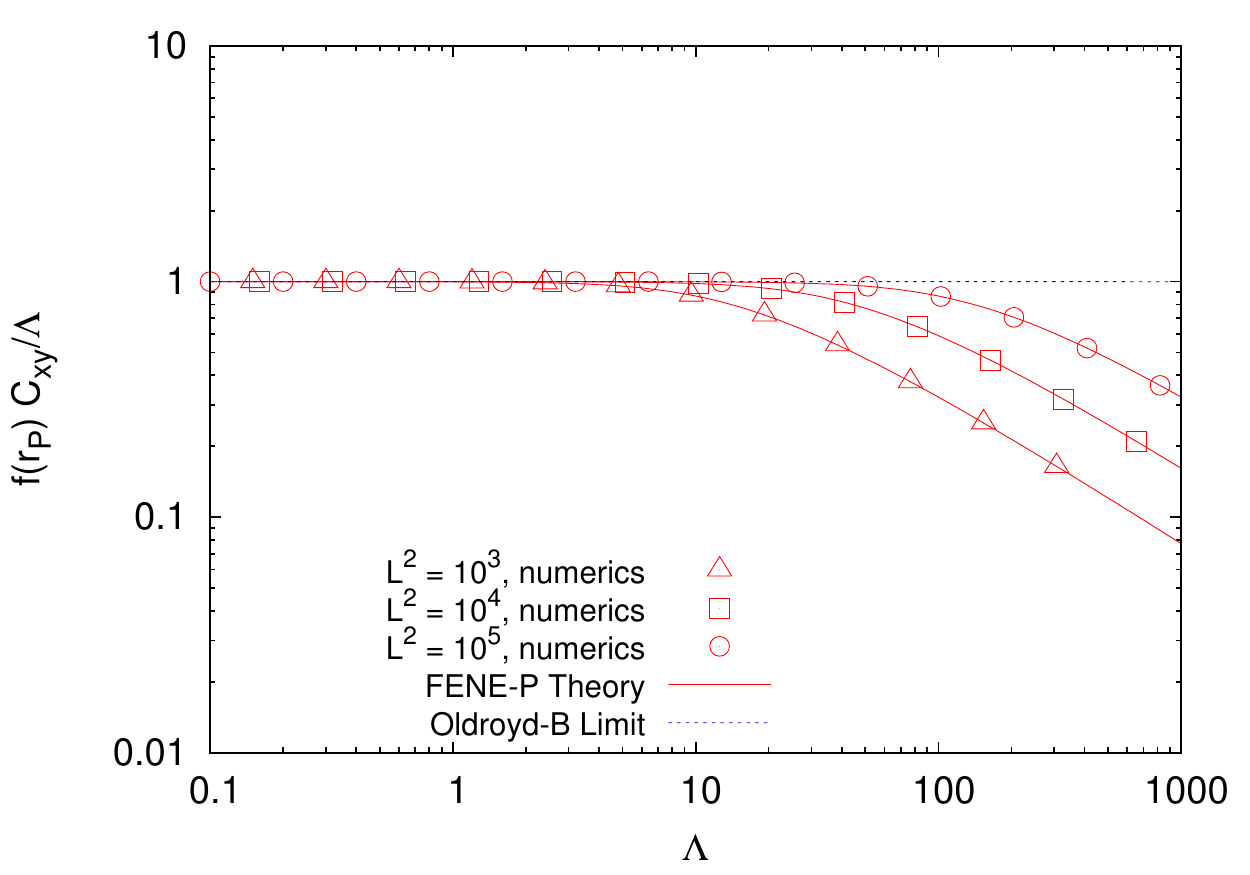}\\
}\\
\subfigure[{\scriptsize Polymer first normal stress difference in shear flow}]
{
\includegraphics[width = 0.45\linewidth]{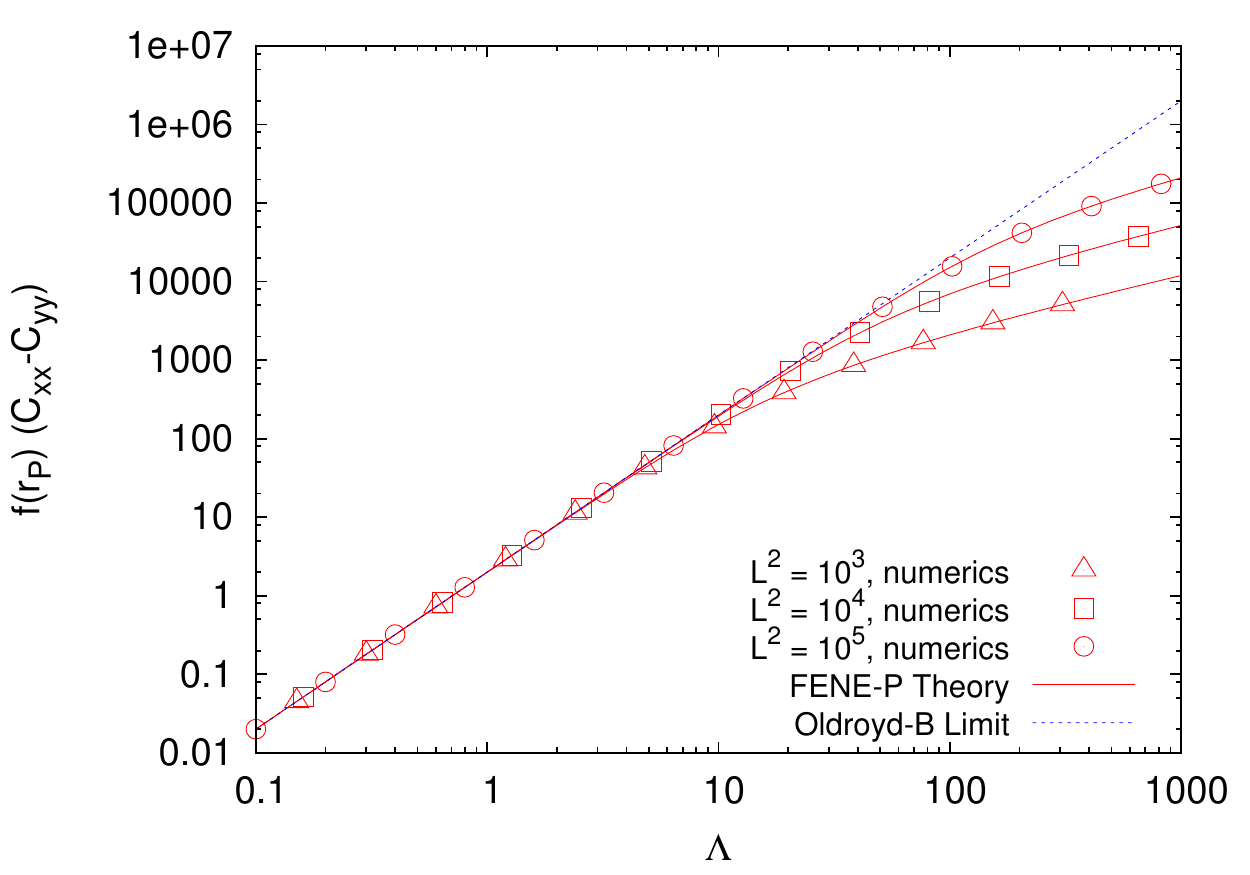}
}
\subfigure[{\scriptsize Polymer first normal stress coefficient in shear flow}]
{
\includegraphics[width = 0.45\linewidth]{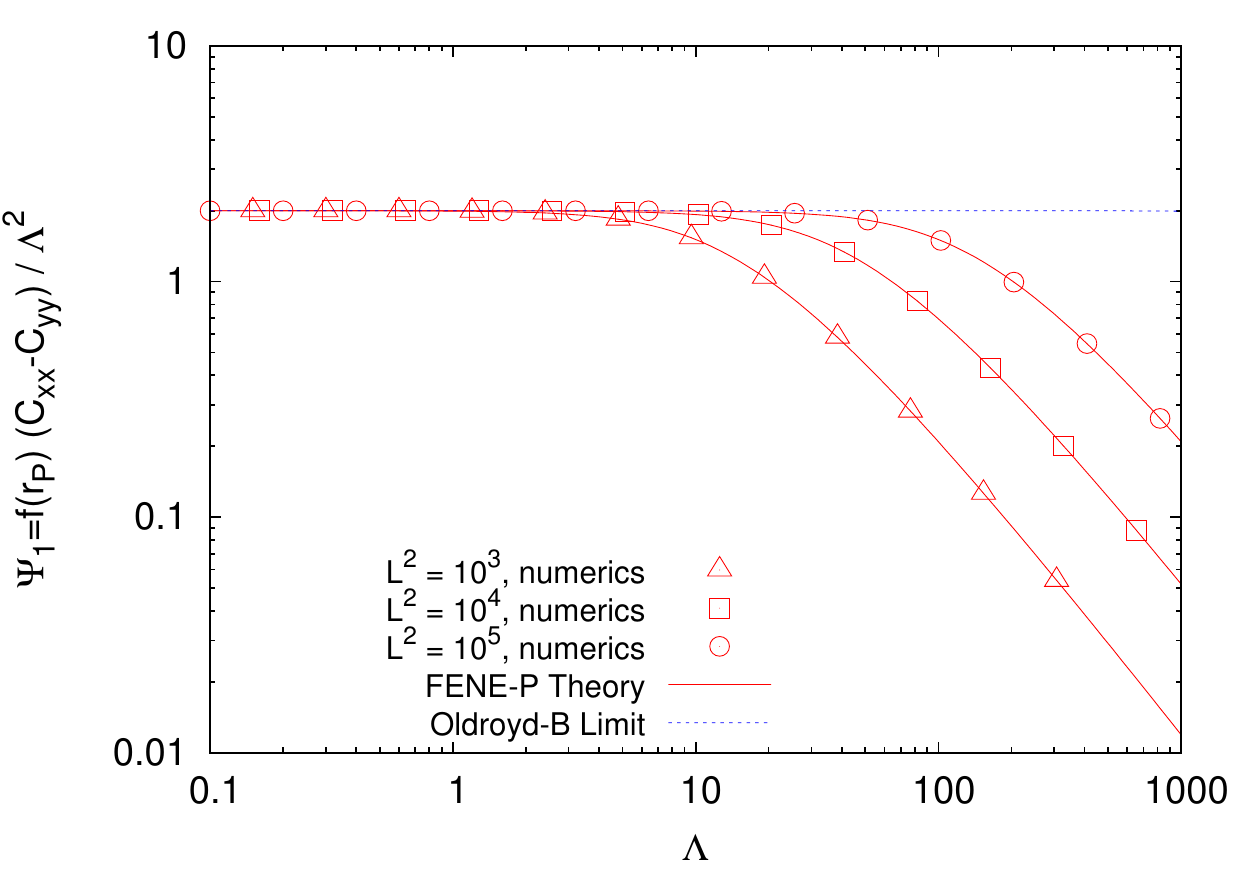}
}\\
\subfigure[{\scriptsize Polymer normal stress difference in elongational flow}]
{
\includegraphics[width = 0.45\linewidth]{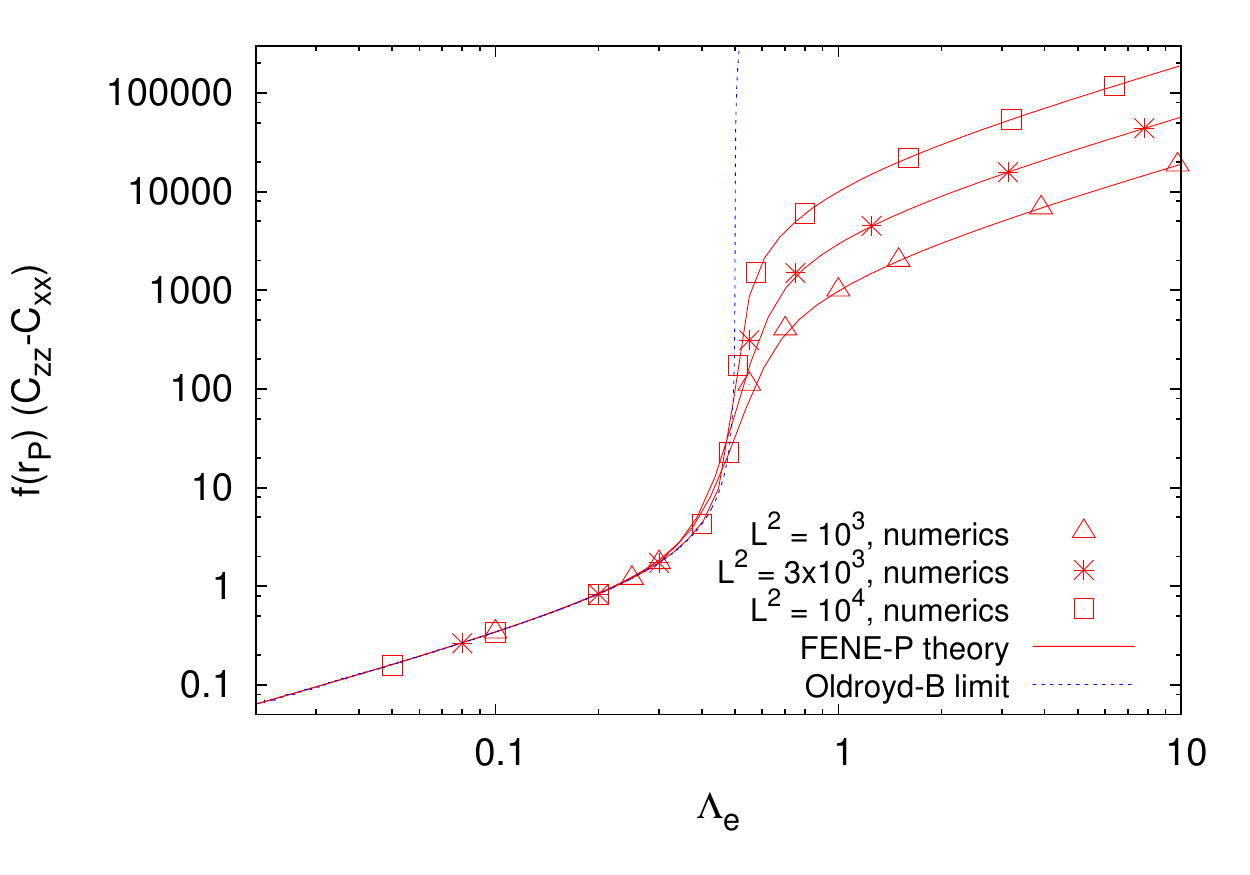}
}
\subfigure[{\scriptsize Polymer elongational viscosity}]
{
\includegraphics[width = 0.45\linewidth]{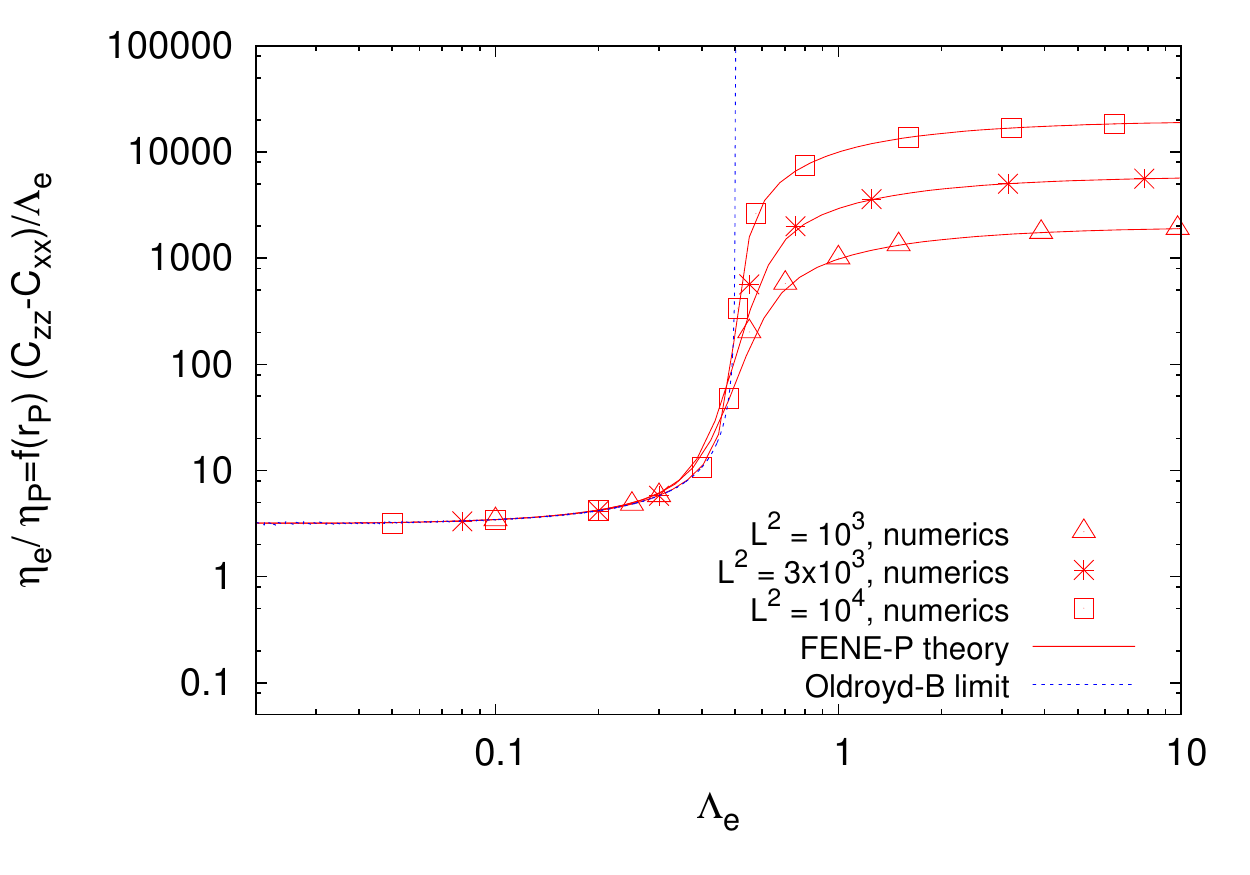}
}\\
\caption{Polymer shear and elongational rheology. Panels (a)-(b): we plot the shear stress and the shear viscosity (both scaled to the polymer viscosity $\eta_P$ and polymer relaxation time $\tau_P$, see \eqref{NSc}-\eqref{FENEP} and text for details) as a function of the dimensionless shear $\Lambda=\tau_P \dot{\gamma}$ in a steady shear flow with intensity $\dot{\gamma}$. Symbols are the results of the numerical simulations~\cite{SbragagliaGuptaScagliarini,SbragagliaGupta} with different imposed shears, different $\tau_P$, and different values of $L^2$. Numerical simulations for the flow-focusing geometry will be performed with $L^2=10^4$ (see text for details), while the other values of $L^2$ are here used to give an idea of how much $L^2$ affects the viscoelastic response. The solid lines are drawn from theoretical predictions based on equations (\ref{lindner}) and (\ref{lindner2}). Panels (c)-(d): we plot the dimensionless first normal stress difference and the first normal stress coefficient. Panels (e)-(f): we plot the dimensionless normal stress difference and elongational viscosity developed in steady elongational flow (see text for details) as a function of the dimensionless elongational rate $\Lambda_e=\tau_P \dot{\epsilon}$. In all cases we also report the prediction for Hookean dumbbells (Oldroyd-B limit, $L^2 \gg 1$).\label{fig:1}}\end{figure*}

%%%%%%%%%%%%%%%%%%%%%%%%%%%%%%%%%%%%%%%%%%%%%%%%%%%%%%%%%%%%%%%%%%%%%%%%%%%%%%%%%%%%%%%%%%%%%%%%%%%%%%%%%%%%%%%%%%%%%%%%%%%%%%%%%%%%%%%%%%%%%%%%%%%%%%%%%%%%%%%%
%%%%%%%%%%%%%%%%%%%%%%%%%%%%%%%%%%%%%%%%%%%%%%%%%%%%%%%%%%%%%%%%%%%%%%%%%%%%%%%%

The viscosity ratio of the two Newtonian phases $\eta_d/\eta_c$~\cite{SbragagliaGuptaScagliarini,SbragagliaGupta} is tunable, a fact that allows to work with unitary viscosity ratio, defined in terms of the total (fluid + polymer) shear viscosity $\lambda=\eta_d/(\eta_c+\eta_{P})=1$ for MV and  $\lambda=(\eta_d+\eta_P)/\eta_c=1$ for DV. Consistently, we will compare the non-Newtonian simulations with the corresponding Newtonian case at $\lambda=\eta_d/\eta_c=1$. The ratio between the polymer viscosity and the total viscosity is chosen as $\eta_P/(\eta_{c,d}+\eta_P) \approx 0.40$. In order to report data with dimensionless numbers, we choose to quantify the degree of viscoelasticity with the Deborah number that we define as $\De=\frac{N_1 H}{2 \sigma}\left(\frac{\sigma}{(\eta_{c,d}+\eta_P) H \dot{\gamma}}\right)^2$, where $N_1$ is the first normal stress difference which develops in the viscoelastic phase in presence of a homogeneous steady shear~\cite{bird,Lindner03}. In the definition of the Deborah number, the viscosity is obviously indicated in the viscoelastic phase, either $\eta_{c}+\eta_P$ for MV or $\eta_{d}+\eta_P$ for DV. The shear rheology of the model can be quantitatively verified in the numerical simulations. There are indeed exact analytical results we can get by solving the constitutive equations for the hydrodynamical problem of steady shear flow, $u_x=\dot{\gamma} y$, $u_y=u_z=0$: both the polymer feedback stress and the first normal stress difference $N_1$ for the FENE-P model~\cite{bird,Lindner03} follow
\begin{eqnarray}\label{lindner}
\frac{\eta_P}{\tau_P} f(r_P) {C}_{xy} &=& \frac{2 \eta_P}{\tau_P} \left(\frac{L^2}{6} \right)^{1/2} \times \nonumber \\ 
&& \sinh \left(\frac{1}{3} \mbox{arcsinh} \left(\frac{\Lambda L^2}{4} \left(\frac{L^2}{6}\right)^{-3/2}\right) \right) \label{S}
\end{eqnarray}
\begin{eqnarray}\label{lindner2}
N_1 &=& \frac{\eta_P}{\tau_P} f(r_P) ({C}_{xx}-{C}_{yy})=8 \frac{\eta_P}{\tau_P} \left(\frac{L^2}{6} \right) \times \nonumber \\ 
&& \sinh^2 \left(\frac{1}{3} \mbox{arcsinh} \left(\frac{\Lambda L^2}{4} \left(\frac{L^2}{6}\right)^{-3/2}\right) \right) \label{N1}
\end{eqnarray}
where $\Lambda=\dot{\gamma} \tau_P$ is the dimensionless shear. The validity of both Eqs. (\ref{lindner}) and (\ref{lindner2}) is benchmarked in Fig. \ref{fig:1}(a)-(d): numerical simulations have been carried out in three dimensional domains with $H \times H \times H=  20 \times 20 \times 20$ cells. Periodic boundary conditions are applied in the stream-flow (x) and in the transverse-flow (z) directions while two walls are located at $y=0$ and $y=H$. The linear shear flow $u_x=\dot{\gamma} y$, $u_y=u_z=0$ is imposed in the numerics by applying two opposite velocities in the stream-flow direction ($u_x(x,y=0,z)=-u_x(x,y=H,z)=U_w$) at the upper ($y=H$) and lower wall ($y=0$) with the bounce-back rule \cite{Gladrow00}. We next change the shear in the range $10^{-6} \le 2U_w/H \le 10^{-2}$ lbu (lattice Boltzmann units) and the polymer relaxation time in the range $10^1 \le \tau_P \le 10^5$ lbu at fixed $\eta_P$. The various quantities are made dimensionless with the viscosity $\eta_P$ and the relaxation time $\tau_P$, and they are plotted as a function of the dimensionless shear $\Lambda=\tau_P \dot{\gamma}$. The values of the conformation tensor are taken when the simulation has reached the steady state.  As we can see from the figures, all the numerical results well agree with equation (\ref{lindner}) and (\ref{lindner2}). In particular, both the shear stress and normal stress difference increase at large $\Lambda$ to exhibit a sublinear behaviour (see figures \ref{fig:1}(a) and \ref{fig:1}(c)) which directly relates to thinning effects in the dimensionless polymer shear viscosity, $f(r_P) {C}_{xz}/\Lambda$, and first normal stress coefficient, $\Psi_1=f(r_P) ({C}_{xx}-{C}_{yy})/\Lambda^2$ (see figures \ref{fig:1}(b) and \ref{fig:1}(d)). In the limit of Hookean dumbbells (Oldroyd-B limit, $L^2 \gg 1$) we can use the asymptotic expansion of the hyperbolic functions and we get $N_1 \approx 2 \tau_P \eta_P \dot{\gamma}^2$, so that 
\be\label{Desimple}
\De=\frac{\tau_P}{\tau_{H}} \frac{\eta_P}{\eta_{c,d}+\eta_P}.
\ee
Equation (\ref{Desimple}) shows that $\De$ is clearly dependent on the ratio between the polymer relaxation time $\tau_P$ and the time $\tau_H$ defined as
\be\label{emulsiontime}
\tau_{H}=\frac{H (\eta_{c,d}+\eta_P)}{\sigma}
\ee
which represents the relaxation time of a droplet with characteristic size $H$, determined by viscous and Capillary forces. Finally, in figures \ref{fig:1}(e)-(f) we benchmark our numerical simulations under the effect of a steady elongational flow, $u_z=\dot{\epsilon} z$, $u_x=-\dot{\epsilon}x/2$, $u_y=-\dot{\epsilon} y/2$, with $\dot{\epsilon}$ the elongation rate. The associated normal stress difference $f(r_P) ({C}_{zz}-{C}_{xx})$ developed in steady conditions can also be evaluated analytically~\cite{bird,Lindner03} and exact expressions are here omitted for simplicity. Similarly to the shear rheology, we have carried out numerical simulations in a three dimensional cubic domain of edge $H$ consisting of $H \times H \times H = 20 \times 20 \times 20$ cells. Periodic conditions are now applied in all directions. The elongational rate is changed in the range $10^{-6} \le \dot{\epsilon} \le 10^{-2}$ lbu and the polymer relaxation time in the range $10^3 \le \tau_P \le 10^5$ lbu. Again, the various quantities are made dimensionless with the viscosity $\eta_P$ and the relaxation time $\tau_P$, and they are plotted as a function of the dimensionless extensional rate $\Lambda_e=\tau_P \dot{\epsilon}$ for different values of $L^2$: for small $\Lambda_e$ the elongational viscosity is three times the polymer viscosity, while at large $\Lambda_e$ we approach another constant value dependent on the finite extensibility parameter $L^2$. \\ 
Numerical simulations for the flow-focusing geometry will be performed with $L^2=10^4$: we note that the value of $L^2$ chosen rules out important thinning effects up to $\Lambda \approx \dot{\gamma} \tau_P \approx 1-10$, which are above the values achieved in our simulations. We emphasize that the choice of a finite (but large) $L^2$ is the one that maximizes the effects of viscoelasticity, while keeping the computation stable. Such choice is indeed  instrumental to ensure upper bounds for the polymer elongational viscosity in steady elongational flows, that would tend to diverge for the Oldroyd-B limit as soon as dimensionless elongational rates of order one are used (see figures \ref{fig:1}(e)-(f)). Obviously, inspecting the importance of $L^2$ by performing simulations of the flow-focusing geometry with other values of $L^2$ \cite{Arratia08,Arratia09} is surely worth of investigation, but outside the scope of the present paper.

\begin{table*}[t!]
\begin{center}
   \begin{tabular}{@{\extracolsep{\fill}} |c|c|c|c|c|c|c|c|c|c|c|c|}
    \hline
    $\Ca$ & $Q$ &$L_{x} \times L_y \times H$ &  $\eta_d$ & $\eta_c$  & $\eta_P$  & $\tau_P$ & $\De$ \\
   & & cells & lbu & lbu & lbu & lbu & \\
   \hline \hline
    $0.0056$ & $ 0.5-4$ & $1150 \times 250 \times 50$ & $1.75$ & $1.75$ & $0.00$  &                 $   $ & $ $ \\
    $0.0056$ & $ 0.5-4$ & $1150 \times 250 \times 50$ & $1.05$ & $1.75$ & $0.69$  &  $5-80 \times 10^2$ & $0.2-3.2$ \\
    $0.0056$ & $ 0.5-4$ & $1150 \times 250 \times 50$ & $1.75$ & $1.05$ & $0.0$   &  $5-80 \times 10^2$ & $0.2-3.2$ \\
\hline
    $0.007$ & $ 0.5-4$ & $1150 \times 250 \times 50$ & $1.75$ & $1.75$ & $0.00$   &                 $   $ & $ $ \\
    $0.007$ & $ 0.5-4$ & $1150 \times 250 \times 50$ & $1.05$ & $1.75$ & $0.69$ &  $5-80 \times 10^2$ & $0.2-3.2$ \\
    $0.007$ & $ 0.5-4$ & $1150 \times 250 \times 50$ & $1.75$ & $1.05$ & $0.0$ &   $5-80 \times 10^2$ & $0.2-3.2$ \\
\hline
   \end{tabular}
\end{center}
\caption{\small
Parameters for the numerical simulations with flow-focusing geometry: $\Ca$ is the Capillary number, which quantifies the importance of the viscous forces with respect to the surface tension forces, and $Q$ is the flow-rate ratio of the two fluids. The flow-focusing geometry is embedded in a rectangular parallelepiped with size $L_{x}  \times L_y \times H$, where $H$ is the characteristic edge of the channels with square ($H \times H$) cross section (see also figure \ref{fig:sketch} for a sketch). The dynamic viscosity of the Newtonian solvent inside the dispersed phase is $\eta_d$, while $\eta_c$ is the dynamic viscosity of the Newtonian solvent inside the continuous phase. $\eta_P$ is the viscosity of the polymers in the dispersed or continuous phase. The polymer relaxation time is $\tau_P$, while $\De$ is the Deborah number based on definition \eqref{Desimple}. \label{table:para}}
\end{table*}

%%%%%%%%%%%%%%%%FIG 2%%%%%%%%%%%%%%%%%%%%%%%%%%%%%%%%%%%%%%%%%%%%%%%%%%%%%%%%%%%%%%%%
\begin{figure}
%\begin{center}
\includegraphics[width = 0.9\linewidth]{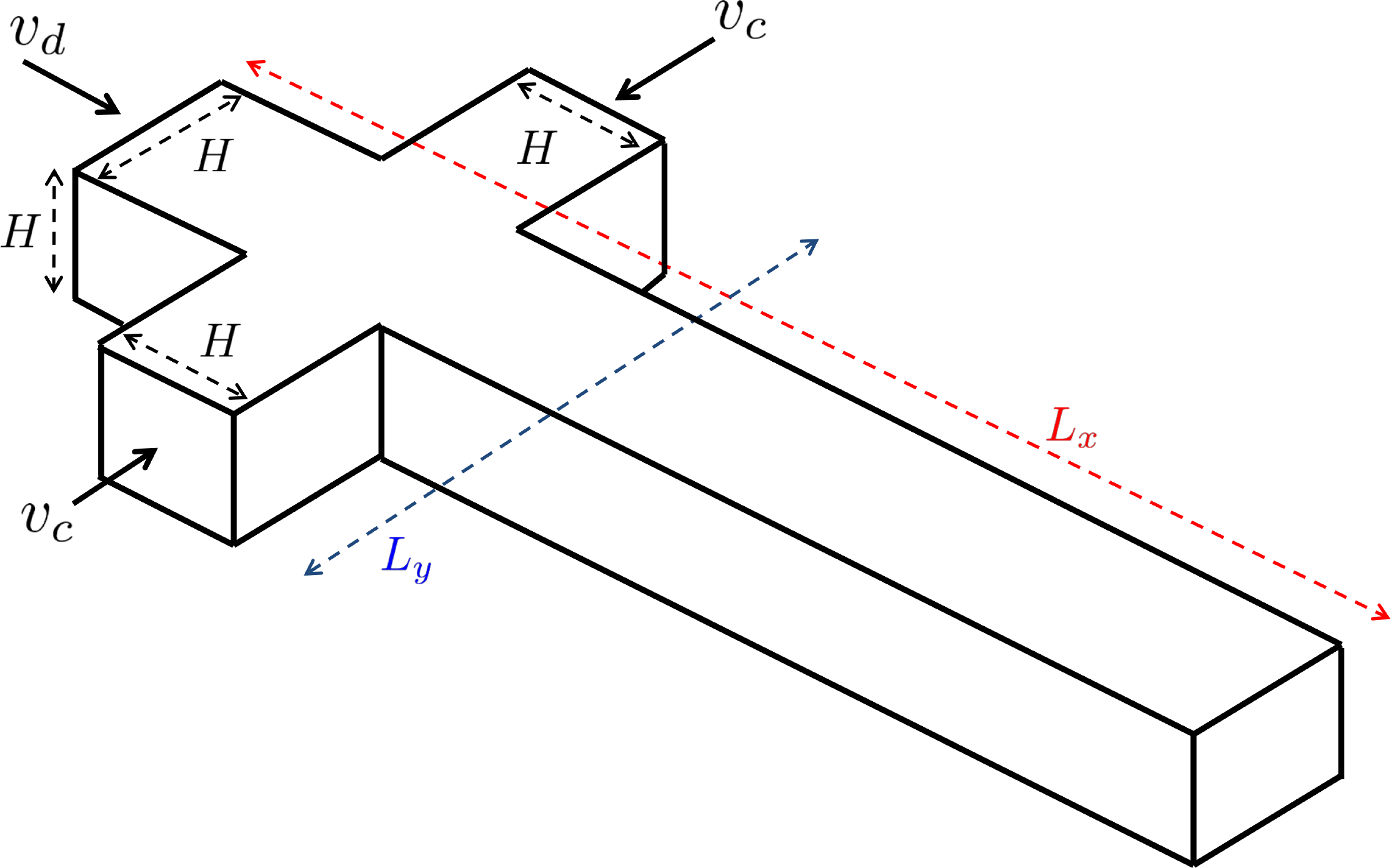}
\caption{A sketch for the flow-focusing geometry used \label{fig:sketch}.}
%\end{center}
\end{figure}

\section{Results and Discussions}\label{sec:results} 

\begin{figure*}[t!]
\makeatletter
\def\@captype{figure}
\makeatother
\subfigure[\,\,$t=t_0+1.6 \tshear$, $Q = 1.0$]
{
\includegraphics[width=.235\textwidth]{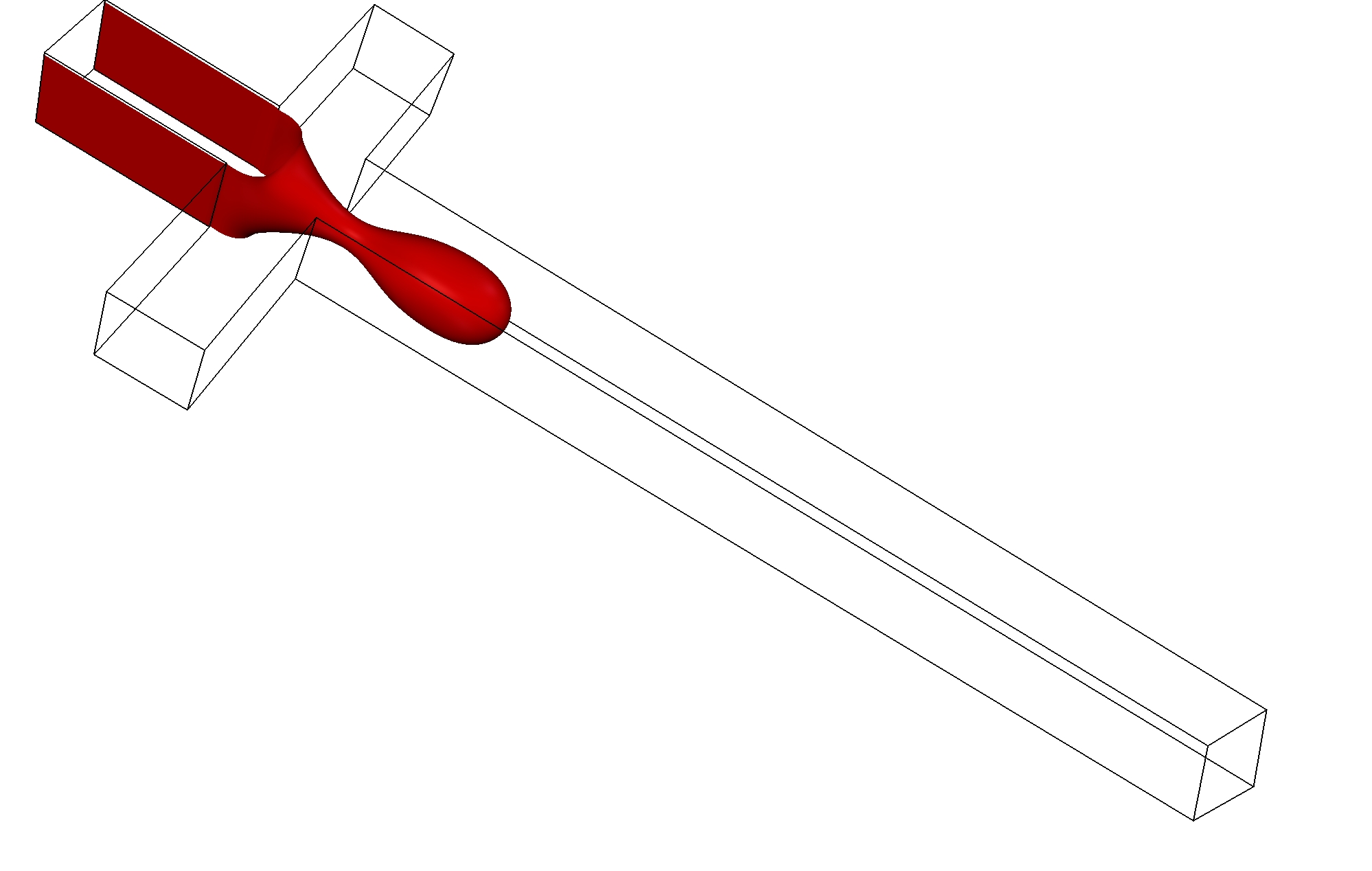}
}
\subfigure[\,\,$t=t_0+2.3 \tshear$, $Q = 1.0$]
{
\includegraphics[width=.235\textwidth]{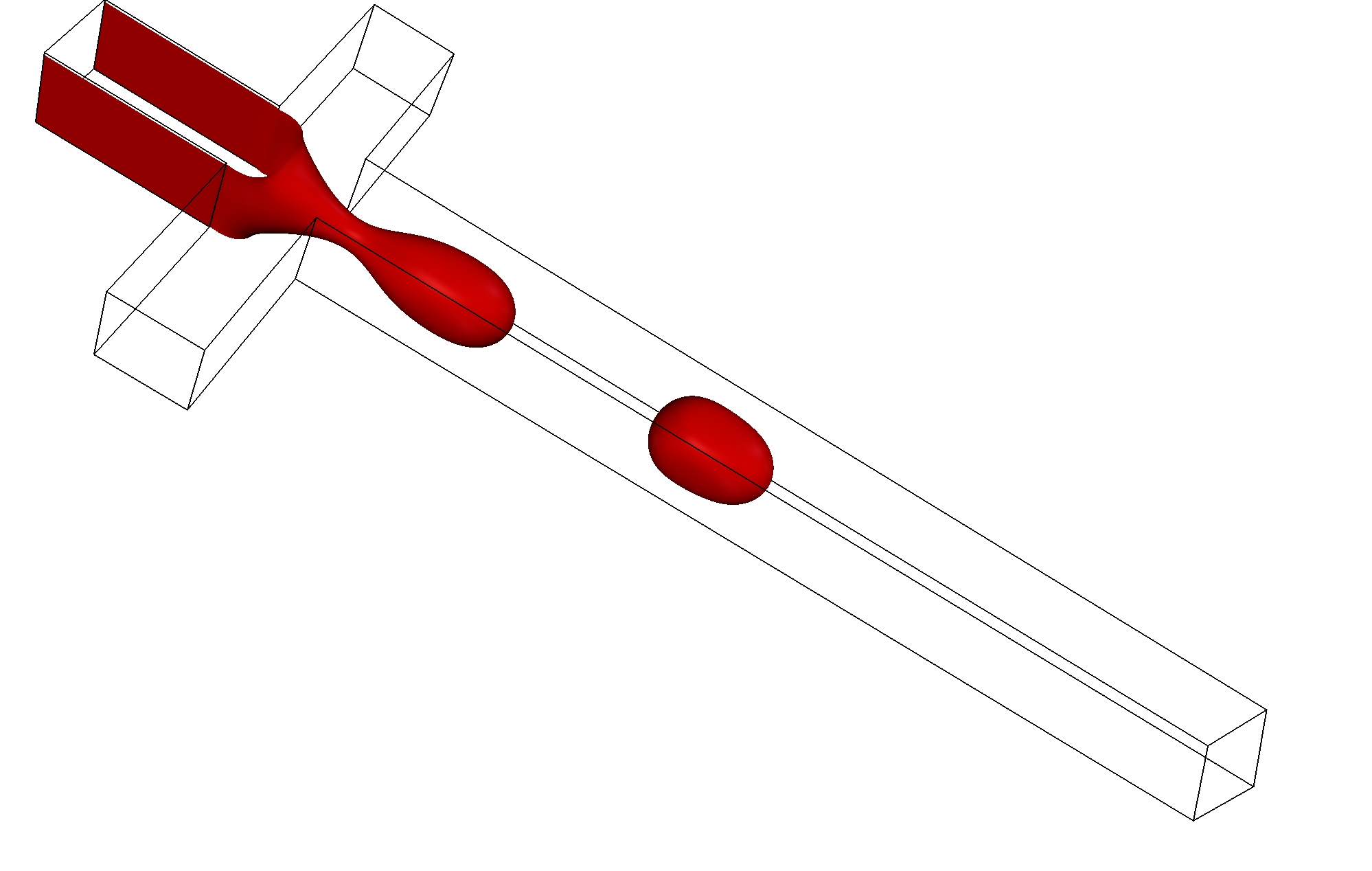}
}
\subfigure[\,\,$t=t_0+3.0 \tshear$, $Q = 1.0$]
{
\includegraphics[width=.235\textwidth]{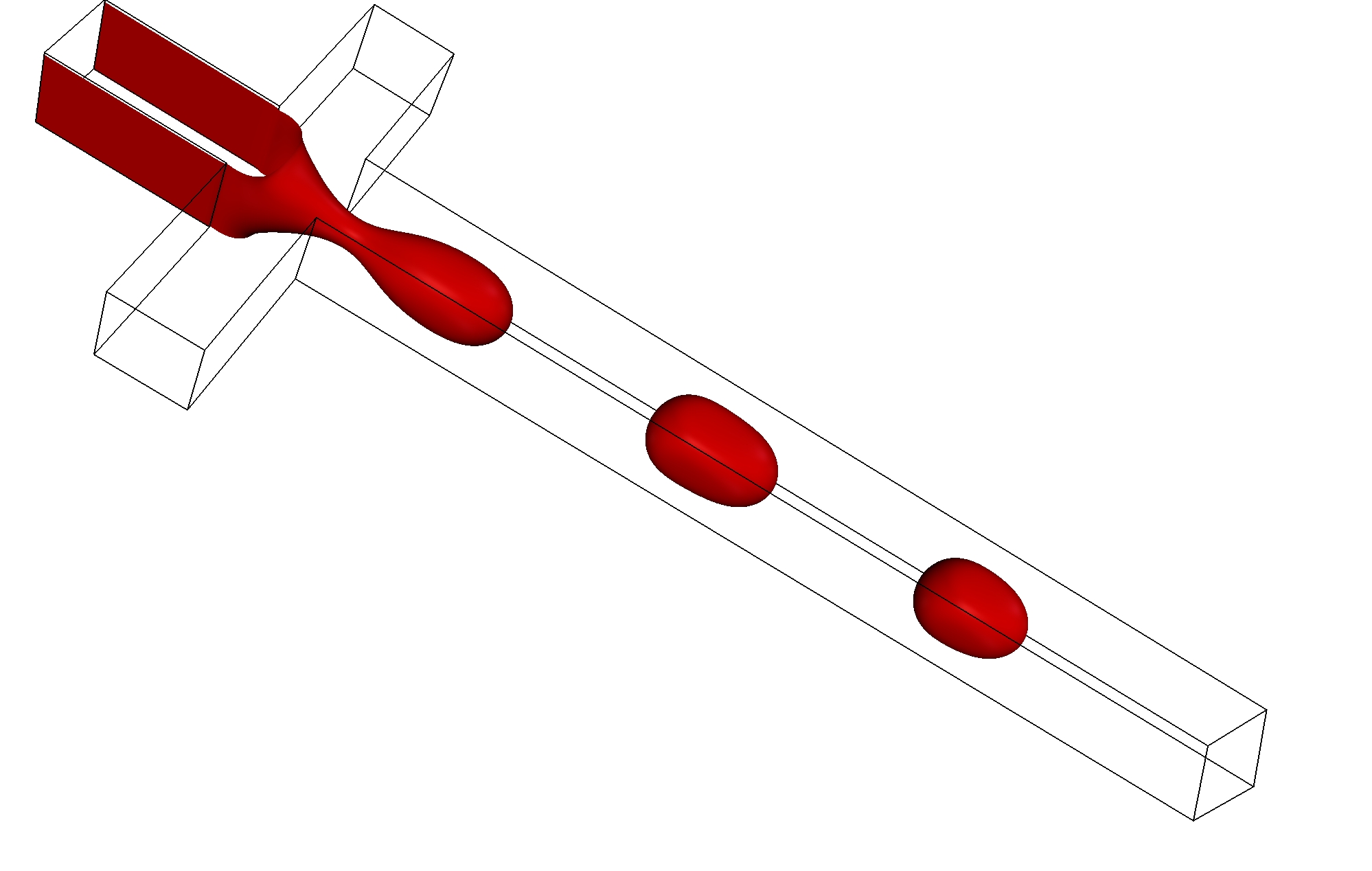}
}
\subfigure[\,\,$t=t_0+3.8 \tshear$, $Q = 1.0$]
{
\includegraphics[width=.235\textwidth]{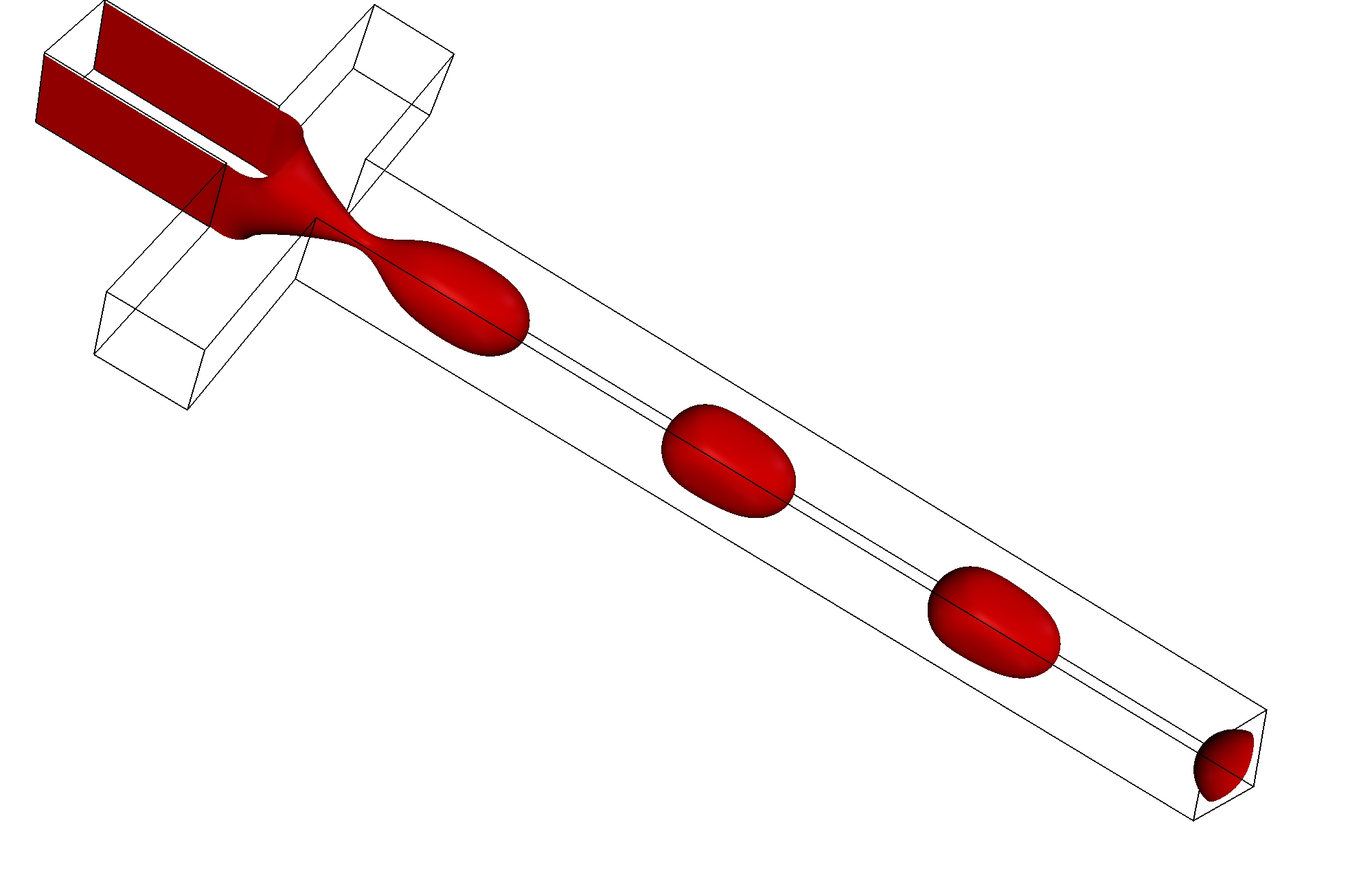}
}
\\
\subfigure[\,\,$t=t_0+1.1 \tshear$, $Q = 3.0$]
{
\includegraphics[width=.235\textwidth]{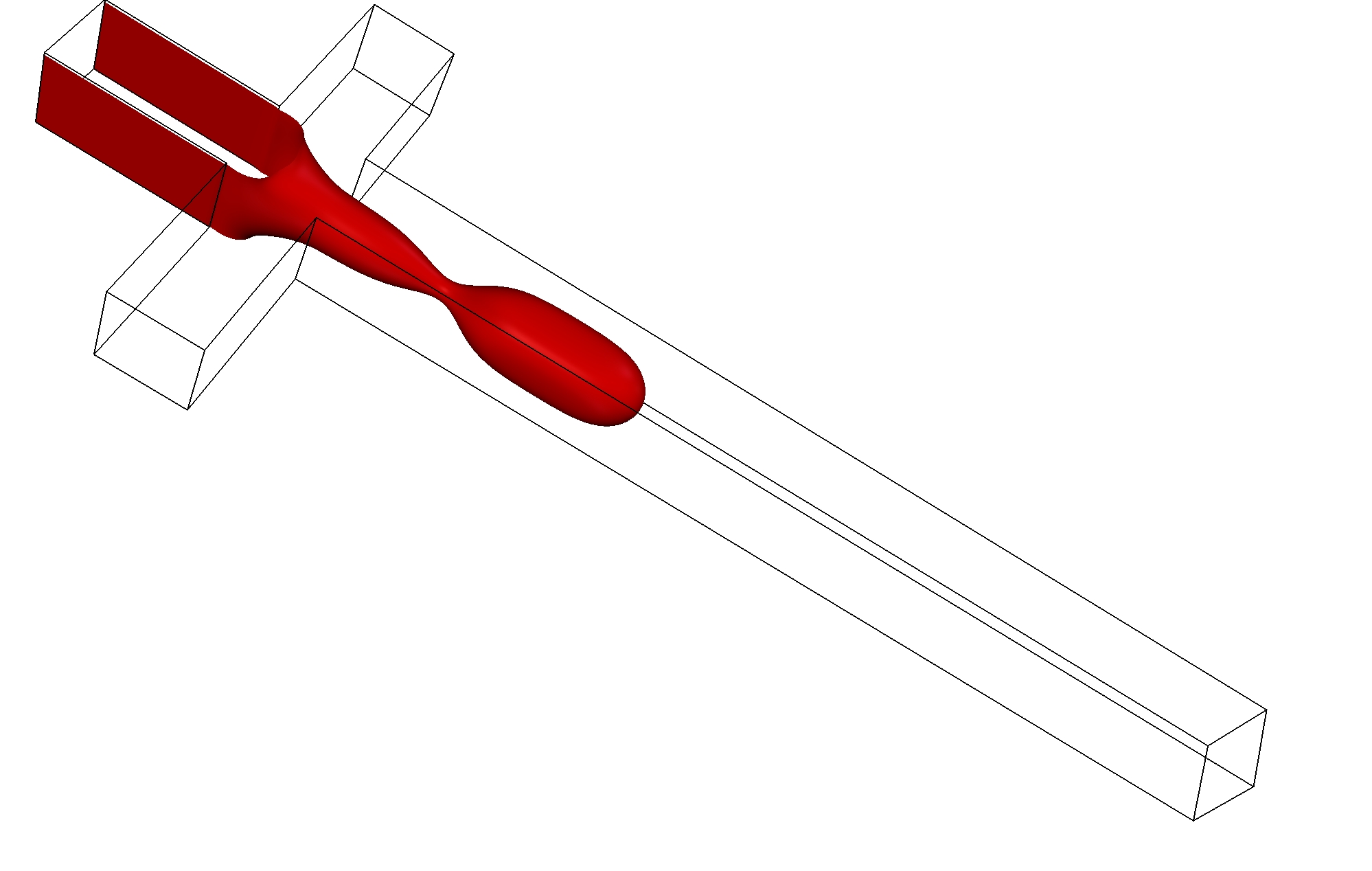}
}
\subfigure[\,\,$t=t_0+1.6 \tshear$, $Q = 3.0$]
{
\includegraphics[width=.235\textwidth]{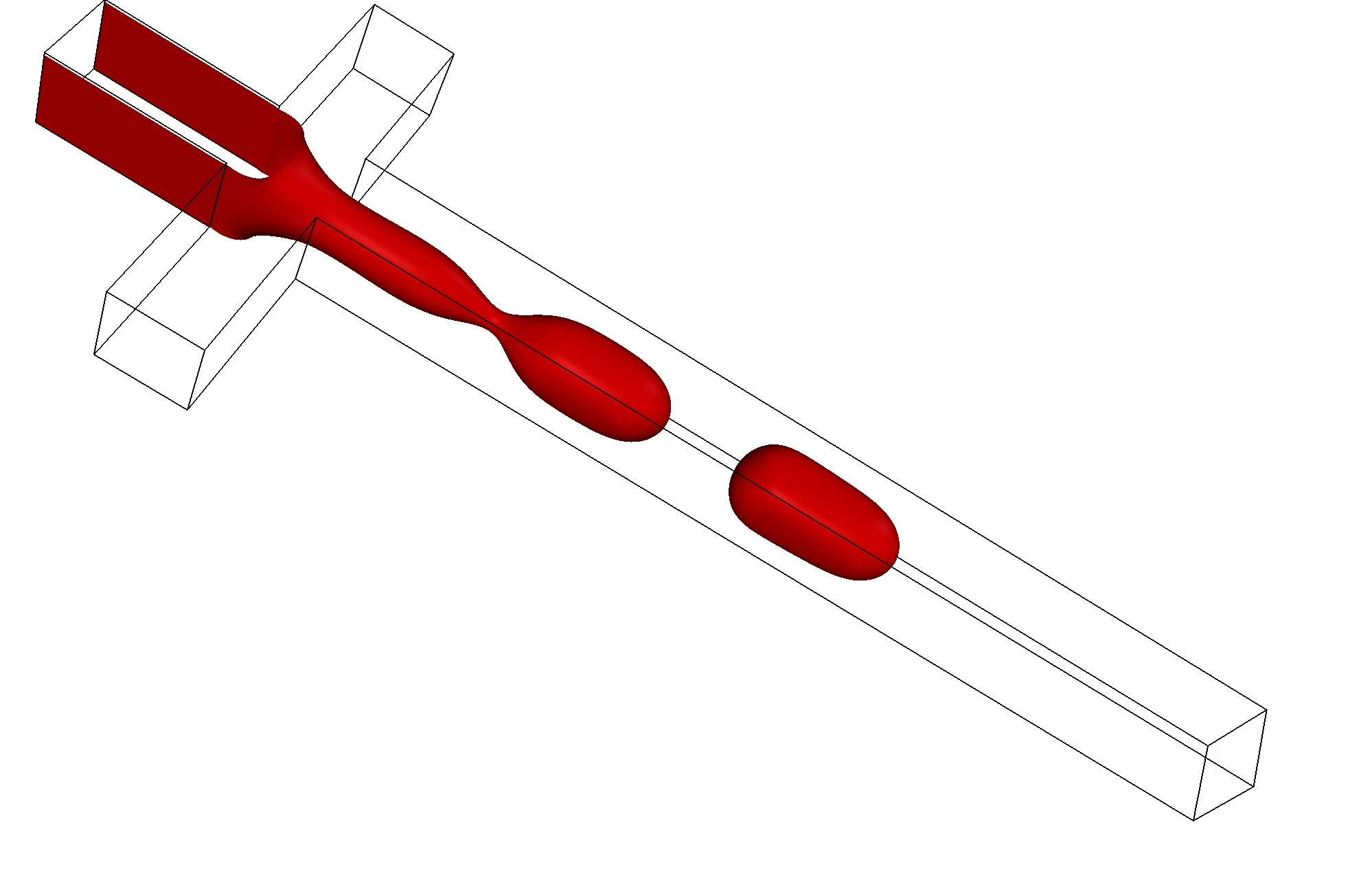}
}
\subfigure[\,\,$t=t_0+2.1 \tshear$, $Q = 3.0$]
{
\includegraphics[width=.235\textwidth]{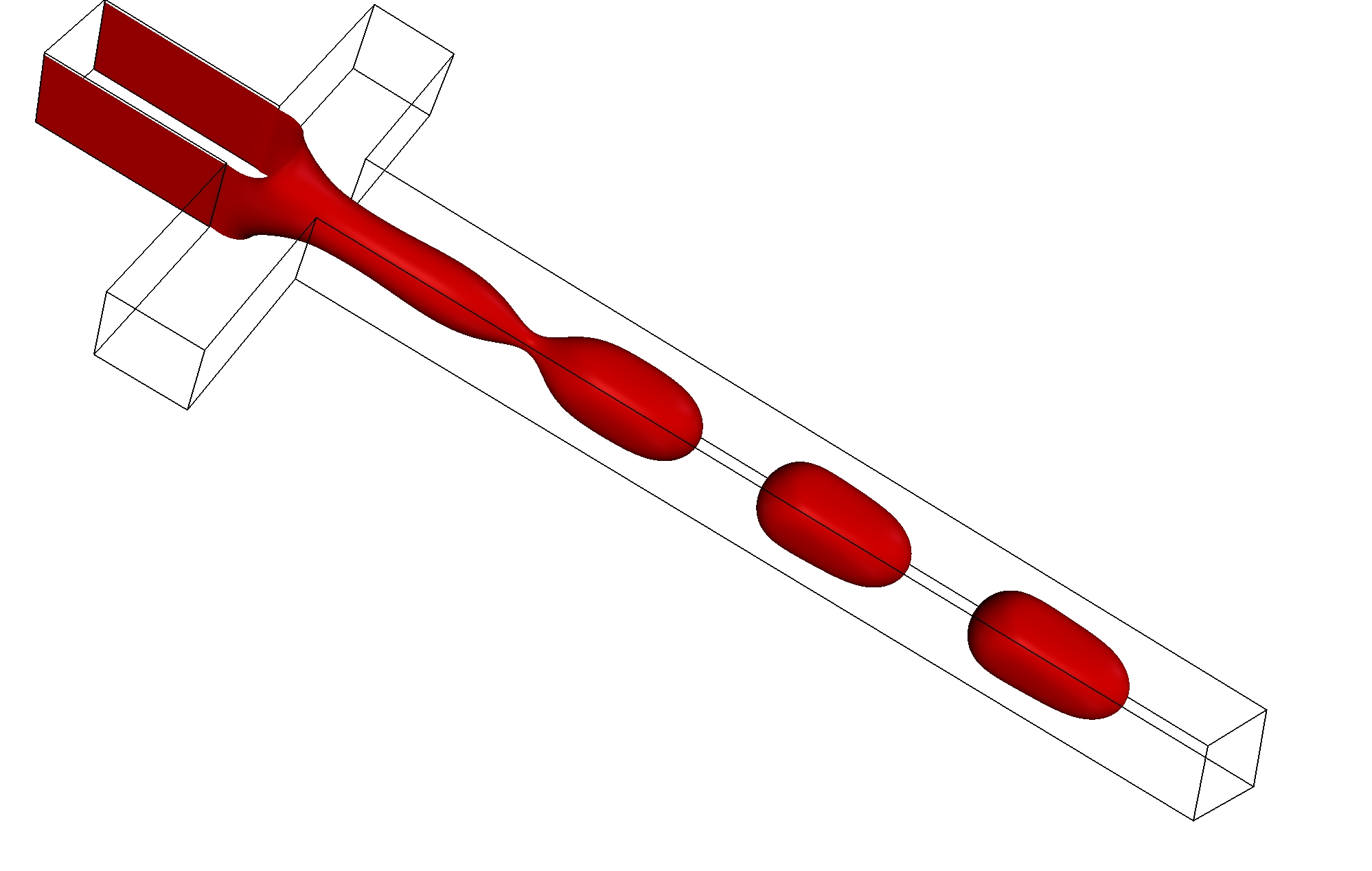}
}
\subfigure[\,\,$t=t_0+2.5 \tshear$, $Q = 3.0$]
{
\includegraphics[width=.235\textwidth]{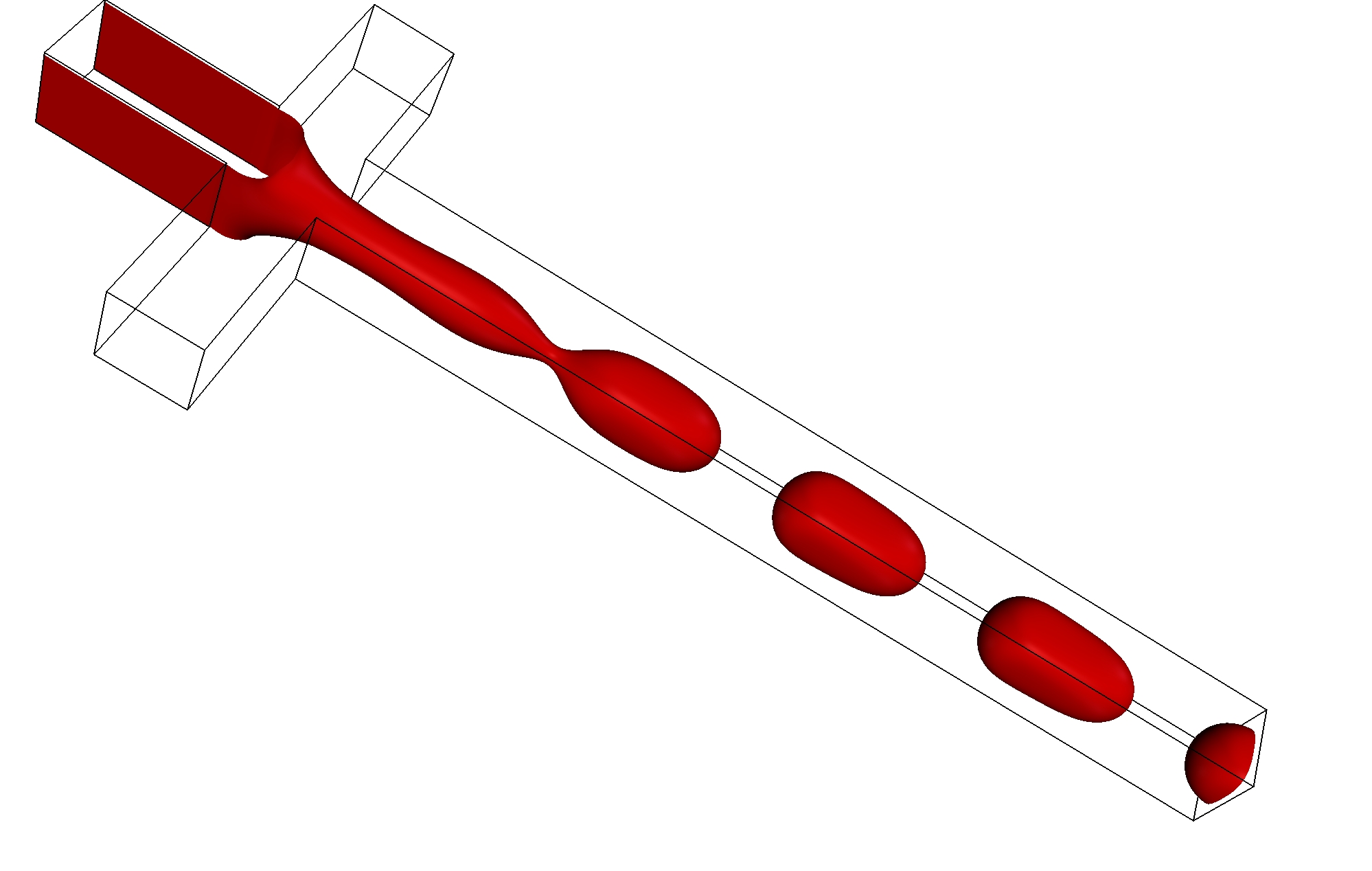}
}
\\
\subfigure[\,\,$t=t_0+0.6 \tshear$, $Q = 4.0$]
{
\includegraphics[width=.235\textwidth]{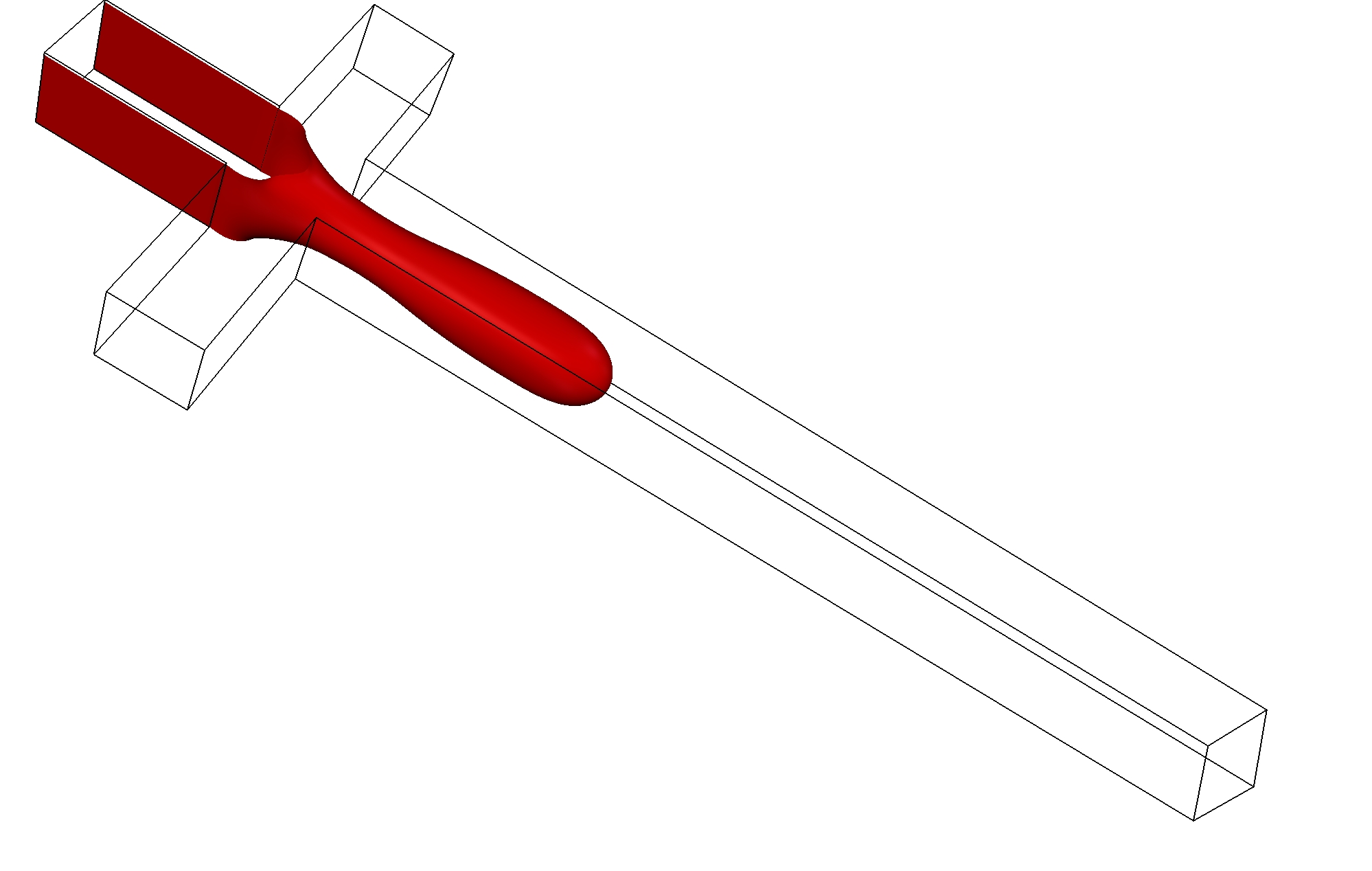}
}
\subfigure[\,\,$t=t_0+1.2 \tshear$, $Q = 4.0$]
{
\includegraphics[width=.235\textwidth]{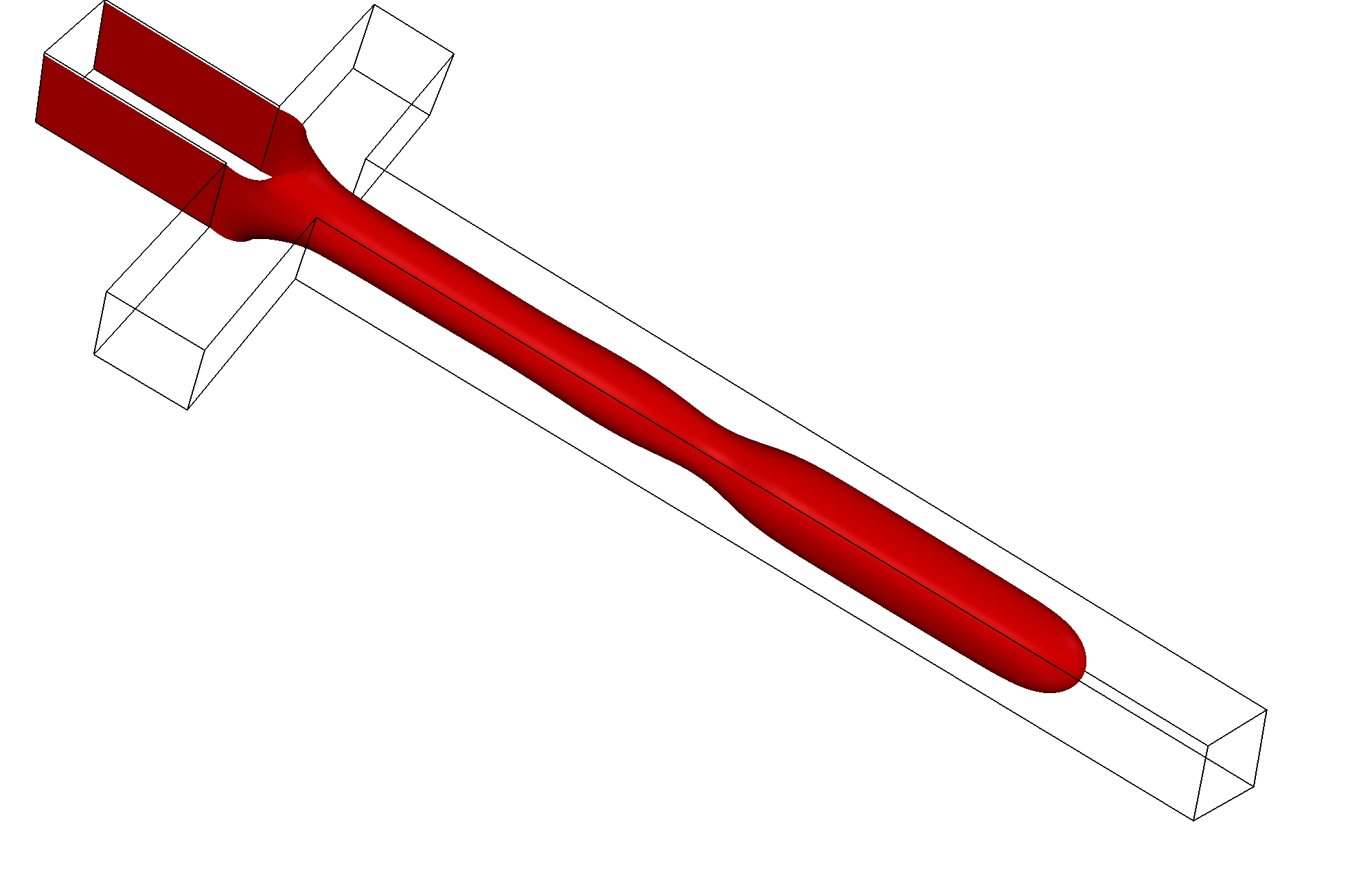}
}
\subfigure[\,\,$t=t_0+1.8 \tshear$, $Q = 4.0$]
{
\includegraphics[width=.235\textwidth]{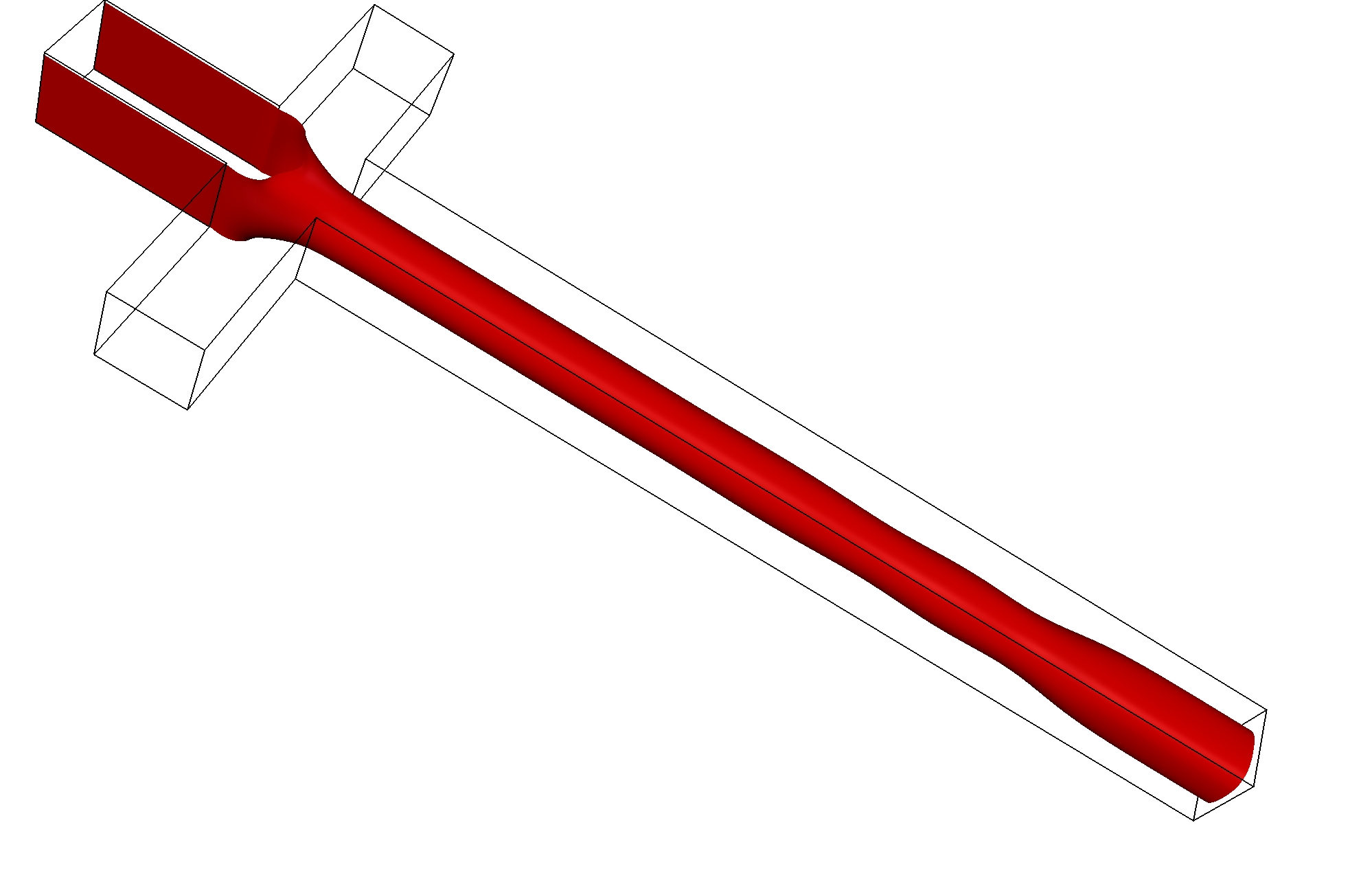}
}
\subfigure[\,\,$t=t_0+2.4 \tshear$, $Q = 4.0$]
{
\includegraphics[width=.235\textwidth]{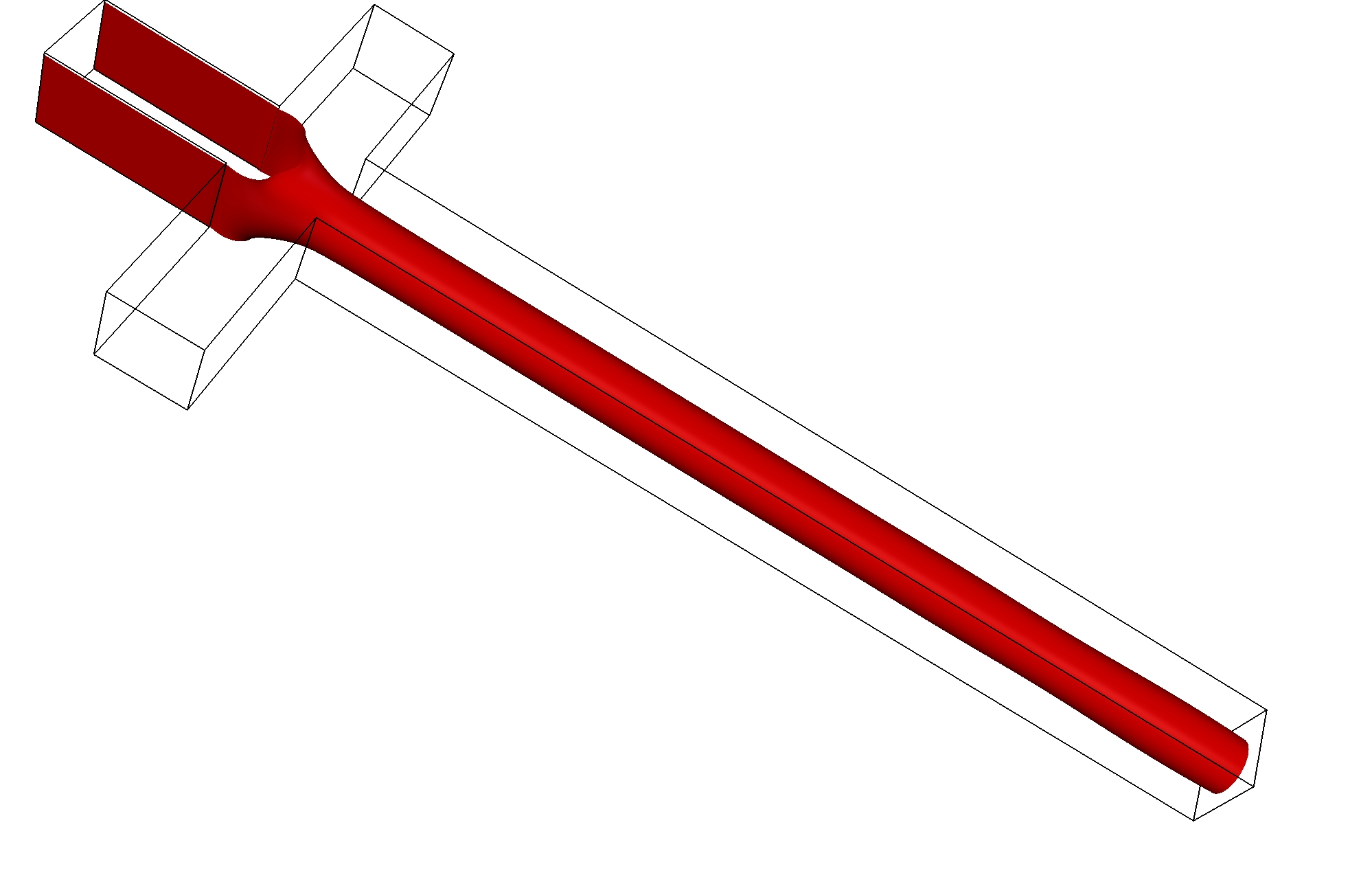}
}
\caption{Dynamics and Break-up of liquid threads in the flow-focusing geometry from simulations with Newtonian phases. The dispersed (d) phase (with viscosity $\eta_d$) enters at the inlet of the main channel with average velocity $v_d$ and periodically breaks due to forces created by the cross-flowing continuous (c) phase from the side channels, the latter characterized by a viscosity $\eta_c$ and average velocity $v_c$. We consider the simplest case of square ducts with edge $H$. The various hydrodynamical parameters can be grouped into the following dimensionless numbers: the Capillary number calculated for the continuous phase, $\Ca =\eta_c v_c/\sigma$, the Reynolds number $\Ren =\rho v_c H/ \eta_c$, the viscosity ratio $\lambda=\eta_d/\eta_c$, and the flow-rate ratio $Q=v_d/v_c=Q_d/Q_c$, where $Q_d=v_d H^2$ and $Q_c =v_c H^2$ are the flow-rates at the two inlets. The case reported in this figure corresponds to $\Ca=0.007$, $\Ren=0.018$, $\lambda=1$. The first row shows results for $Q=1.0$ (DCJ regime); the second row corresponds to $Q=3.0$ (DC regime); the third row corresponds to $Q=4.0$ (PF regime). The various regimes (DCJ, DC, PF) are described in the text. In all cases we have used the characteristic shear time $\tshear=H/v_c$ as a unit of time, while $t_0$ is a reference time (the same for all simulations).}
\label{fig:2}
\end{figure*}

In this section we report the results for the break-up of liquid threads in the confined flow-focusing geometry at fixed Capillary number. We aim at exploring the effects of both MV and DV on the scenarios which are known for Newtonian phases. In particular, Liu \& Zhang~\cite{LiuZhang11} performed LB simulations and systematically characterized three typical flow patterns for Newtonian phases, dependently on the flow-rate ratio $Q$. These flow patterns are actually well reproduced by our numerical simulations, as shown in figure~\ref{fig:2} for $\Ca = 0.007$. At very low flow-rate ratio $Q$, the droplets are formed at the cross-junction (DCJ) due to the squeezing mechanism~\cite{Demenech07,Garstecki06}. Panels (a)-(d) show the liquid thread prior to the first, second, third and fourth break-up at time $t = t_0 + 1.6 \tshear$, $t = t_0 + 2.3 \tshear$, $t = t_0 + 3.0 \tshear$, and $t = t_0 + 3.8 \tshear$, respectively.  Notice that we have used the characteristic shear time $\tshear=H/v_c$ as a unit of time, while $t_0$ is a reference time (the same for all simulations). As we can see, the action of the co-flowing liquid periodically breaks droplets and the break-up point is stable in time. Upon increasing  $Q$, droplets are found to pinch-off downstream of the cross-junction (DC). Panels (e)-(h) refer to a flow-rate ratio $Q=3.0$ and show the liquid thread at time $t = t_0 + 1.1 \tshear$, $t = t_0 + 1.6 \tshear$, $t = t_0 + 2.1 \tshear$, and $t = t_0 + 2.5 \tshear$, respectively. It is evident that the dispersed thread actually becomes unstable after a distance $L_b$ from the cross-junction and this distance increases as a function of time, which is a distinctive signature of the DC regime. Practically, $L_b$ is computed as the distance between the break-up point and the corner up-stream at the cross-junction. As the flow-rate ratio $Q$ increases to a critical value, stable parallel flows (PF) are observed, where the three incoming streams co-flow in parallel downstream of the cross junction without pinching. This is evidenced by Panels (i)-(l), showing the liquid thread dynamics at time $t = t_0 + 0.6 \tshear$, $t = t_0 + 1.2 \tshear$, $t = t_0 + 1.8 \tshear$, and $t = t_0 + 2.4 \tshear$, respectively. In this latter case, the flow-rate ratio is $Q=4.0$. In addition, the transitions from DCJ to DC and from DC to PF are influenced by the Capillary number. As $\Ca$ increases, the threshold value of flow-rate ratio at which the transition occurs decreases, and the width of the DC regime also decreases~\cite{LiuZhang11}. 

%%%%%%%%%%%%%%%%%%%%%%%%%%%%%%%%%%%%%%%%%%%%%%%%%%%%%%%%%%%%%%%%%%%%%%%%%%%%%%%%
%%%%%%%%%%%%%%%%%%%%%%%%%%%%%%%%%%%%%%%%%FIG 3%%%%%%%%%%%%%%%%%%%%%%%%%%%%%%%%%%%%%%%%%%%%%%%%%%%%%%%%%%%%%%%%%%%%%%%%%%%%%%%%%%%%%%%%%%%%%%%%%%%%%%%%%%%%%%%%%%

%%%%%%%%%%%%%%%%%%%%%%%%%%%%%%%%%%%%%%%%%%%%%%%%%%%%%%%%%%%%%%%%%%%%%%%%%%%%%%%%%%%%%%%%%%%%%%%%%%%%%%%%%%%%%%%%%%%%%%%%%%%%%%%%%%%%%%%%%%%%%%%%%%%%%%%%%%%%%%%%

\subsection{Effects of Matrix Viscoelasticity (MV)}\label{sec:MV}

%%%%%%%%%%%%%%%%%%%%%%%%%%%%%%%%%%%%%%%%%%%%%%%%%%%%%%%%%%%%%%%%%%%%%%%%%%%%%%%%
%%%%%%%%%%%%%%%%%%%%%%%%%%%%%%%%%%%%FIG 4%%%%%%%%%%%%%%%%%%%%%%%%%%%%%%%%%%%%%%%%%%%%%%%%%%%%%%%%%%%%%%%%%%%%%%%%%%%%%%%%%%%%%%%%%%%%%%%%%%%%%%%%%%%%%%%%%%%%%%%

\begin{figure*}[tbh!]
\makeatletter
\def\@captype{figure}
\makeatother
\subfigure[\,\,$t=t_0+3.0 \tshear$, $\De = 0$]
{
\includegraphics[width = 0.330\linewidth]{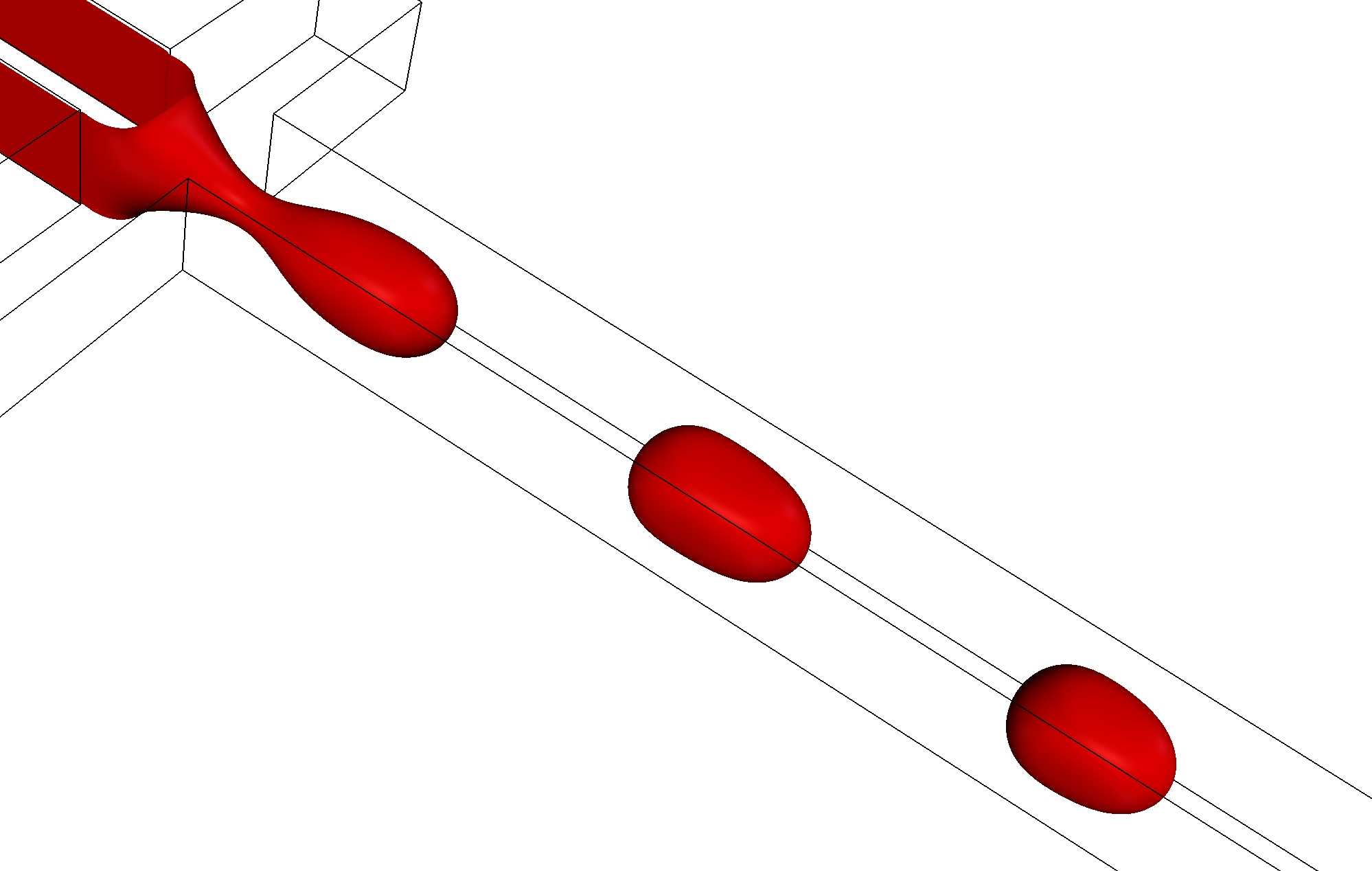}
}
\subfigure[\,\,$t=t_0+2.8 \tshear$, $\De = 0.2$]
{
\includegraphics[width = 0.330\linewidth]{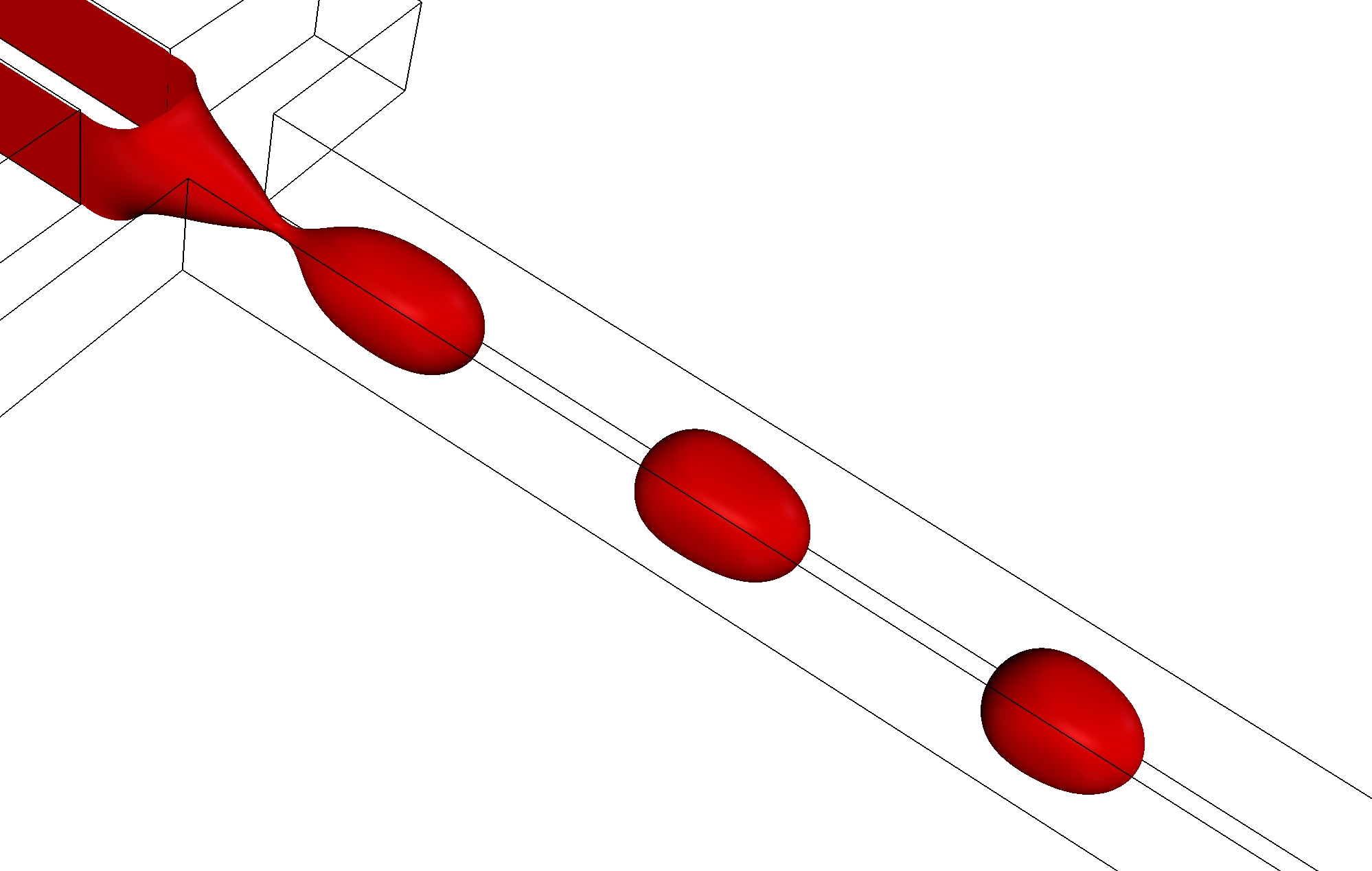}
}
\subfigure[\,\,$t=t_0+2.8 \tshear$, $\De = 2.0$]
{
\includegraphics[width = 0.330\linewidth]{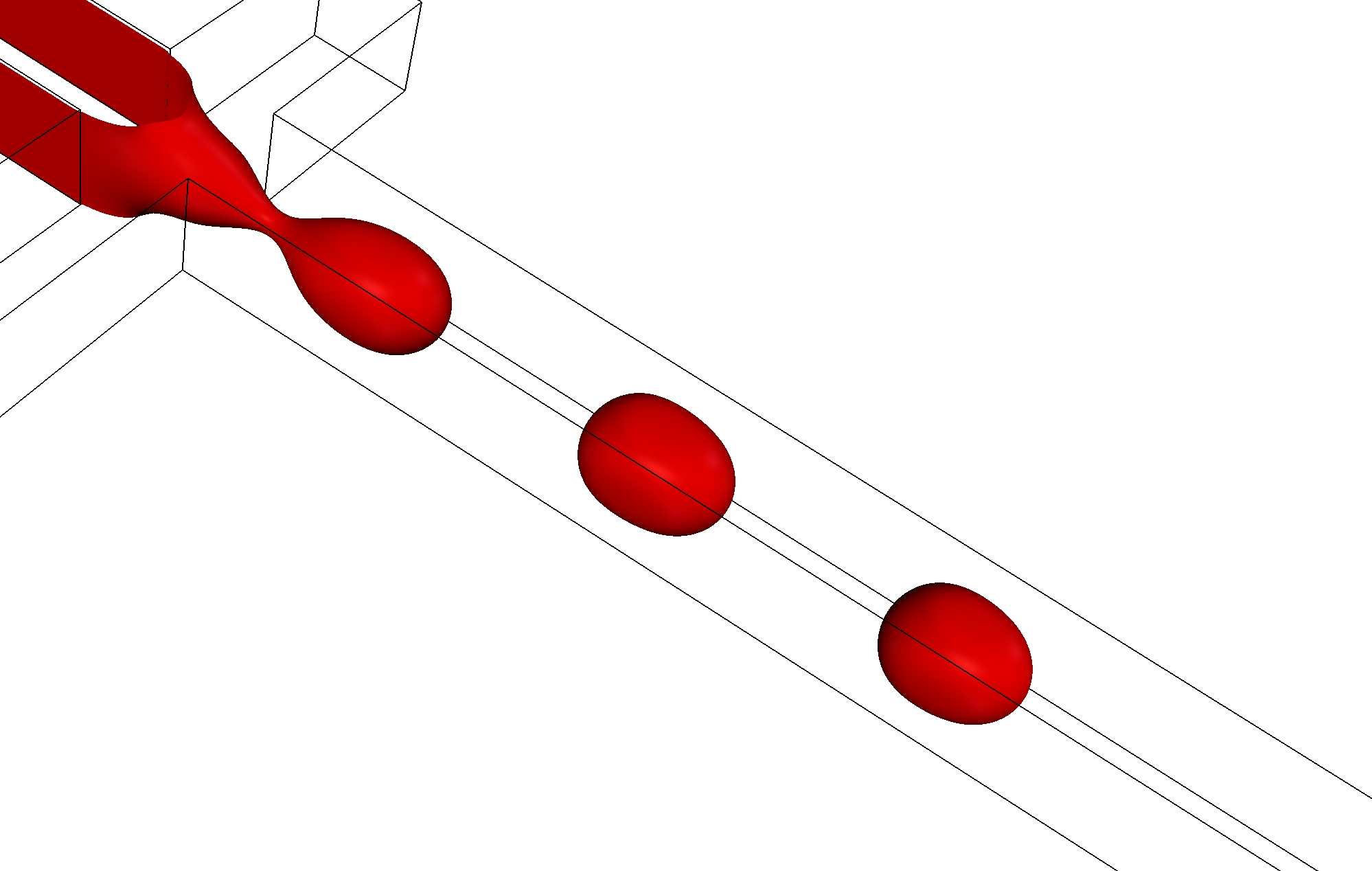}
}
\caption{Effects of matrix viscoelasticity (MV) on droplet formation in the flow-focusing geometry at fixed $\Ca=0.007$, $\Ren=0.018$, $\lambda=1$ and $Q=1.0$ (DCJ regime, see text for details). Three cases are compared: (a) Newtonian case ($\De = 0.0$) at time $t = t_0 + 3.0 \tshear$; (b) slightly MV case with $\De = 0.2$ at time $t = t_0 + 2.8 \tshear$; (c) MV case with $\De = 2.0$ at time $t = t_0 + 2.8 \tshear$. In all cases we have used the characteristic shear time $\tshear=H/v_c$ as a unit of time, while $t_0$ is a reference time (the same for all simulations).\label{fig:3}}
\end{figure*}

%%%%%%%%%%%%%%%%%%%%%%%%%%%%%%%%%%%%%%%%%%%%%%%%%%%%%%%%%%%%%%%%%%%%%%%%%%%%%%%%%%%%%%%%%%%%%%%%%%%%%%%%%%%%%%%%%%%%%%%%%%%%%%%%%%%%%%%%%%%%%%%%%%%%%%%%%%%%%%%%

%%%%%%%%%%%%%%%%%%%%%%%%%%%%%%%%%%%%%%%%%%%%%%%%%%%%%%%%%%%%%%%%%%%%%%%%%%%%%%%%%%%%%%%%%%%%%%%%%%%%%%%%%%%%%%%%%%%%%%FIG 5 %%%%%%%%%%%%%%%%%%%%%%%%%%%%%%%%%%%%%%%%%%%%%%%%%%%%%%%%%%%%%%%%%%%%%%%%%%%%%%%%%%%%%%%%%%%%%%%%%%%%%%%%%%%%%%%%%%%%%

\begin{figure*}[tbh!]
\makeatletter
\def\@captype{figure}
\makeatother
\subfigure[\,\,$t=t_0+3.0 \tshear$, $\De = 0.0$]
{
\includegraphics[width = 0.33\linewidth]{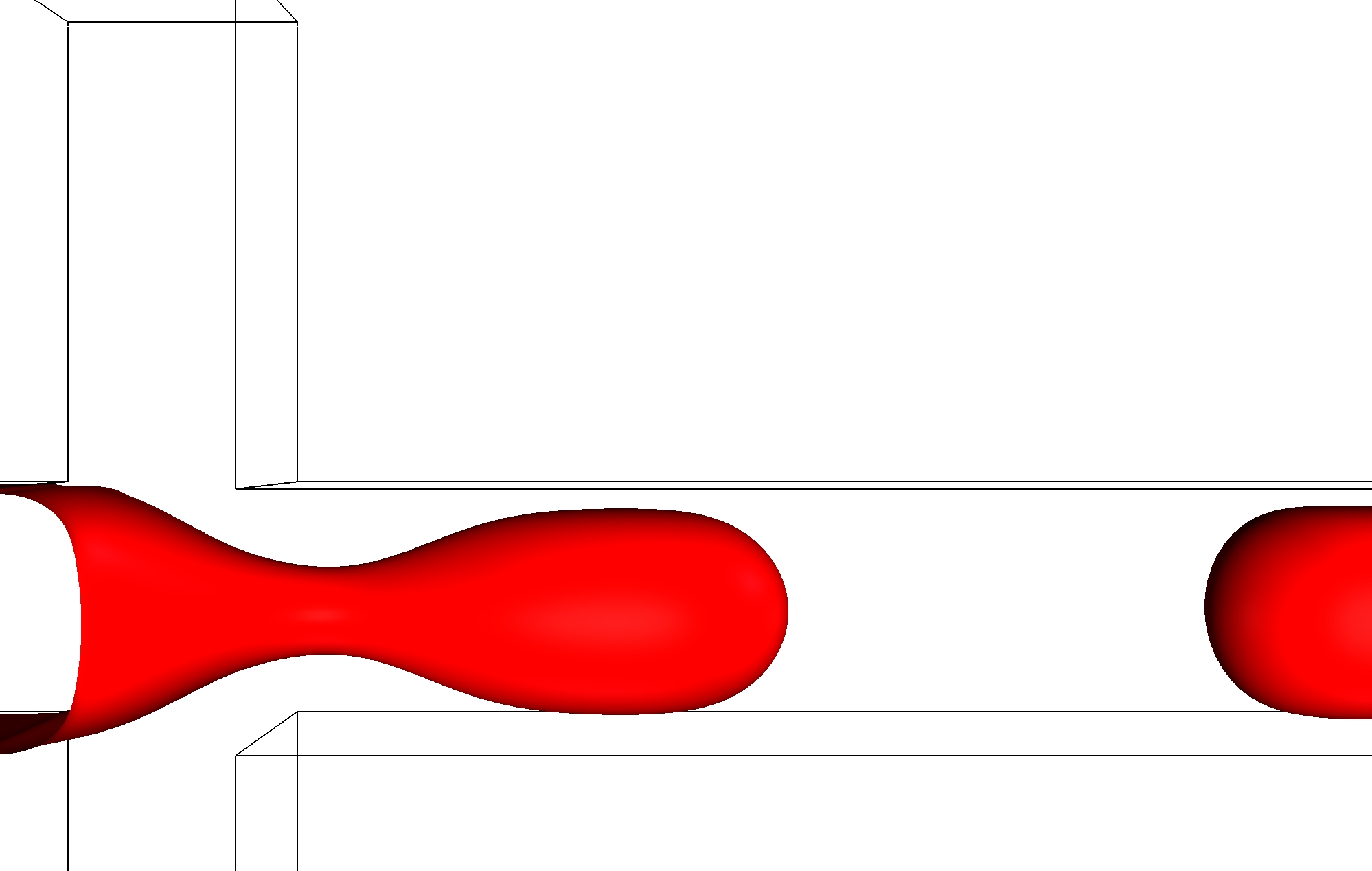}
}
\subfigure[\,\,$t=t_0+2.8 \tshear$, $\De = 0.2$]
{
\includegraphics[width = 0.33\linewidth]{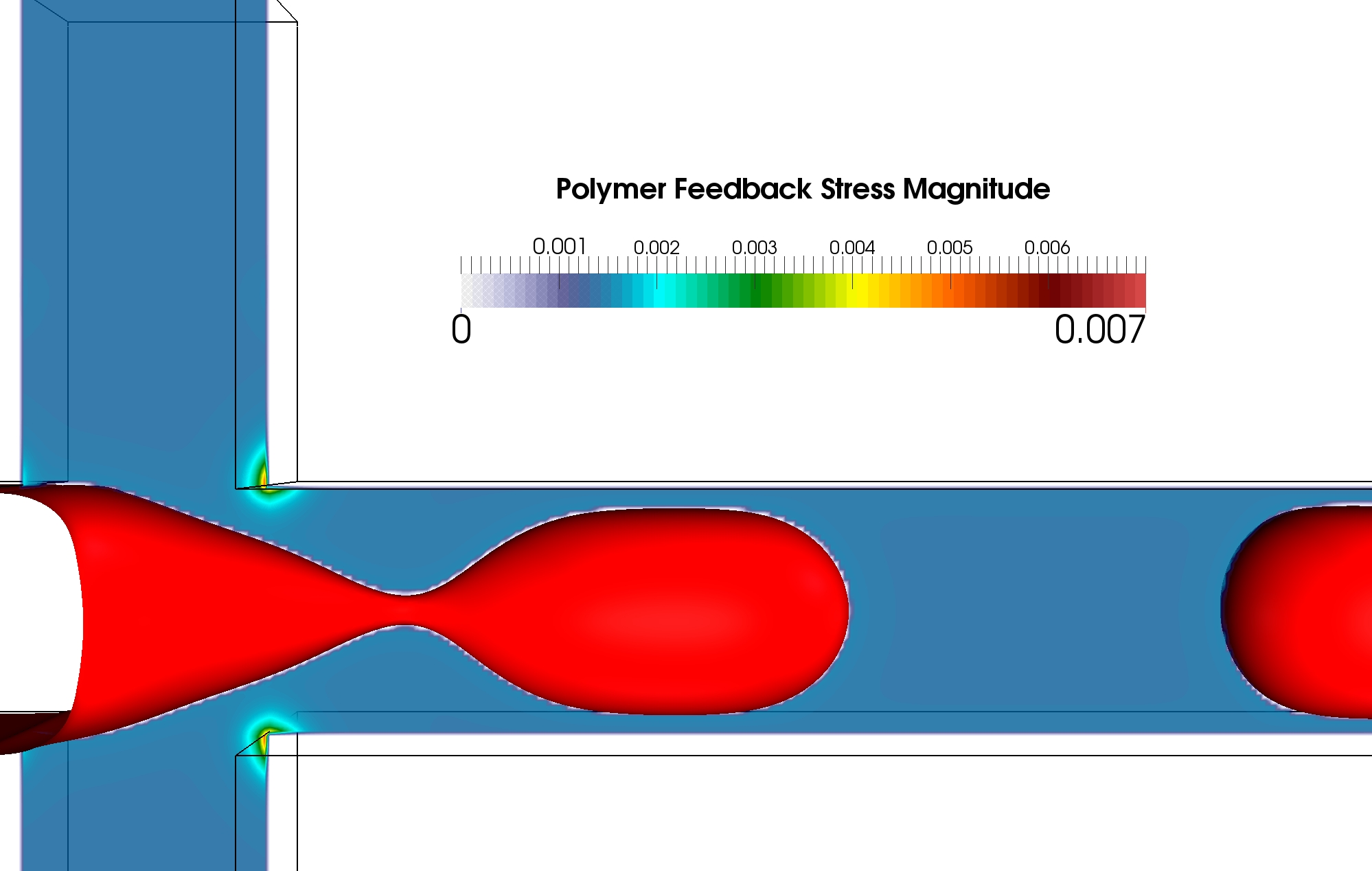}
}
\subfigure[\,\,$t=t_0+2.8 \tshear$, $\De = 2.0$]
{
\includegraphics[width = 0.33\linewidth]{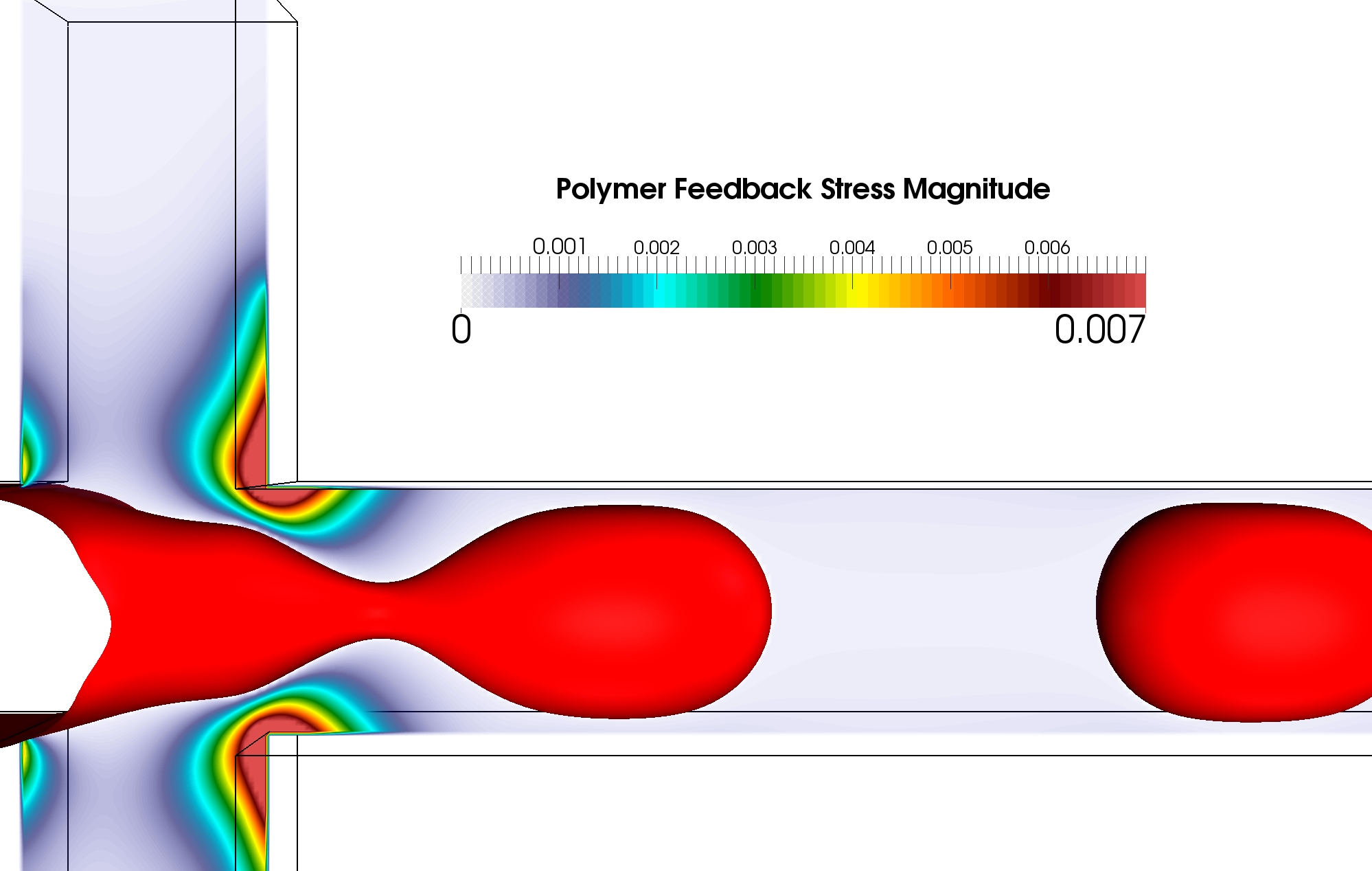}
}
\caption{Effects of matrix viscoelasticity (MV) in the DCJ regime. Density contours of the dispersed phase are overlaid on the polymer feedback stress magnitude (see Eq. \eqref{NSc}) for different values of the Deborah number $\De$ at fixed $\Ca=0.007$, $\Ren=0.018$, $\lambda=1$ and $Q=1.0$. The value of the flow-rate ratio $Q$ is such that the squeezing dominated-DCJ regime \cite{LiuZhang11} drives the break-up process. By definition, the Newtonian case (Panel (a)) has no feedback stress. As $\De$ is increased, we see that the flow in the continuous phase develops enhanced polymer feedback stress in the corners where the side channels meet the main channel. The feedback is higher downstream at the cross-junction, causing enhanced droplet deformation and quantitatively changing the break-up process (see also figure \ref{fig:6}). In all cases we have used the characteristic shear time $\tshear=H/v_c$ as a unit of time, while $t_0$ is a reference time (the same for all simulations).\label{fig:4}}
\end{figure*}

%%%%%%%%%%%%%%%%%%%%%%%%%%%%%%%%%%%%%%%%%%%%%%%%%%%%%%%%%%%%%%%%%%%%%%%%%%%%%%%%%%%%%%%%%%%%%%%%%%%%%%%%%%%%%%%%%%%%%%%%%%%%%%%%%%%%%%%%%%%%%%%%%%%%%%%%%%%%%%%%

%%%%%%%%%%%%%%%%%%%%%%%%%%%%%%%%%%%%%%%%%%%%%%%%%%%%%%%%%%%%%%%%%%%%%%%%%%%%%%%%%%%%%%%%%%%%%%%%%%%%%%%%%%%%%%%%%%%%%%FIG 6 %%%%%%%%%%%%%%%%%%%%%%%%%%%%%%%%%%%%%%%%%%%%%%%%%%%%%%%%%%%%%%%%%%%%%%%%%%%%%%%%%%%%%%%%%%%%%%%%%%%%%%%%%%%%%%%%%%%%%

\begin{figure*}[tbh!]
\makeatletter
\def\@captype{figure}
\makeatother
\hspace{0.7cm}
\subfigure{\includegraphics[width = 0.35\linewidth]{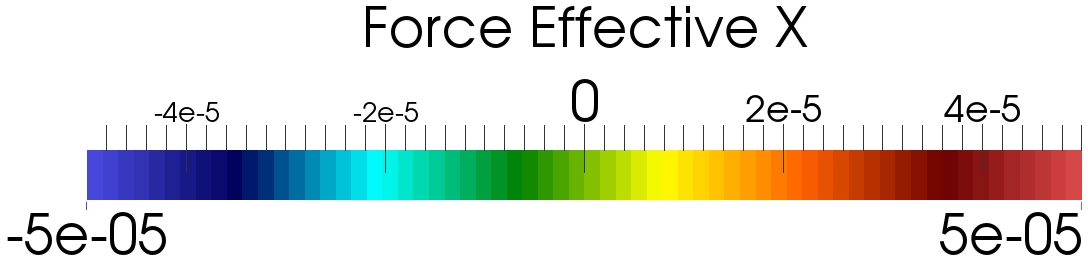}}
\hspace{2.4cm}
\subfigure{\includegraphics[width = 0.35\linewidth]{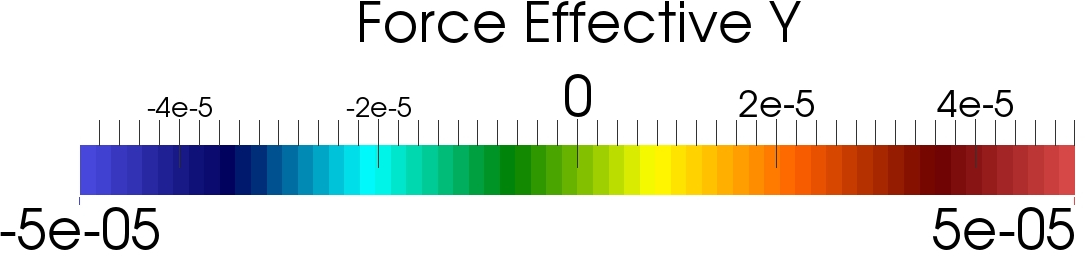}}
\hspace{2.4cm}\\
\setcounter{subfigure}{0}% Reset subfigure counter
\subfigure[\,\,$t=t_0+2.8 \tshear$, $\De = 2.0$]
{
\includegraphics[width = 0.48\linewidth]{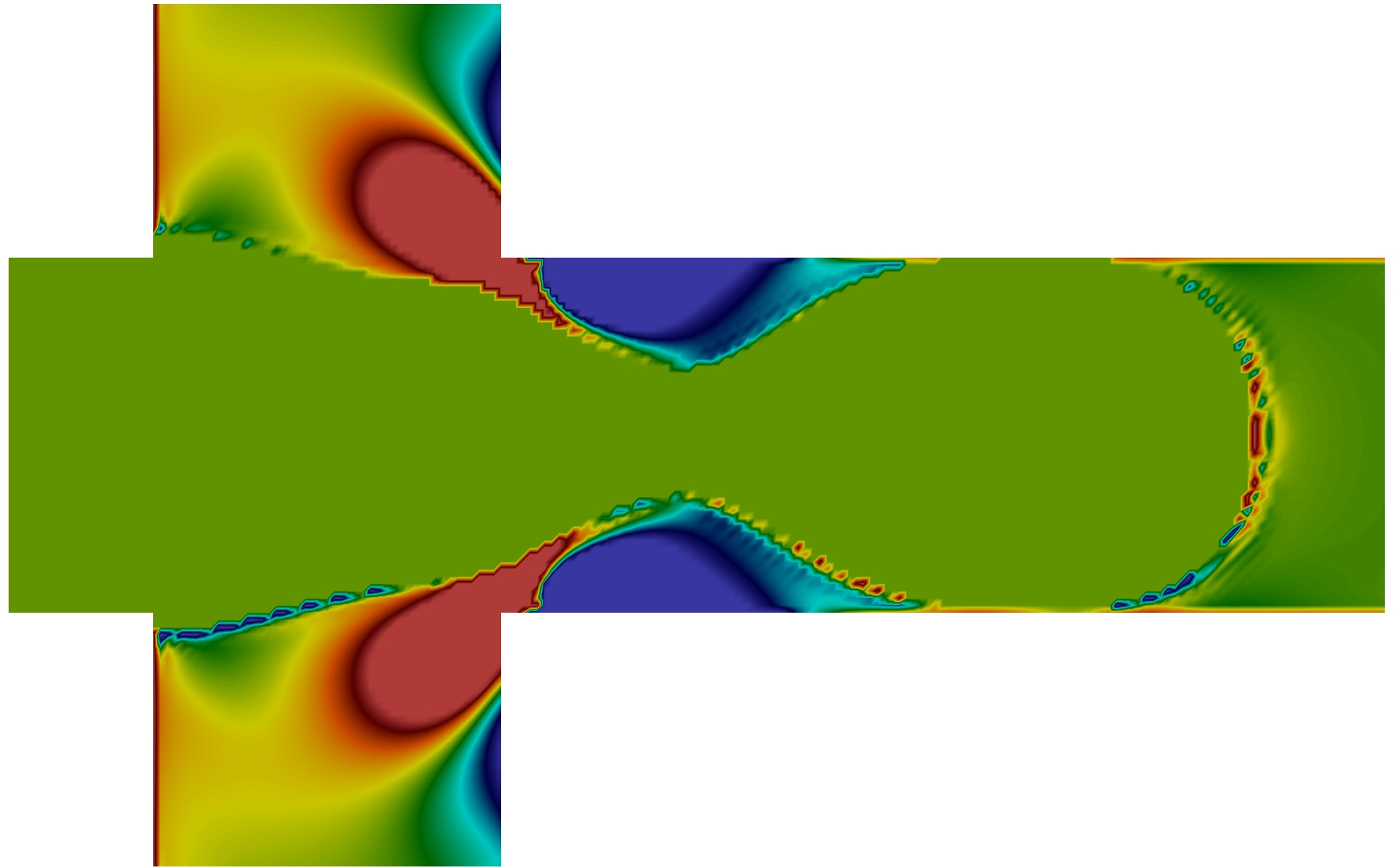}
}
\subfigure[\,\,$t=t_0+2.8 \tshear$, $\De = 2.0$]
{
\includegraphics[width = 0.48\linewidth]{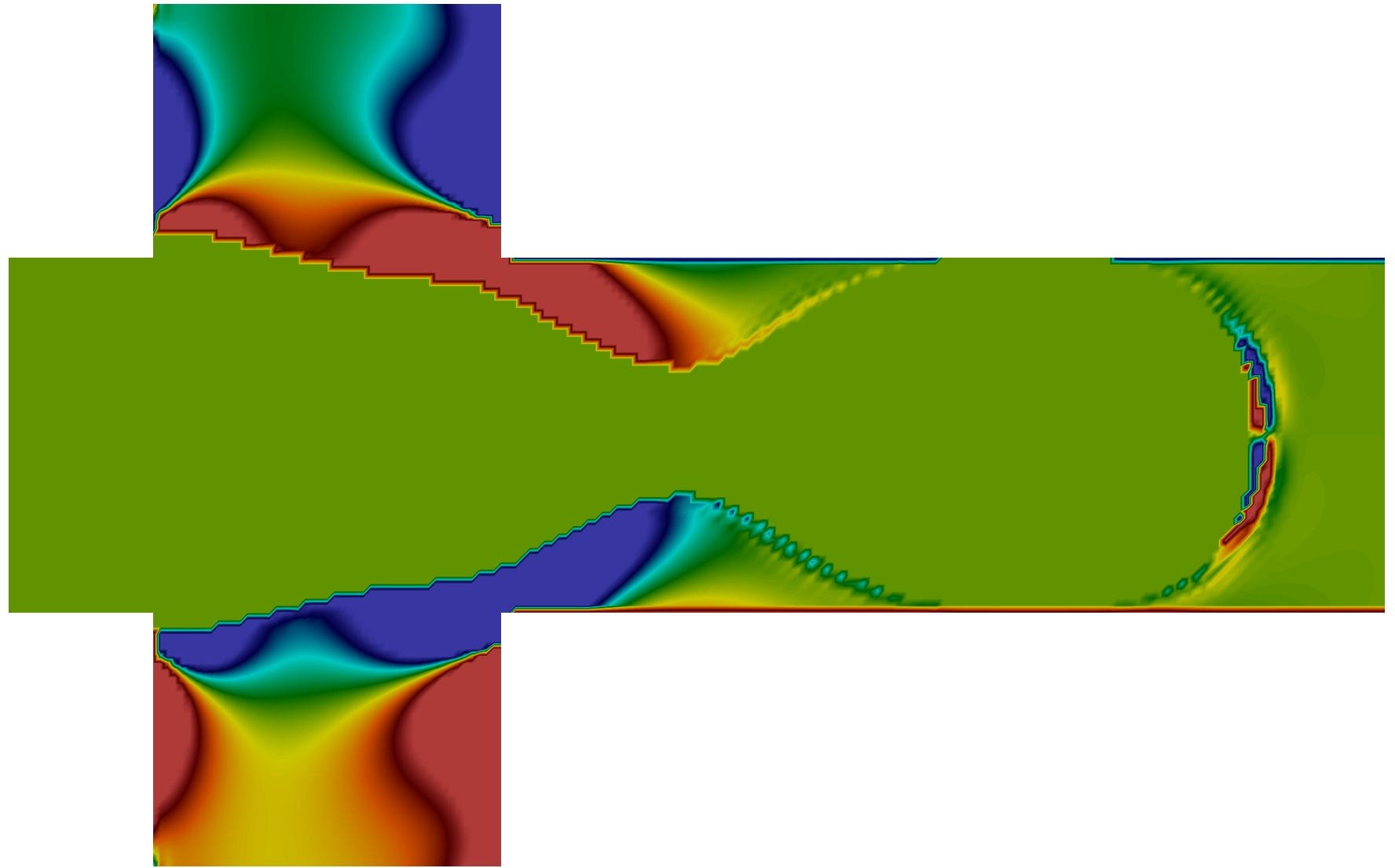}
}
\caption{Panels (a)-(b): x and y component of the effective force ${\bm F}_{\mbox{\tiny{eff}}}$ (see Eq. \eqref{eq:effectiveforce}) for a matrix viscoelasticity (MV) case at $t=t_0+2.8 \tshear$, $\De = 2.0$, $\Ca=0.007$, $\lambda=1$ and $Q=1.0$. When the liquid thread obstructs the main channel, we observe the formation of an elastic boundary layer around the emerging thread. This promotes an extra force which points towards the corners downstream of the emerging thread and triggers an extra deformation at the cross-junction. In all cases we have used the characteristic shear time $\tshear=H/v_c$ as a unit of time, while $t_0$ is a reference time (the same for all simulations). A vectorial plot of the effective force is provided in figure \ref{fig:5_bis}. \label{fig:5}}
\end{figure*}

%%%%%%%%%%%%%%%%%%%%%%%%%%%%%%%%%%%%%%%%%%%%%%%%%%%%%%%%%%%%%%%%%%%%%%%%%%%%%%%%%%%%%%%%%%%%%%%%%%%%%%%%%%%%%%%%%%%%%%%%%%%%%%%%%%%%%%%%%%%%%%%%%%%%%%%%%%%%%%%%

\begin{figure}
\begin{center}
\includegraphics[width = 1.1\linewidth]{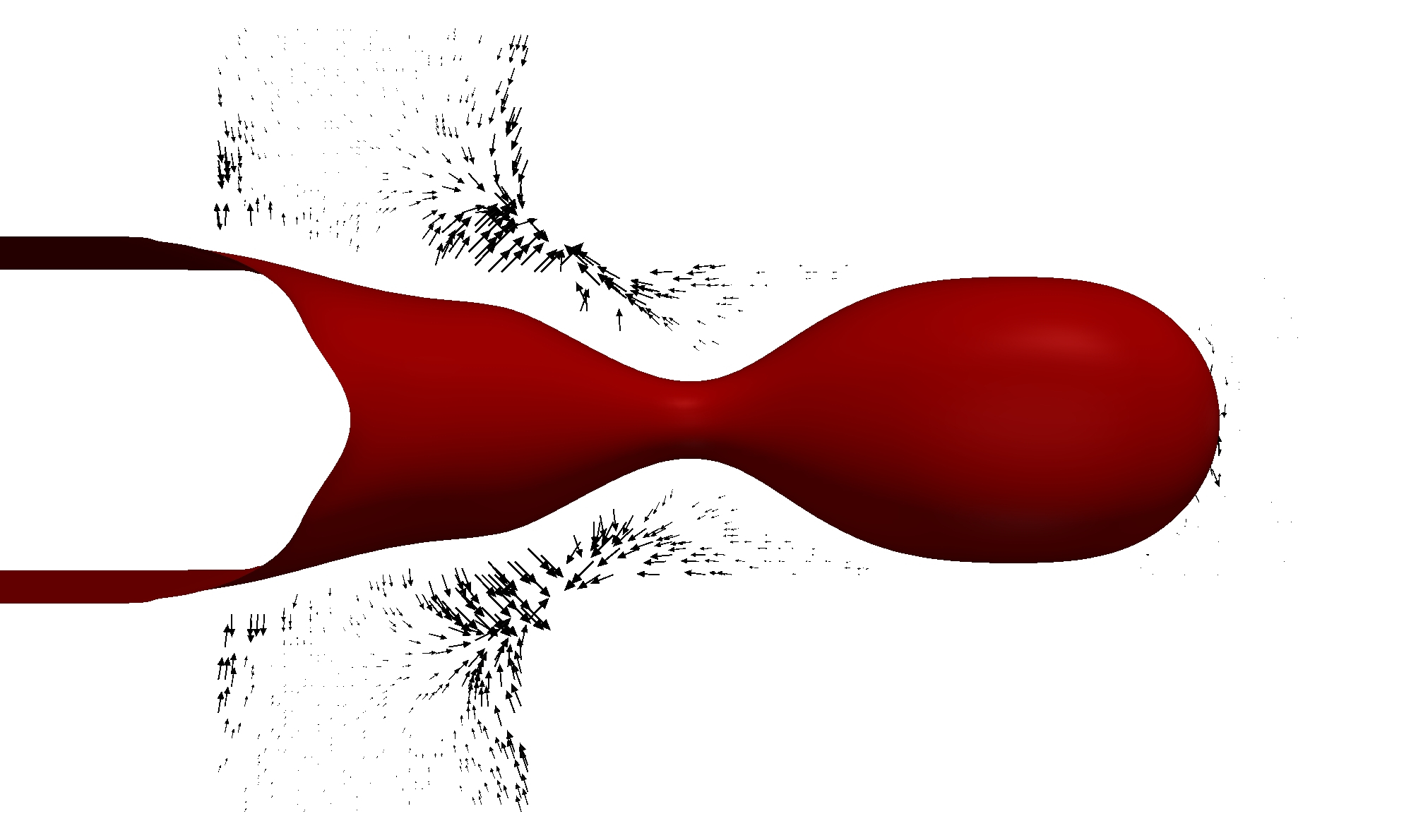}
\caption{A vectorial plot of the effective force given in figure \ref{fig:5} is provided. \label{fig:5_bis}}
\end{center}
\end{figure}

To appreciate the effects of matrix viscoelasticity in the DCJ regime, in figure \ref{fig:3} we show the break-up process for different $\De$. We report the density contours of the dispersed phase for three cases: (a) Newtonian matrix (no feedback stresses) at time $t=t_0+ 3.0 \tshear$; (b) slightly viscoelastic matrix with $\De = 0.2$ at time $t=t_0+2.8 \tshear$;  (c) viscoelastic matrix with Deborah number just above unity ($\De = 2.0$) at time $t=t_0+ 2.8 \tshear$. All the other flow parameters are kept fixed, $\Ca=0.007$, $\Ren=0.018$, $\lambda=1$ and $Q=1.0$. The mechanism of break-up is essentially the same, since droplets are formed at the cross-junction due to the squeezing mechanism. We observe, however, that as soon as the degree of viscoelasticity sensibly increases ($\De=2.0$), droplets pinch off with a slightly smaller size (figure \ref{fig:3}(c)) and the shape of the thread prior to break-up is different from that of the Newtonian case (figure \ref{fig:3}(a)). The opportunity offered by numerical studies is to allow for a systematic analysis of the various forces present in the equations of motion (see Eqs. (\ref{NSc}) and (\ref{FENEP})). We can use this advantage to go deeper into the problem of the break-up in presence of viscoelasticity and visualize the magnitude of the polymer feedback stress ($\frac{\eta_{P}}{\tau_{P}} f(r_{P}){\bm {\bm {\mathcal C}}}$  in Eq. \eqref{NSc}) for the two MV cases presented in figure \ref{fig:3} at time $t=t_0+2.8 \tshear$. In figure \ref{fig:4} we report the density contours of the dispersed phase overlaid on snapshots of the magnitude of the polymer feedback stress. Working with MV, the feedback stress is obviously non zero only in the continuous phase. These images clearly show that the stretching effect on the polymers, and consequently the feedback stress on the fluid, are well pronounced in correspondence of the corners where the side channels meet the main channel. This is a consequence of the flow structure and the associated streamlines which provide stretching of the polymers in those regions. As $\De$ is increased, we see that the flow in the continuous phase develops enhanced polymer feedback stress. The enhancement of the polymer feedback stress has quantitative effects on the velocity field, as predicted by equation \eqref{NSc}. One has also to remark that viscoelastic forces provide a contribution to the shear forces. This happens in simple shear flows and also for weak viscoelasticity~\cite{bird,Herrchen}, where we expect that the viscoelastic stresses closely follow the viscous stresses, i.e. $\frac{\eta_P}{\tau_P}{\bm \nabla} \cdot [f(r_P){\bm {\bm {\mathcal C}}}] \approx {\bm \nabla} \cdot \left(\eta_{P} ({\bm \nabla} {\bm u}_c+({\bm \nabla} {\bm u}_c)^{T})\right)$. Obviously, this cannot be the case when viscoelasticity is enhanced and the Deborah number is above unity. For this reason, to better visualize the importance of the viscoelastic forces in comparison with the Newtonian case, one can define an {\it effective} force (${\bm F}_{\mbox{\tiny{eff}}}$) as \cite{SbragagliaGupta}
\be\label{eq:effectiveforce}
{\bm F}_{\mbox{\tiny{eff}}}=\frac{\eta_P}{\tau_P}{\bm \nabla} \cdot [f(r_P){\bm {\bm {\mathcal C}}}]-{\bm \nabla} \left(\eta_{P} ({\bm \nabla} {\bm u}_c+({\bm \nabla} {\bm u}_c)^{T})\right).
\ee
Since all our simulations are performed with the same shear viscosity, the effective force gives an idea of how much the viscoelastic system differs from the corresponding Newtonian system with the same viscosity. If present (${\bm F}_{\mbox{\tiny{eff}}} \neq 0$), this change is solely attributed to viscoelasticity. For the case reported in figure \ref{fig:4}(c), we have analyzed the effective force in the xy-plane at $z=H/2$. Results are reported in figure \ref{fig:5}, where we show two distinct plots for the x and y component of ${\bm F}_{\mbox{\tiny{eff}}}$. The figure reveals the formation of an elastic boundary layer and the emergence of extra forces which point towards the corners downstream of the emerging thread. A vectorial plot of the effective force is provided in figure \ref{fig:5_bis} to highlight how the direction of the force varies across the junction.

%%%%%%%%%%%%%%%%%%%%%%%%%%%%%%%%%%%%%%%%%%%%%%%%%%%%%%%%%%%%%%%%%%%%%%%%%%%%%%%%%%%%%%%%%%%%%%%%%%%%%%%%%%%%%%%%%%%%%%FIG 7 %%%%%%%%%%%%%%%%%%%%%%%%%%%%%%%%%%%%%%%%%%%%%%%%%%%%%%%%%%%%%%%%%%%%%%%%%%%%%%%%%%%%%%%%%%%%%%%%%%%%%%%%%%%%%%%%%%%%%

%%%%%%%%%%%%%%%%%%%%%%%%%%%%%%%%%%%%%%%%%%%%%%%%%%%%%%%%%%%%%%%%%%%%%%%%%%%%%%%%%%%%%%%%%%%%%%%%%%%%%%%%%%%%%%%%%%%%%%%%%%%%%%%%%%%%%%%%%%%%%%%%%%%%%%%%%%%%%%%%

%%%%%%%%%%%%%%%%%%%%%%%%%%%%%%%%%%%%%%%%%%%%%%%%%%%%%%%%%%%%%%%%%%%%%%%%%%%%%%%%%%%%%%%%%%%%%%%%%%%%%%%%%%%%%%%%%%%%%%FIG 8%%%%%%%%%%%%%%%%%%%%%%%%%%%%%%%%%%%%%%%%%%%%%%%%%%%%%%%%%%%%%%%%%%%%%%%%%%%%%%%%%%%%%%%%%%%%%%%%%%%%%%%%%%%%%%%%%%%%%
\begin{figure}
%\begin{center}
\includegraphics[width = 0.95\linewidth]{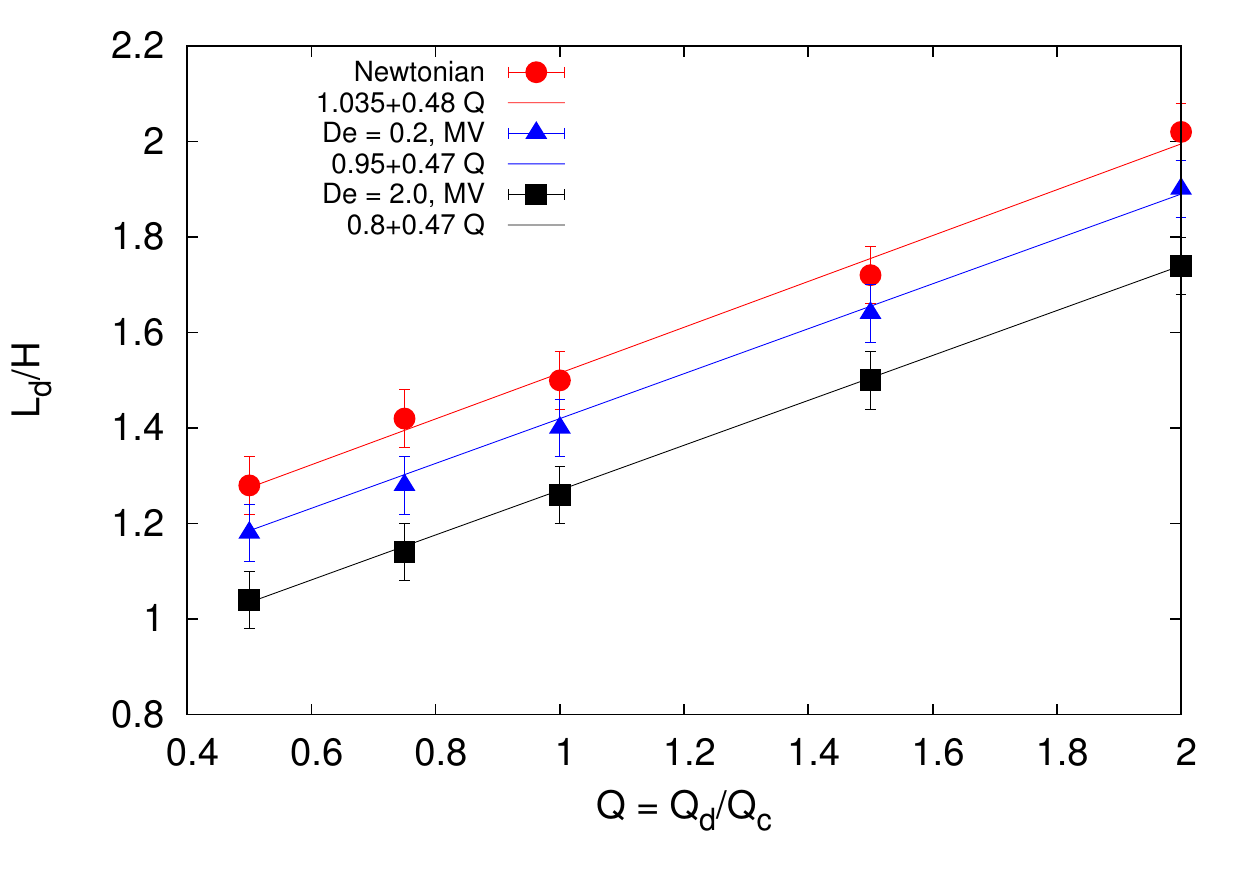}
\caption{Effects of matrix viscoelasticity (MV) on the droplet length. We report the droplet length ($L_{d}$) normalized to the characteristic edge of the channels ($H$) versus the flow-rate ratio $Q$ for different values of $\De$ for fixed $\Ca=0.007$, $\Ren=0.018$, and $\lambda=1$. We see that the droplet length is linearly increasing with flow-rate ratio $Q$ for all the three cases presented in the figure (Newtonian, slightly viscoelastic ($\De = 0.2$) and viscoelastic ($\De = 2.0$) case). Half of the interface thickness has been used as errorbar. Linear fits are drawn from the expected behaviour \eqref{eq:lin}, with $\alpha_{1,2}$ two fitting parameters \label{fig:6}.}
%\end{center}
\end{figure}

%%%%%%%%%%%%%%%%%%%%%%%%%%%%%%%%%%%%%%%%%%%%%%%%%%%%%%%%%%%%%%%%%%%%%%%%%%%%%%%%%%%%%%%%%%%%%%%%%%%%%%%%%%%%%%%%%%%%%%%%%%%%%%%%%%%%%%%%%%%%%%%%%%%%%%%%%%%%%%%%

To proceed further and complement the results of figures \ref{fig:3} and \ref{fig:4}, we have characterized the effect of MV and the flow-rate ratio on the droplet length $L_d$. In figure \ref{fig:6} we study $L_{d}$ normalized to the characteristic edge of the channels ($H$) for fixed $\Ca=0.007$, $\Ren=0.018$, and $\lambda=1$. We report $L_d/H$ as a function of $Q$, for various values of the Deborah number. In the squeezing-dominated DCJ regime, a linear relation between the droplet length $L_d$ and flow-rate ratio $Q$ is expected~\cite{Garstecki06}
\begin{equation}\label{eq:lin}
\frac{L_d}{H}=(\alpha_1+\alpha_2 Q)
\end{equation}
where $\alpha_{1,2}$ are coefficients of order one. The specific values of these coefficients depend on the Capillary number and channel geometry \cite{Christopher08,LiuZhang11,Tan,Xu}. The study that is probably best useful for us is that of Liu \& Zhang \cite{LiuZhang11}, who computed the coefficients $\alpha_{1,2}$ for a wide range of channel geometries and Capillary numbers in the Newtonian case. For a square duct with $\Ca=0.007$, their prediction (p. 11 in \cite{LiuZhang11}) is $\alpha_1=1.17$ and $\alpha_2=0.85$. From a best fit of our data, we obtain $\alpha_1=1.035$ and $\alpha_2=0.48$. The value of $\alpha_1$ is actually in the ballpark of that found by Liu \& Zhang \cite{LiuZhang11}, whereas our $\alpha_2$ is smaller. As for the value of $\alpha_2$, we notice that Liu \& Zhang define $Q$ with half of the average velocity in one of the two side channels (see figure 3 in \cite{LiuZhang11}), which can explain why our fitted value of $\alpha_2$ is roughly a factor 2 smaller than what they provide. Other discrepancies may be attributed to different factors, including a different Reynolds number $\Ren$ \cite{RenardyCristini01} and viscosity ratio $\lambda$, although we do not expect a major influence of the latter in the squeezing regime. Moreover, the resolution used by  Liu \& Zhang \cite{LiuZhang11} is smaller than ours, although the authors acknowledge a grid independence with several different flow conditions and find that the mesh refinement produces variations not more than 5\%. \\
For the viscoelastic cases, the relation between the droplet length $L_d$ and flow-rate ratio $Q$ is still linear, but an overall decrease is observed when comparing with the corresponding Newtonian data. In particular, the coefficient $\alpha_2$ is not sensibly different, whereas we acknowledge a decrease in $\alpha_1$. We go from $\alpha_1=1.035$ in the Newtonian case to $\alpha_1=0.8$ when $\De=2.0$. This behaviour can be better understood if we explore the basic ingredients which lie at the core of the observed linear scaling law \cite{Garstecki06}. The break-up is the result of two distinct physical processes: first, a {\it blocking process}, where the thread of the dispersed phase grows until it effectively blocks the cross-section of the side channel. At this particular moment, the blocking length $L_{\mbox{\tiny{block}}}$ of the plug is of the order of the channel width, say  $\alpha_1 H$ (with $\alpha_1$ a constant of order unity). Afterwards, the {\it necking process} starts. For a neck with a characteristic width $\alpha_2 H$ ($\alpha_2$ is a constant, again, of order unity) and squeezing at a rate approximately equal to the average velocity ($Q_c/H^2$), it takes a time $\tau \approx \alpha_2 H H^2/Q_c$ to complete the squeezing process. During this time, the plug continues to elongate with a rate $Q_d/H^2$. The resulting ``squeezing length'' of the plug is therefore $L_{\mbox{\tiny{squeeze}}} \approx \tau Q_d/H^2 = \alpha_2 H Q_d/Q_c$. Consequently, the final dimensionless length $L_d/H$ of the droplet can be expressed as $L_d/H \approx \alpha_1+ \alpha_2 Q$. It is clear from figure \ref{fig:6} that viscoelasticity directly affects the offset constant $\alpha_1$, i.e. it changes the {\it blocking process} of the squeezing regime. This is due to the extra force generated by the polymers feedback stresses which is visible in figure \ref{fig:5}. This force effectively pulls the droplet interface downstream at the cross junction, so that the blocking length, i.e. the distance that the droplet has to travel to obstruct the side channels, is effectively smaller. 

%%%%%%%%%%%%%%%%%%%%%%%%%%%%%%%%%%%%%%%%%%%%%%%%%%%%%%%%%%%%%%%%%%%%%%%%%%%%%%%%
%%%%%%%%%%%%%%%%%%%%%%%%%%%%%%%%%%%%FIG 9%%%%%%%%%%%%%%%%%%%%%%%%%%%%%%%%%%%%%%%%%%%%%%%%%%%%%%%%%%%%%%%%%%%%%%%%%%%%%%%%%%%%%%%%%%%%%%%%%%%%%%%%%%%%%%%%%%%%%%%

\begin{figure*}[t!]
\makeatletter
\def\@captype{figure}
\makeatother
\subfigure[\,\,$t=t_0+3.0 \tshear$, $\De = 0$]
{
\includegraphics[width = 0.330\linewidth]{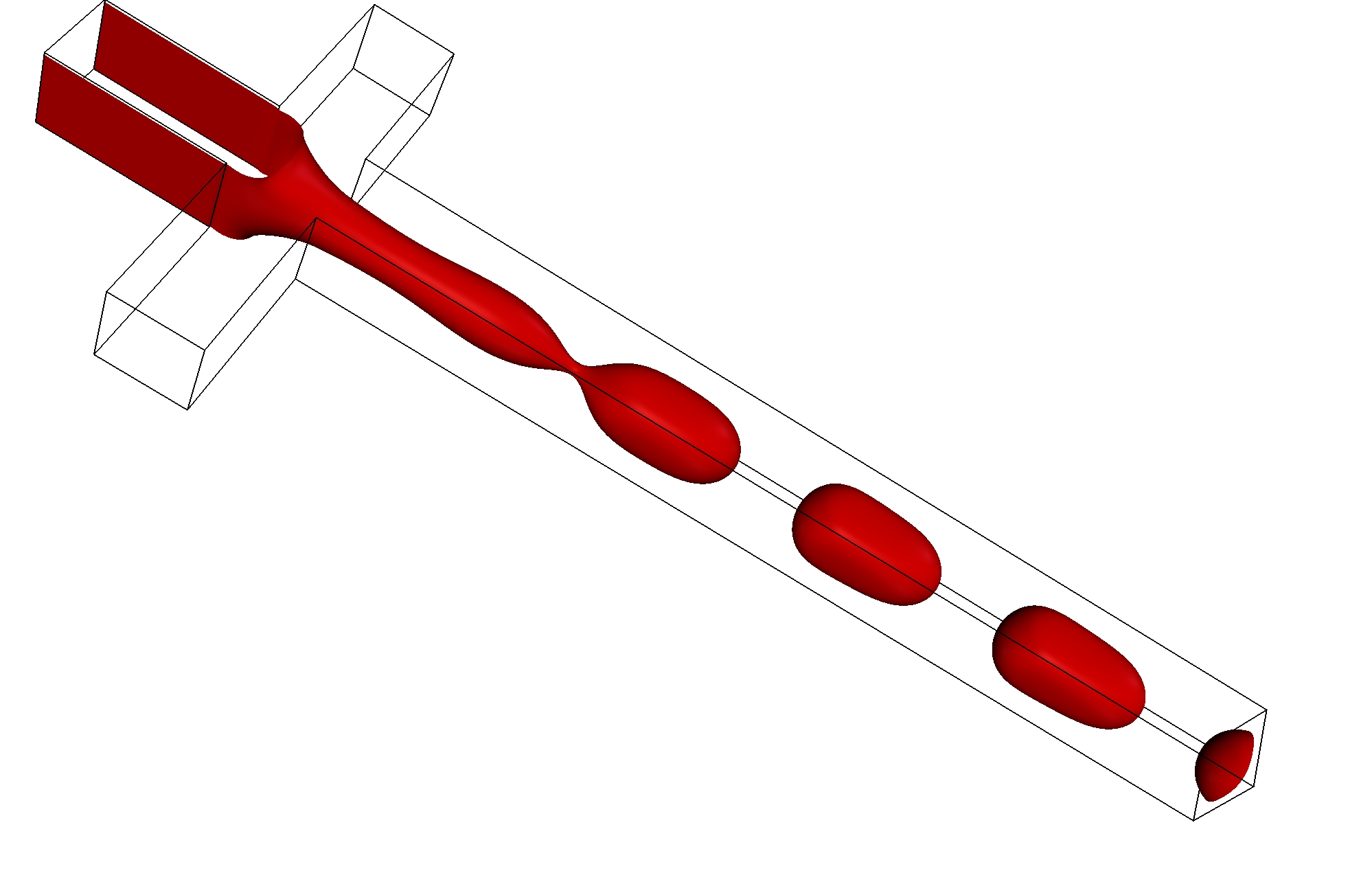}
}
\subfigure[\,\,$t=t_0+2.8 \tshear$, $\De = 0.2$]
{
\includegraphics[width = 0.330\linewidth]{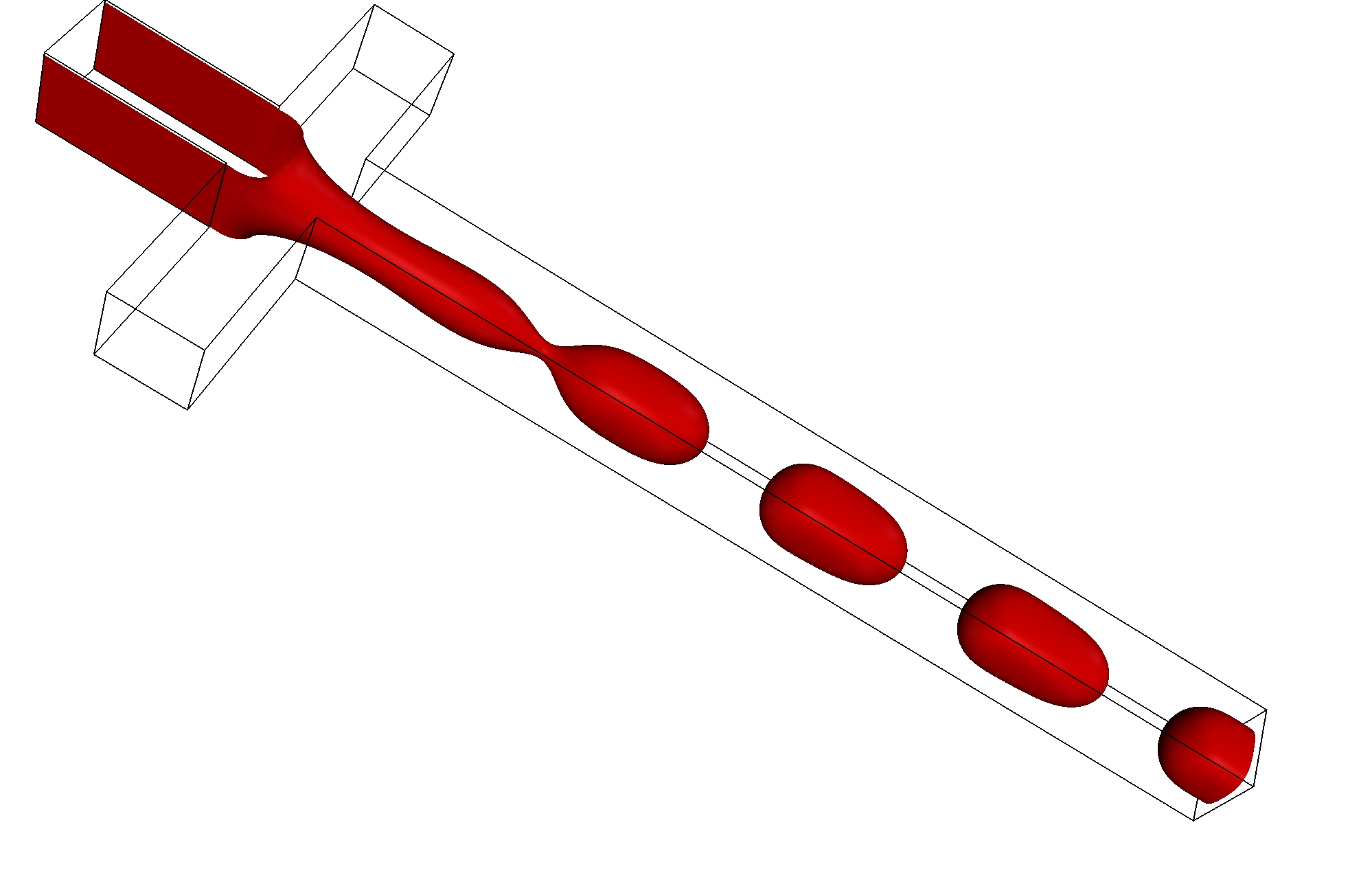}
}
\subfigure[\,\,$t=t_0+2.9 \tshear$, $\De = 2.0$]
{
\includegraphics[width = 0.330\linewidth]{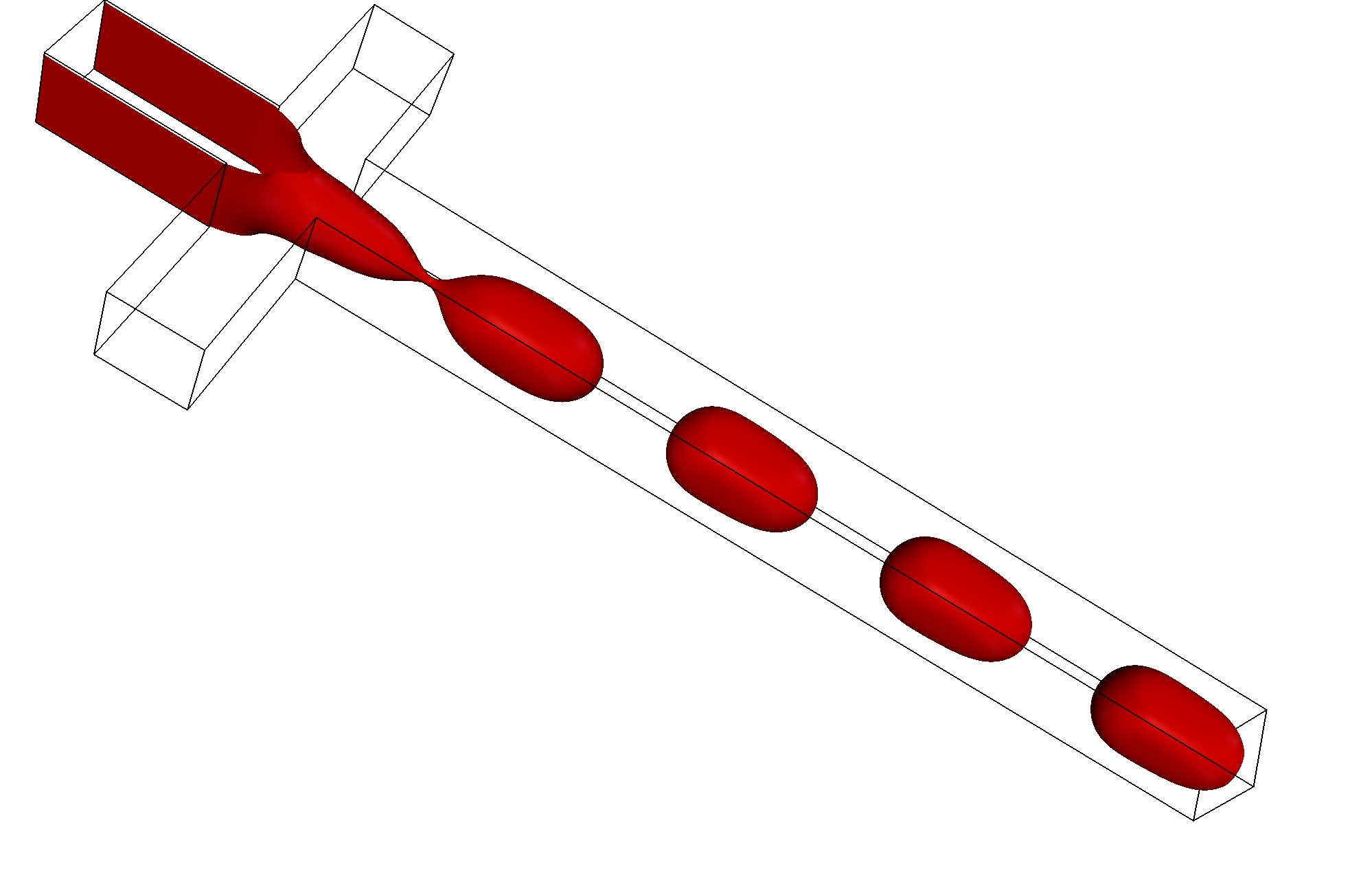}
}
\caption{Effects of matrix viscoelasticity (MV) in the DC regime. Droplet formation in the flow-focusing geometry at fixed $\Ca=0.007$, $\Ren=0.018$, $\lambda=1$ and $Q=3.0$. Three cases are compared: (a) Newtonian case ($\De = 0.0$) at time $t = t_0 + 3.0 \tshear$; (b) slightly viscoelastic case with $\De = 0.2$ at time $t = t_0 + 2.8 \tshear$; (c) MV case with $\De = 2.0$ at time $t = t_0 + 2.9 \tshear$.  In all cases we have used the characteristic shear time $\tshear=H/v_c$ as a unit of time, while $t_0$ is a reference time (the same for all simulations). \label{fig:7}}
\end{figure*}

%%%%%%%%%%%%%%%%%%%%%%%%%%%%%%%%%%%%%%%%%%%%%%%%%%%%%%%%%%%%%%%%%%%%%%%%%%%%%%%%%%%%%%%%%%%%%%%%%%%%%%%%%%%%%%%%%%%%%%%%%%%%%%%%%%%%%%%%%%%%%%%%%%%%%%%%%%%%%%%%%%%%%%%%%%%

%%%%%%%%%%%%%%%%%%%%%%%%%%%%%%%%%%%%%%%%%%%%%%%%%%%%%%%%%%%%%%%%%%%%%%%%%%%%%%%%%%%%%%%%%%%%%%%%%%%%%%%%%%%%%%%%%%%%%%FIG 10%%%%%%%%%%%%%%%%%%%%%%%%%%%%%%%%%%%%%%%%%%%%%%%%%%%%%%%%%%%

\begin{figure*}[]
\makeatletter
\def\@captype{figure}
\makeatother
\subfigure[\,\,$t=t_0+2.8 \tshear$, $\De = 0.2$]
{
\includegraphics[width = 0.49\linewidth]{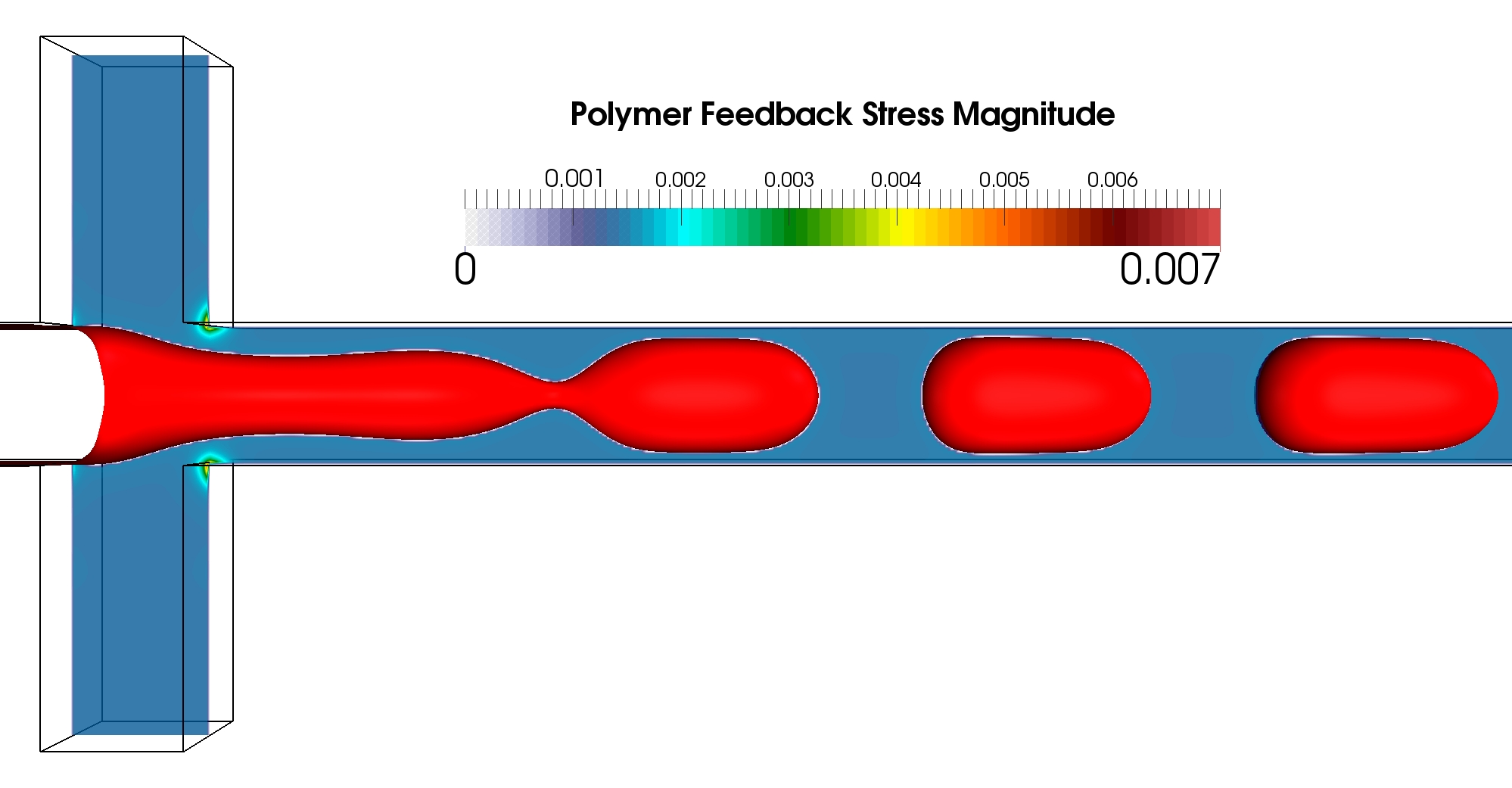}
}
\subfigure[\,\,$t=t_0+2.9 \tshear$, $\De = 2.0$]
{
\includegraphics[width = 0.49\linewidth]{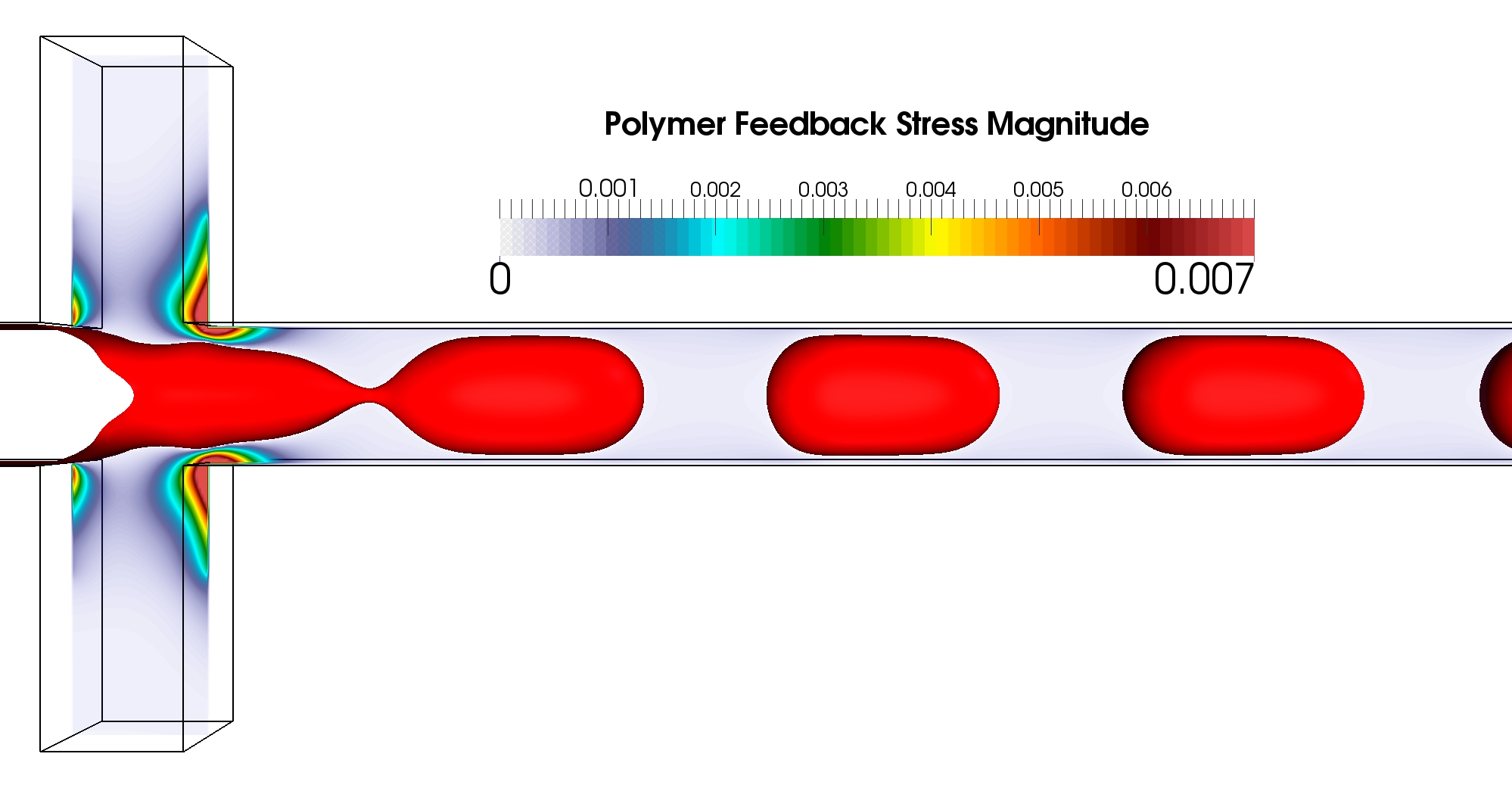}
}
\caption{Effects of matrix viscoelasticity (MV) in the DC regime. Density contours of the dispersed phase overlaid on the polymer feedback stress magnitude (see Eq. \eqref{NSc}) for two characteristic values of the Deborah number $\De$ at fixed $\Ca=0.007$, $\Ren=0.018$, $\lambda=1$ and $Q=3.0$. As $\De$ is increased, we see that the flow in the continuous phase develops enhanced polymer feedback stress at the junction between the side channel and the main channel. The feedback is higher downstream at the cross-junction. The effect of viscoelasticity is to promote a shift of the break-up point from downstream of the cross-junction to the cross junction itself. In all cases we have used the characteristic shear time $\tshear=H/v_c$ as a unit of time, while $t_0$ is a reference time (the same for all simulations).}\label{fig:8}
\end{figure*}

%%%%%%%%%%%%%%%%%%%%%%%%%%%%%%%%%%%%%%%%%%%%%%%%%%%%%%%%%%%%%%%%%%%%%%%%%%%%%%%%%%%%%%%%%%%%%%%%%%%%%%%%%%%%%%%%%%%%%%%%%%%%%%%%%%%%%%%%%%%%%%%%%%%%%%%%%%%%%%%%

%%%%%%%%%%%%%%%%%%%%%%%%%%%%%%%%%%%%%%%%%%%%%%%%%%%%%%%%%%%%%%%%%%%%%%%%%%%%%%%%
%%%%%%%%%%%%%%%%%%%%%%%%%%%%%%%%%%%%FIG 11%%%%%%%%%%%%%%%%%%%%%%%%%%%%%%%%%%%%%%%%%%%%%%%%%%

\begin{figure*}[]
\makeatletter
\def\@captype{figure}
\makeatother
\subfigure[\,\,$t=t_0+2.4 \tshear$, $\De = 0$]
{
\includegraphics[width = 0.330\linewidth]{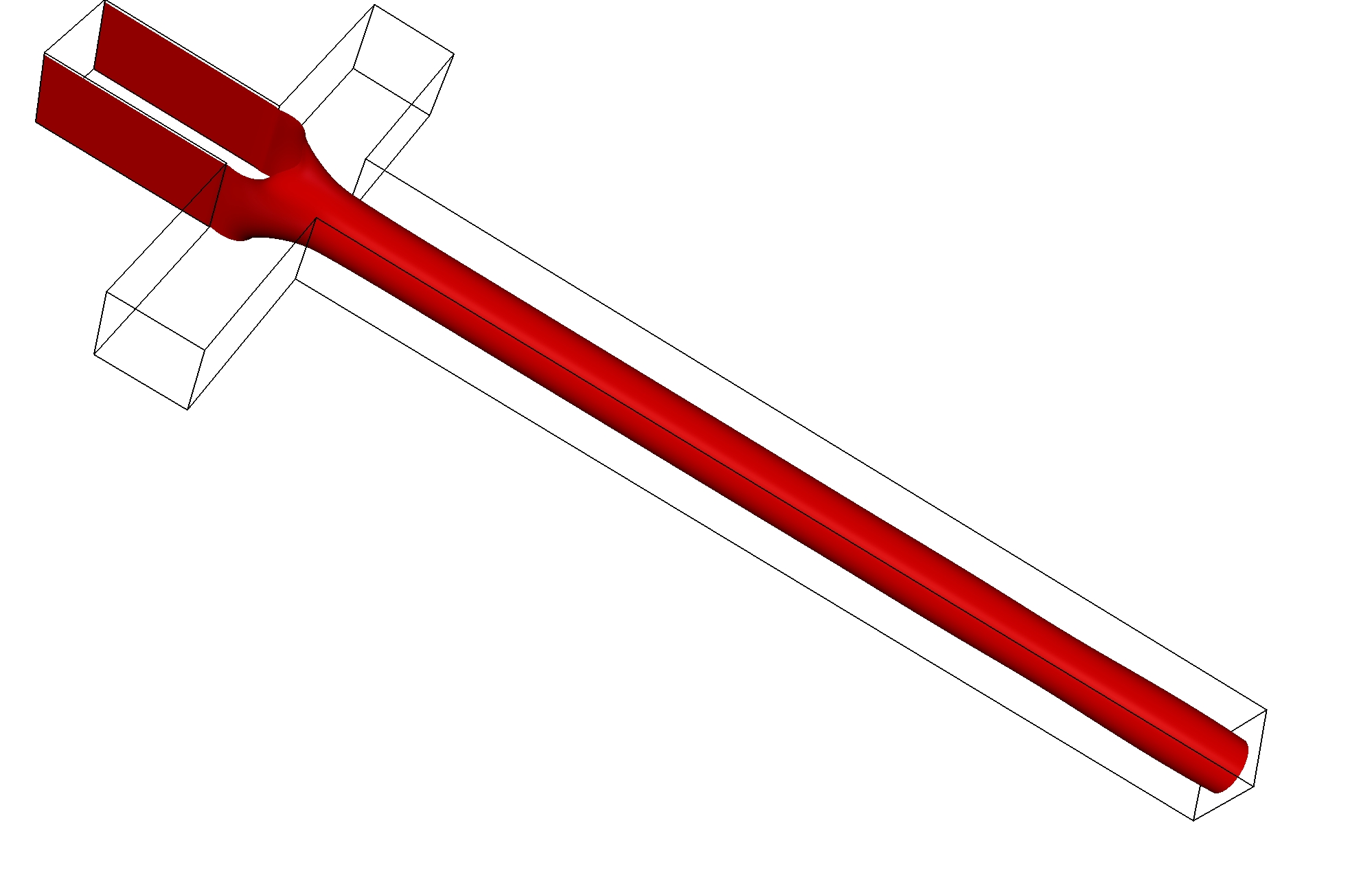}
}
\subfigure[\,\,$t=t_0+2.4 \tshear$, $\De = 0.8$]
{
\includegraphics[width = 0.330\linewidth]{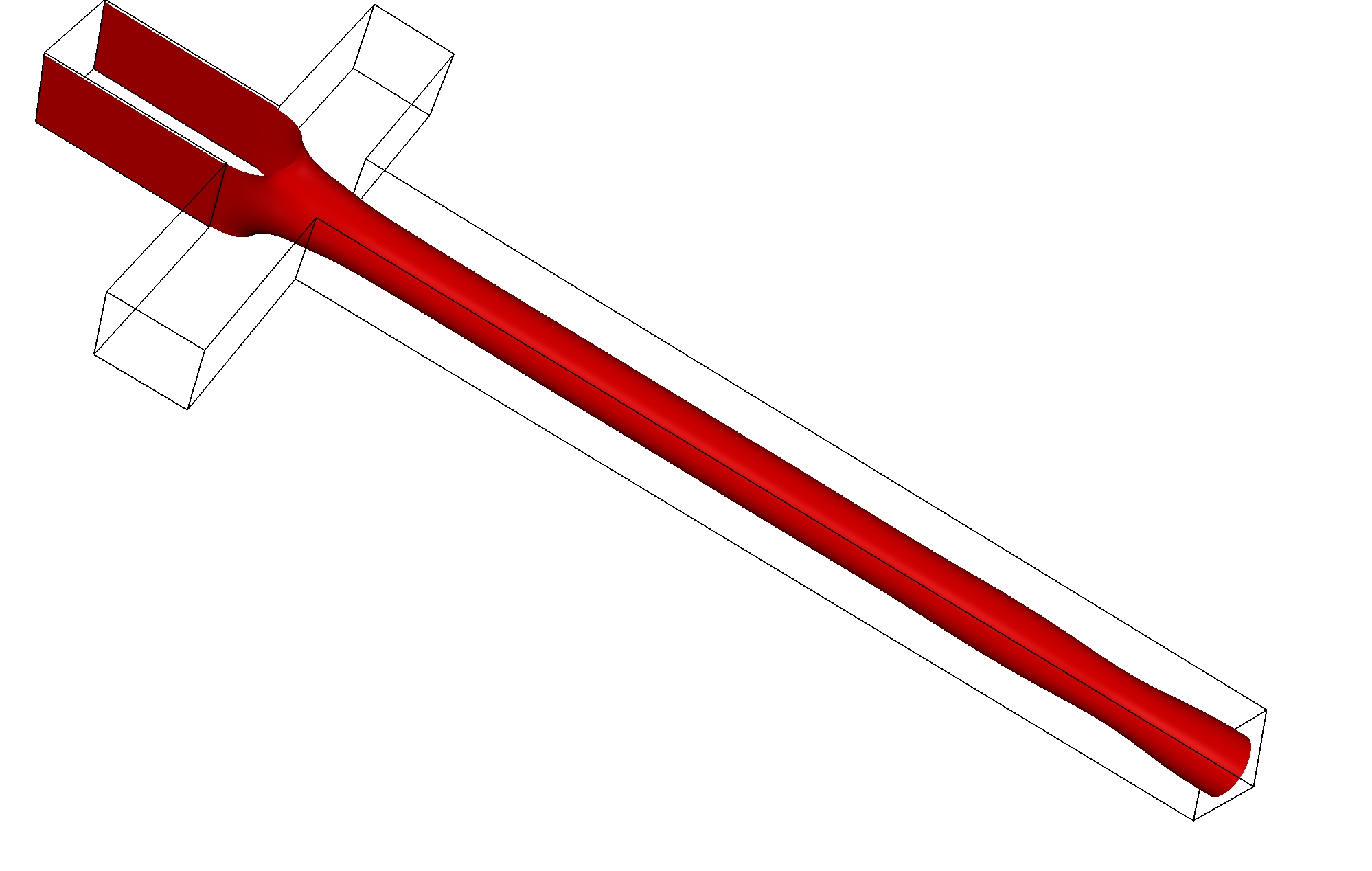}
}
\subfigure[\,\,$t=t_0+2.4 \tshear$, $\De = 2.0$]
{
\includegraphics[width = 0.330\linewidth]{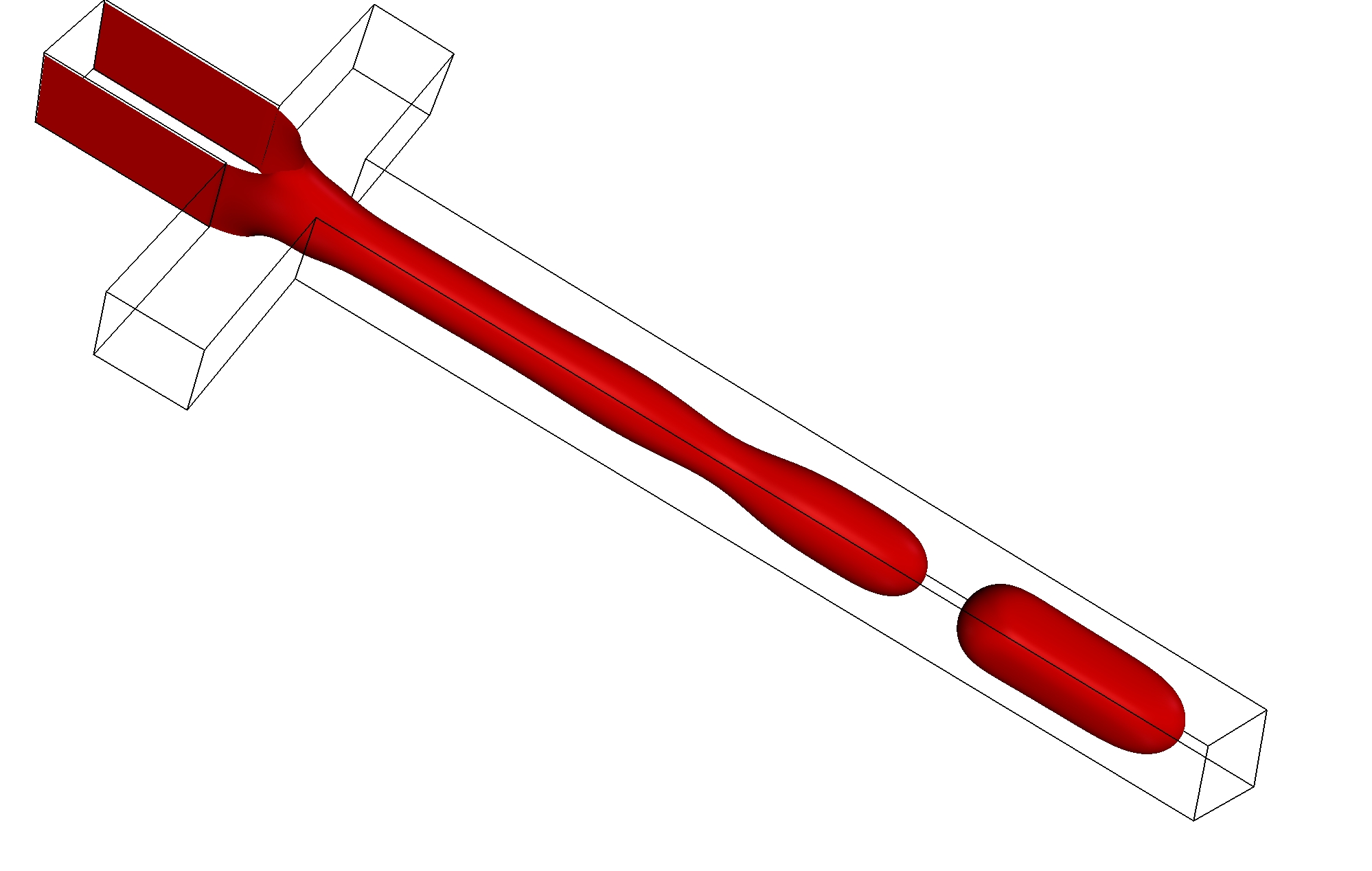}
}
\caption{Effects of matrix viscoelasticity (MV) in the PF regime at fixed $\Ca=0.007$, $\Ren=0.018$, $\lambda=1$ and $Q=4.0$. Three cases are compared: (a) Newtonian case ($\De = 0.0$) at time $t = t_0 + 2.4 \tshear$; (b) slightly viscoelastic case with $\De = 0.8$ at time $t = t_0 + 2.4 \tshear$; (c) viscoelastic case with $\De = 2.0$ at time $t = t_0 + 2.4 \tshear$. In all cases we have used the characteristic shear time $\tshear=H/v_c$ as a unit of time, while $t_0$ is a reference time (the same for all simulations).} \label{fig:9}
\end{figure*}

%%%%%%%%%%%%%%%%%%%%%%%%%%%%%%%%%%%%%%%%%%%%%%%%%%%%%%%%%%%%%%%%%%%%%%%%%%%%%%%%%%%%%%%%%%%%%%%%%%%%%%%%%%%%%%%%%%%%%%%%%%%%%%%%%%%%%%%%%%%%%%%%%%%%%%%%%%%%%%%%
%%%%%%%%%%%%%%%%%%%%%%%%%%%%%%%%%%%%%%%%%%%%%%%%%%%%%%%%%%%%%%%%%%%%%%%%%%%%%%%%

%%%%%%%%%%%%%%%%%%%%%%%%%%%%%%%%%%%%%%%%%%%%%%%%%%%%%%%%%%%%%%%%%%%%%%%%%%%%%%%%%%%%%%%%%%%%%%%%%%%%%%%%%%%%%%%%%%%%%%FIG 12%%%%%%%%%%%%%%%%%%%%%%%%%%%%%%%%%%%%%%%%%%%%%%%%%%%%%%%%%%%%%%%%%%%%%%%%%%%%%%%%%%%%%%%%%%%%%%%%%%%%%%%%%%%%%%%%%%%%%

\begin{figure*}[]
\makeatletter
\def\@captype{figure}
\makeatother
\subfigure[\,\,$t=t_0+2.4 \tshear$, $\De = 0.8$]
{
\includegraphics[width = 0.49\linewidth]{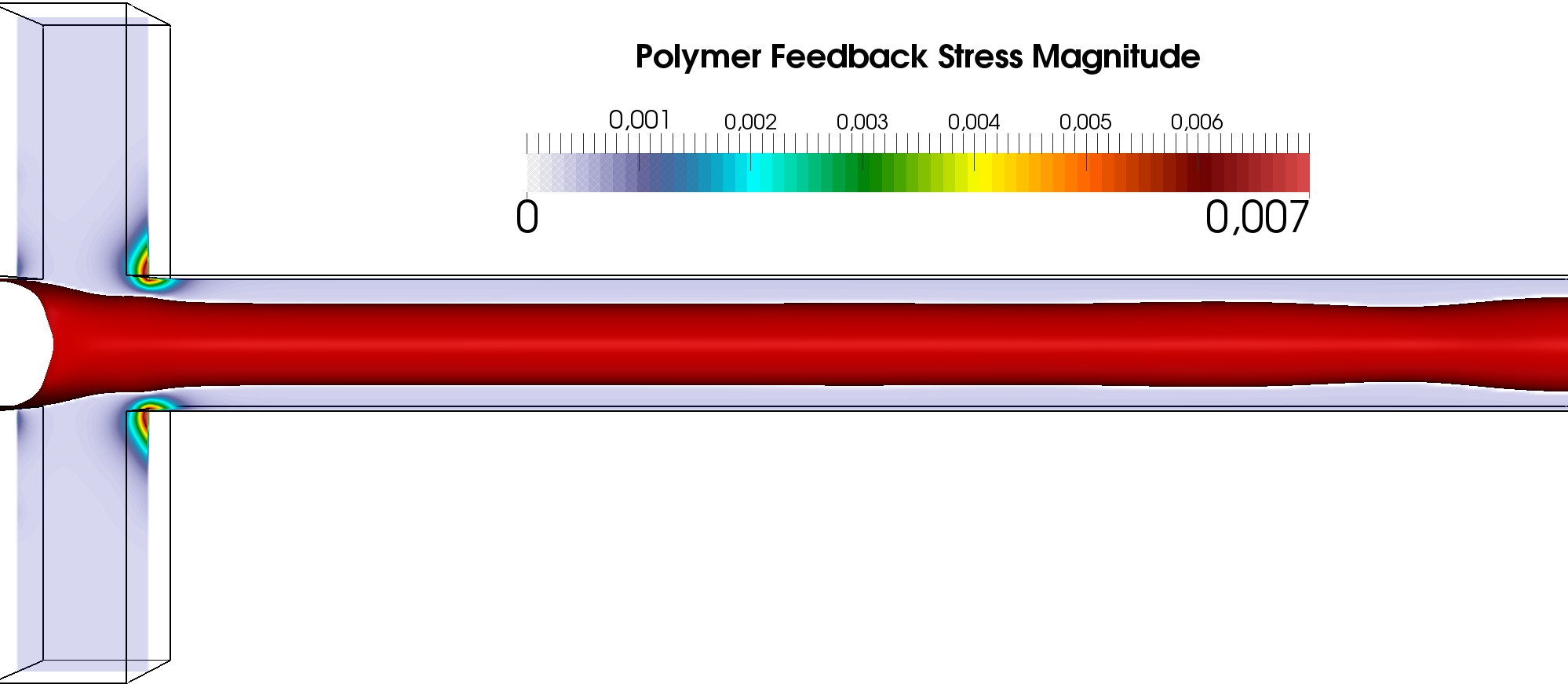}
}
\subfigure[\,\,$t=t_0+2.4 \tshear$, $\De = 2.0$]
{
\includegraphics[width = 0.49\linewidth]{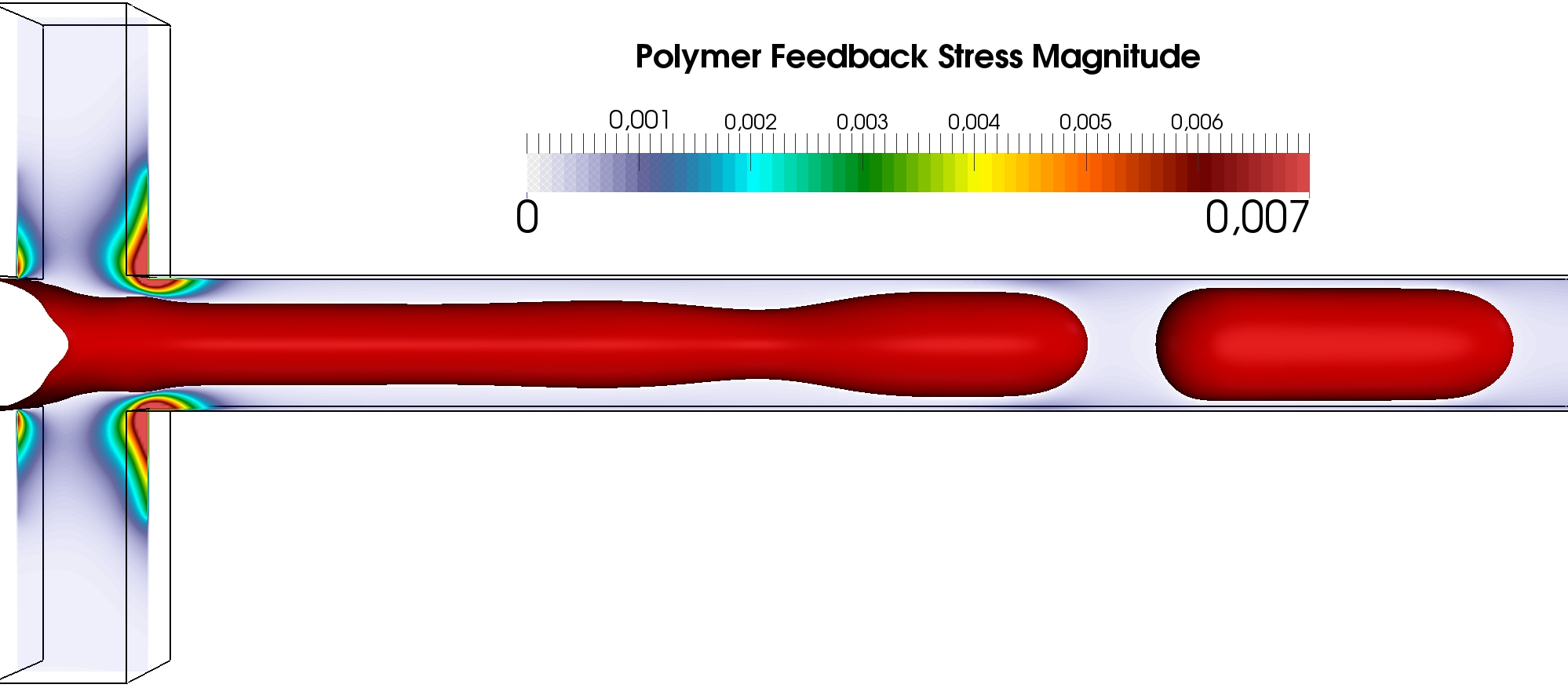}
}
\caption{Effects of matrix viscoelasticity (MV) in the PF regime. Density contours of the dispersed phase overlaid on the polymer feedback stress magnitude (see Eq. \eqref{NSc}) for two characteristic values of the Deborah number $\De$ at fixed $\Ca=0.007$, $\Ren=0.018$, $\lambda=1$ and $Q=4.0$. Similarly to figure \ref{fig:8}, as $\De$ is increased, the flow in the continuous phase develops enhanced polymer feedback stress at the junction between the side channel and the main channel. This results in the appearance of a break-up point downstream at the cross-junction. In all cases we have used the characteristic shear time $\tshear=H/v_c$ as a unit of time, while $t_0$ is a reference time (the same for all simulations).}\label{fig:10}
\end{figure*}

%%%%%%%%%%%%%%%%%%%%%%%%%%%%%%%%%%%%%%%%%%%%%%%%%%%%%%%%%%%%%%%%%%%%%%%%%%%%%%%%%%%%%%%%%%%%%%%%%%%%%%%%%%%%%%%%%%%%%%%%%%%%%%%%%%%%%%%%%%%%%%%%%%%%%%%%%%%%%%%%

%%%%%%%%%%%%%%%%%%%%%%%%%%%%%%%%%%%%%%%%%%%%%%%%%%%%%%%%%%%%%%%%%%%%%%%%%%%%%%%%%%%%%%%%%%%%%%%%%%%%%%%%%%%%%%%%%%%%%%FIG 13%%%%%%%%%%%%%%%%%%%%%%%%%%%%%%%%%%%%%%%%%%%%%%%%%%%%%%%%%%%%%%%%%%%%%%%%%%%%%%%%%%%%%%%%%%%%%%%%%%%%%%%%%%%%%%%%%%%%

\begin{figure}
\includegraphics[width = 0.9\linewidth]{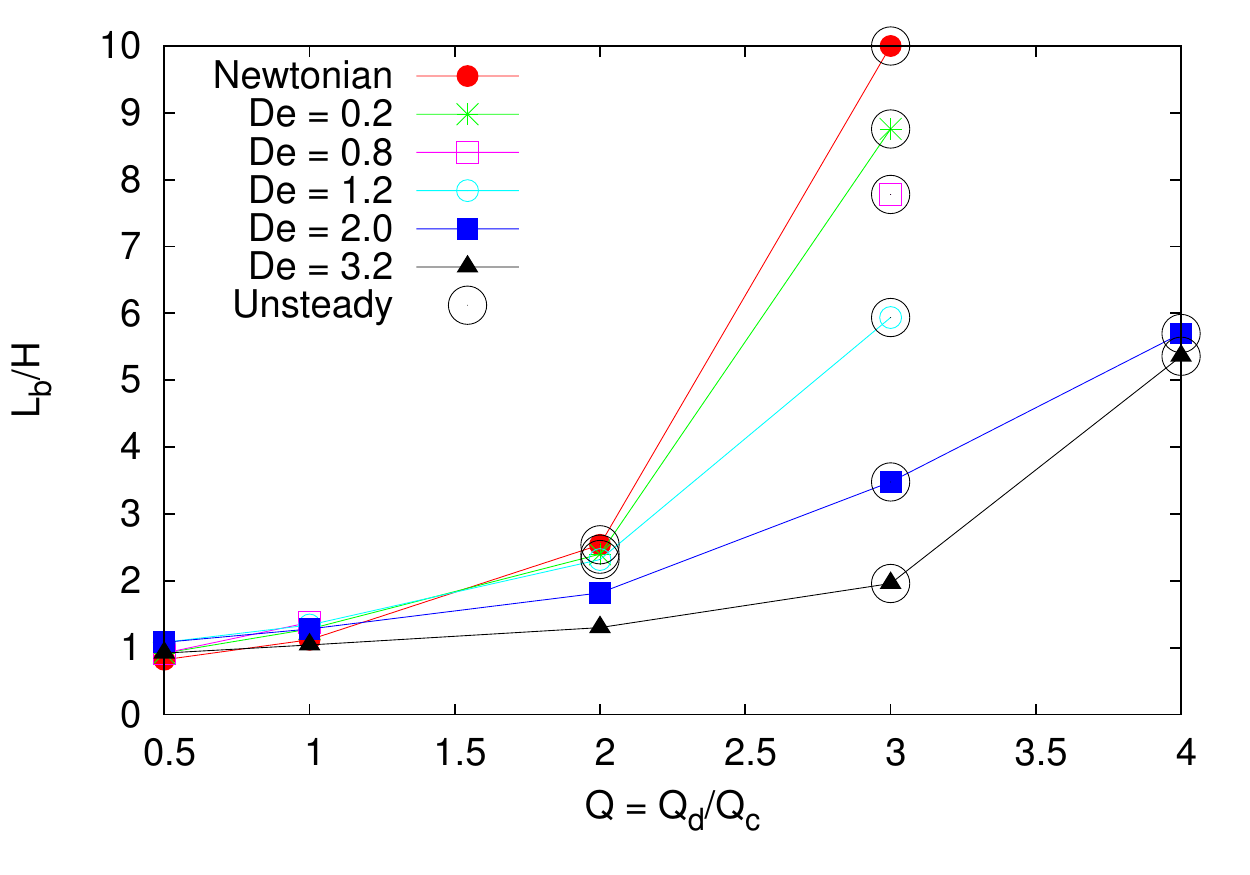}
\caption{Effects of matrix viscoelasticity (MV) on the break-up point. We report the break-up distance of the first break-up ($L_{b}$) normalized to the characteristic edge of the channels ($H$) for fixed $\Ca=0.007$, $\Ren=0.018$, and $\lambda=1$ for different $\De$ and $Q$. As we can see, as soon as $\De$ is just above unity, the flow undergoes a transition and the break-up distance $L_b$ moves towards the cross-junction. Black open circles represent the cases where the break-up length is changing with time and moves progressively downstream of the cross-junction.\label{fig:11}}
\end{figure}

%%%%%%%%%%%%%%%%%%%%%%%%%%%%%%%%%%%%%%%%%%%%%%%%%%%%%%%%%%%%%%%%%%%%%%%%%%%%%%%%%%%%%%%%%%%%%%%%%%%%%%%%%%%%%%%%%%%%%%%%%%%%%%%%%%%%%%%%%%%%%%%%%%%%%%%%%%%%%%%%

\begin{figure*}[tbh!]
\makeatletter
\def\@captype{figure}
\makeatother
\subfigure[\,\,$\Ca = 0.007$, $Q = 2.0$]
{
\includegraphics[width = 0.48\linewidth]{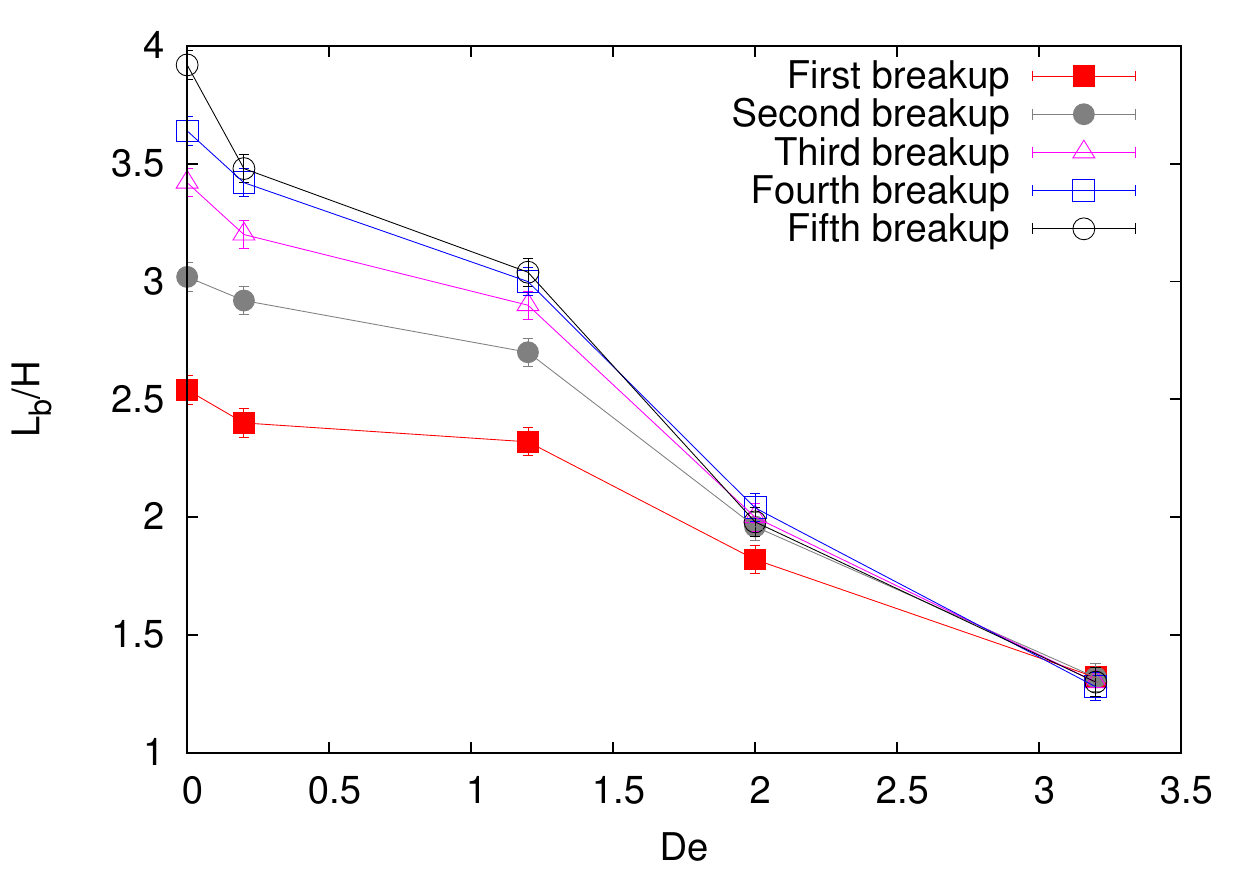}
}
\subfigure[\,\,$\Ca = 0.007$, $Q = 3.0$]
{
\includegraphics[width = 0.48\linewidth]{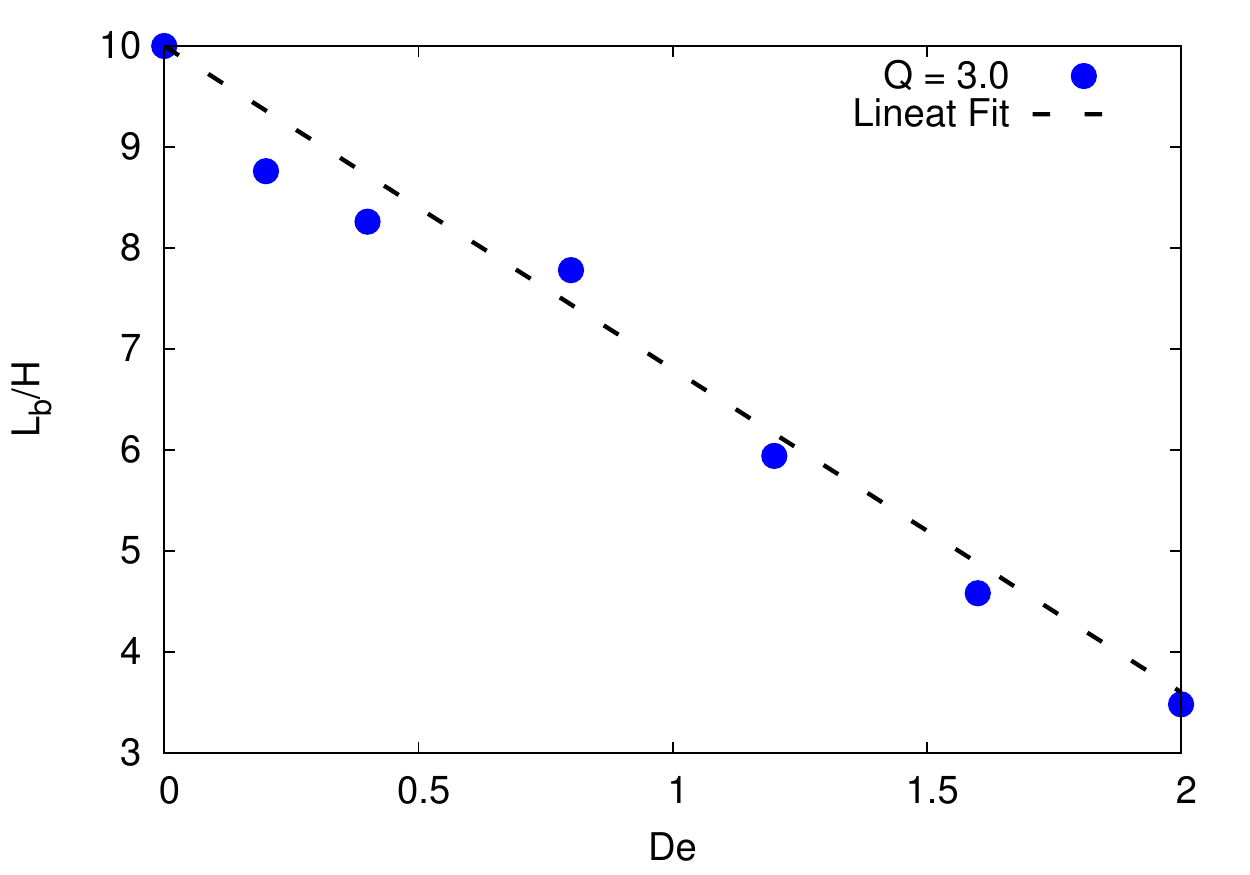}
}
\caption{Effects of matrix viscoelasticity (MV) on the break-up distance. Panel (a): we report the break-up distances up to the fifth break-up as a function of $\De$ for fixed $Q = 2.0$, $\Ca=0.007$, $\Ren=0.018$, and $\lambda=1$. The break-up distance $L_b$ is normalized to the characteristic edge of the channels ($H$). Panel (b): we show the break-up distance of the first break-up as a function of the Deborah Number for a fixed flow-rate ratio $Q = 3.0$.\label{fig:12}}
\end{figure*}

We next go on by quantifying the effects of polymers in the DC regime. As discussed in figure \ref{fig:2}, a distinctive feature of the DC regime, is that the break-up point moves progressively into the main channel where the dispersed phase breaks-up after traveling a distance $L_b$ from the cross-junction. The break-up point is not stable, i.e. the distance $L_b$ increases as a function of time (see figure \ref{fig:2}(e)-(h)). In figure \ref{fig:7} we can appreciate the effects of MV in the DC regime. In particular, we report density contours of the dispersed phase for three cases: (a) Newtonian matrix  at time $t=t_0+3.0 \tshear$; (b) slightly viscoelastic matrix with $\De = 0.2$ at time $t=t_0+2.8 \tshear$;  (c) viscoelastic matrix with Deborah number just above unity ($\De = 2.0$) at time $t=t_0+2.9 \tshear$. All the other flow parameters are kept fixed, $\Ca=0.007$, $\Ren=0.018$, $\lambda=1$ and $Q=3.0$. The time of the observation is essentially the same ($\approx t_0+2.8 \tshear$) and the effect of viscoelasticity is to promote a shift of the break-up point from downstream of the cross-junction (figure \ref{fig:7}(a)) to the cross junction itself (figure \ref{fig:7}(c)). The density contours of the dispersed phase overlaid on snapshots of the magnitude of the polymer feedback stress are reported in figure \ref{fig:8}. Again, we attribute the observed behaviour to the extra force generated by the polymer feedback stress at the cross-junction: this is effectively perturbing the droplet shape, provoking the break-up immediately after the cross junction. In some sense, this effect may be translated in an enhanced stability of the DCJ regime, a fact that will be better quantified in figure \ref{fig:11}.\\
Finally, we investigate the effect of MV on the PF regime. This is shown in figures \ref{fig:9} and \ref{fig:10}, where we can appreciate how the perturbation induced by viscoelasticity to the droplet shapes at the cross-junction are enough to spoil the PF itself leading to droplet break-up (see figure \ref{fig:9}(c)).\\
To provide an overview of the results for the various regimes, in figure \ref{fig:11} we report the first break-up distance normalized to the characteristic edge of the channel ($L_b/H$) as a function of the flow-rate ratio $Q$ for various Deborah numbers. The other parameters are kept fixed to $\Ca=0.007$, $\Ren=0.018$, and $\lambda=1$. In the Newtonian case, the DCJ regime ($L_b/H \approx 1$) is valid for small $Q$ up to $Q \approx 2.0$~\cite{LiuZhang09,LiuZhang11}, whereas at larger $Q$ we observe the emergence of the DC regime, with the break-up distance $L_b$ that is not stably located at the cross-junction anymore (circles indicate unstable $L_b$), but moves progressively downstream of the cross-junction. For the viscoelastic cases, the emergence of the DC regime is delayed. Just to give some numbers, an increase of $\De$ from $\De=0.0$ to $\De=2.0$, results in the emergence of the DC regime for $Q \approx 3.0$ instead of $Q \approx 2.0$. To stress even more the stabilizing character of the MV in the transition from DCJ to DC, at fixed flow-rate ratio $Q=2.0$, we report in figure \ref{fig:12}(a) the break-up distances up to the fifth break-up as a function of $\De$. A clear distinction emerges by comparing the situation at $\De=0$ (Newtonian case) and that with $\De$ just above unity, where the break-up point is manifestly stabilized. For fixed $Q=3.0$, we report in figure \ref{fig:12}(b) the first break-up distance as a function of $\De$, showing a linear decrease, $L_{b} = L_b(\De=0)-\alpha \De$, with $\alpha$ a constant of order unity.

\subsection{Effects of Droplet Viscoelasticity (DV)}\label{sec:DV}

\begin{figure*}[th!]
\makeatletter
\def\@captype{figure}
\makeatother
\subfigure[\,\,$\Ca = 0.007$]
{
\includegraphics[width=.48\linewidth]{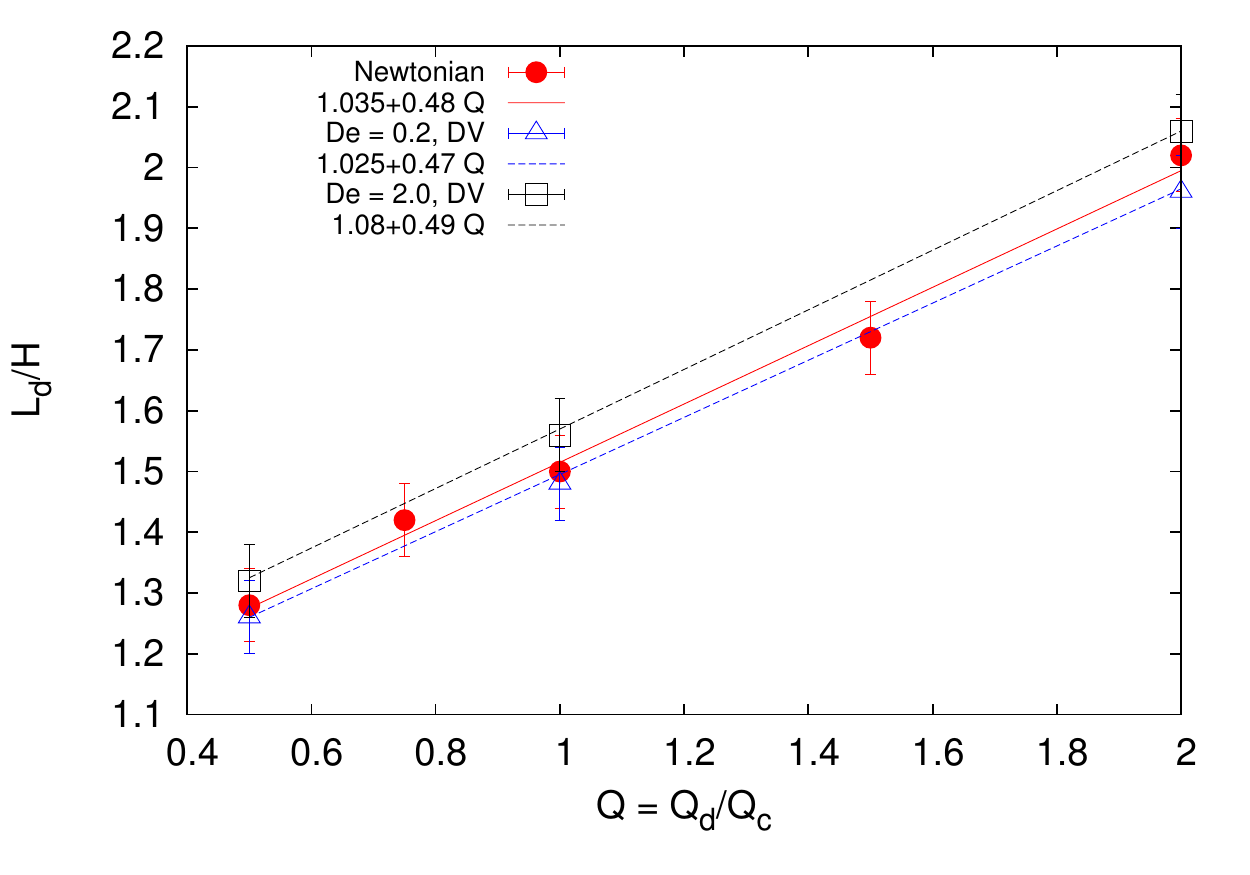}
}
\subfigure[\,\,$\Ca = 0.007$]
{
\includegraphics[width=0.48\linewidth]{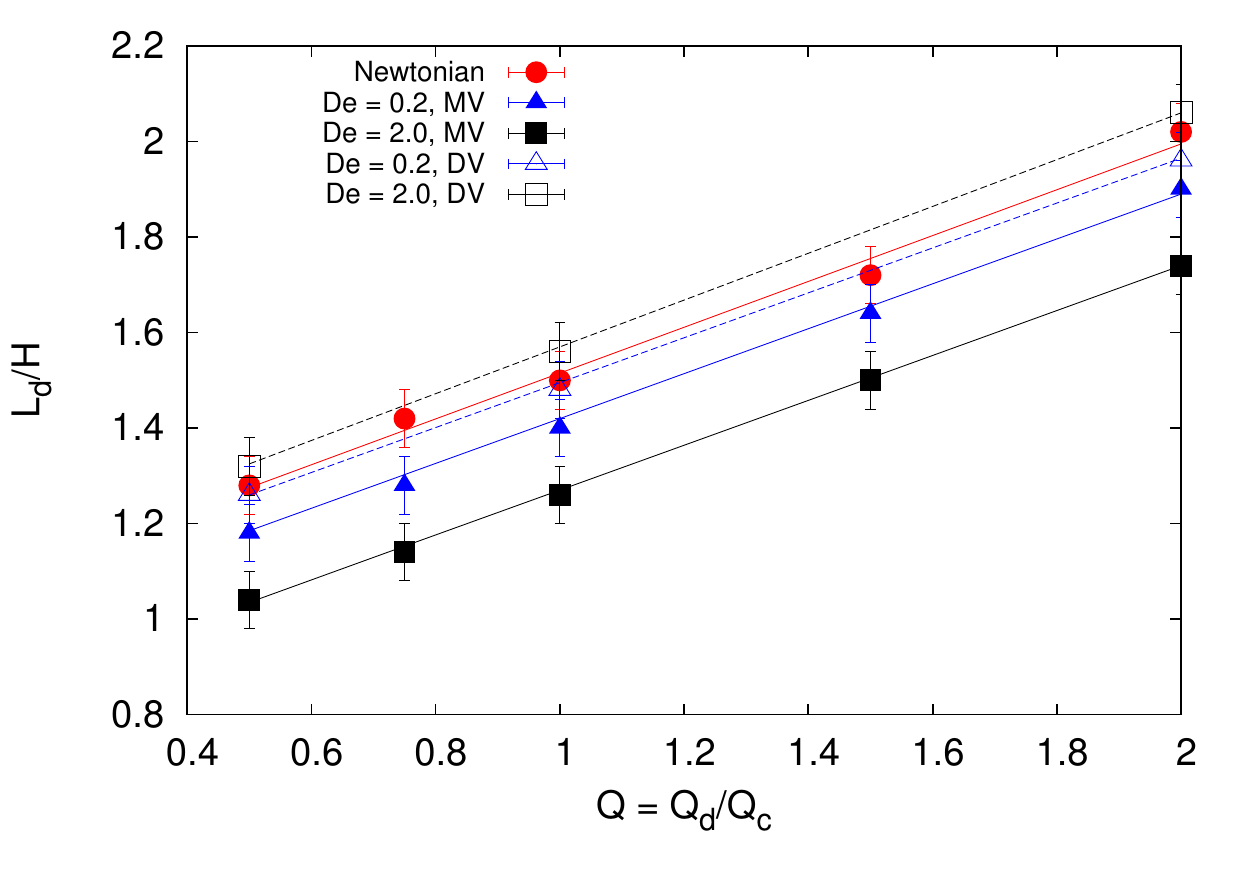}
}
\caption{Effects of matrix viscoelasticity (MV) and droplet viscoelasticity (DV) on the droplet length. Panel (a): we report the droplet length ($L_{d}$) normalized to the characteristic edge of the channels ($H$) versus the flow-rate ratio $Q$ for different values of $\De$ in the DV case. All the other parameters are kept fixed to $\Ca=0.007$, $\Ren=0.018$, and $\lambda=1$. Panel (b): both the cases with MV and DV are compared. Half of the interface thickness has been used as errorbar. Linear fits are drawn from the expected behaviour \eqref{eq:lin}, with $\alpha_{1,2}$ used as fitting parameters. \label{fig:13}}
\end{figure*}

\begin{figure}[t!]
\includegraphics[width=.9\linewidth]{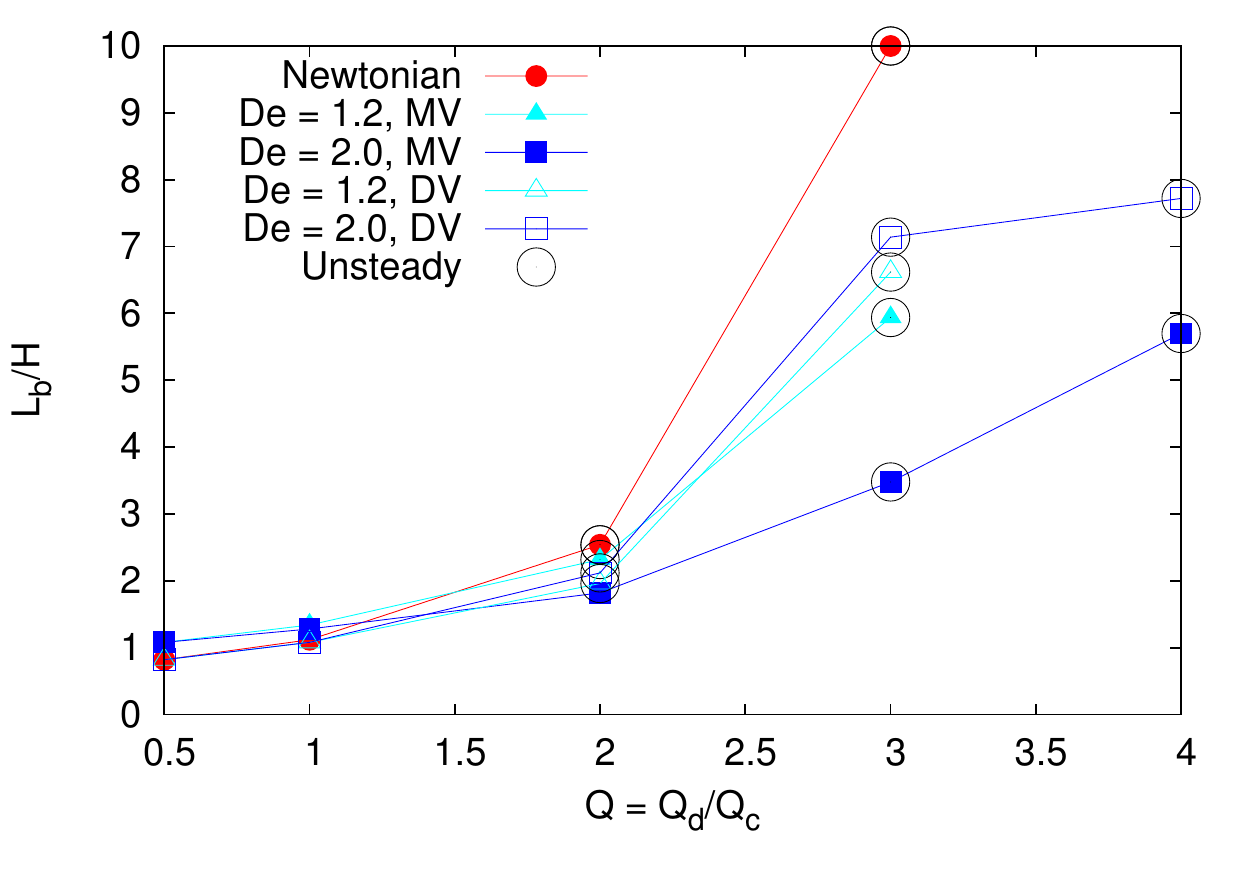}
\caption{We report the break-up distance of the first break-up ($L_{b}$) normalized to the characteristic edge of the channels ($H$) versus the flow-rate ratio $Q$ for different values of $\De$. Results for droplet viscoelasticity (DV) are here compared with those of matrix viscoelasticity (MV) already reported in figure \ref{fig:11}. Black open circles represent the cases where the break-up length is changing with time and moves progressively downstream of the cross-junction. All the cases reported have fixed $\Ca=0.007$, $\Ren=0.018$, and $\lambda=1$.\label{fig:14}}
\end{figure}

In this subsection we will discuss the effect of DV for the break-up of threads in the confined flow-focusing geometry. We will be mainly addressing the role of DV in the DCJ regime, where quantitative results for the MV have been obtained as a function of the flow-rate ratio $Q$ (see figure \ref{fig:6}). In figure \ref{fig:13}(a) we study droplet length $L_{d}$ normalized to the characteristic edge of the channels ($H$) for fixed $\Ca=0.007$, $\Ren=0.018$, and $\lambda=1$. A linear relation between droplet length $L_d$ and flow-rate ratio $Q$ is again observed, which echoes the results for MV cases in figure \ref{fig:6}. For a given flow-rate ratio $Q$, initially the droplet length $L_d$ decreases with $\De$ but, when $\De$ is above unity ($\De = 2$), we find that $L_d$ is slightly larger compared to the Newtonian case. These changes are however small if compared to those achieved for the MV cases. To quantitatively compare the effect of DV and MV we have collected in figure \ref{fig:13}(b) data from figures \ref{fig:6} and \ref{fig:13}(a). In figure \ref{fig:14} we compare both DV and MV, by studying $L_{b}$ normalized to the characteristic edge of the channels ($H$) for fixed $\Ca=0.007$, $\Ren=0.018$, and $\lambda=1$. In both cases, the effect of viscoelasticity is a stabilizing one but, again, it is more pronounced in the case of MV.

\section{Conclusions}\label{sec:conclusion}

A crucial point for designing, developing and exploiting micro- and nanofluidic devices is to achieve a systematic control over the process of formation of droplets as a function of the materials and flow parameters. The confinement that naturally accompanies flows in small devices has significant qualitative and quantitative effects on the droplet dynamics and break-up. Moreover, relevant constituents have commonly a viscoelastic - rather than Newtonian - nature. One of the most common droplet generator is the flow-focusing geometry, based on the focusing of a stream of one liquid in another co-flowing immiscible liquid. In this paper we have presented results based on three dimensional mesoscale numerical simulations to highlight the non trivial role played by confinement and non-Newtonian effects. We have worked in the flow conditions previously studied by other authors~\cite{LiuZhang09,LiuZhang11,Guillot}, where Newtonian droplets in a Newtonian matrix are squeezed either at the cross-junction, or form downstream of the cross-junction (DC). The transition between these two regimes has been found to be affected by viscoelasticity, and in particular by those situations where viscoelastic properties are confined in the continuous phase. This is due to the fact that the action of the co-flowing liquid that periodically breaks droplets is supplemented with the polymer feedback stresses that are well pronounced in correspondence of the corners where the side channels meet the main channel. The enhancement of the polymer feedback stress promotes an extra force which points towards the corners and inevitably perturbs the flow properties. Specifically, an extra deformation of the droplet at the cross-junction is introduced and is able to trigger the break-up immediately after the cross junction itself, thus resulting in a delayed DCJ to DC transition.\\ 
We remark that the results presented in this paper are obtained for a fixed $L^2$, i.e. for a fixed maximum elongation of the polymers. The parameter $L^2$ is actually related to the extensional viscosity of the polymers~\cite{bird,Lindner03}. Similarly to the work that we recently developed for characterizing droplet deformation and break-up in confined shear flow~\cite{SbragagliaGupta}, it would be important to study the impact of a change in the finite extensibility parameter $L^2$ for the geometries studied in this paper~\cite{Arratia08,Arratia09}. Finally, we wish to underscore the role played by numerical simulations: the numerical simulations are crucial to elucidate the relative importance of the free parameters in the model, and to visualize the distribution of the polymer feedback stresses during the motion of the droplets. Thanks to these insights, it was possible to correlate the distribution of the stresses to the droplet shape and the corresponding break-up morphology. \\

We kindly acknowledge funding from the European Research Council under the Europeans Community's Seventh Framework Programme (FP7/2007-2013) / ERC Grant Agreement  N. 279004. We acknowledge support from the COST-Action MP1305 and the computing hours from ISCRA B project (POLYDROP), CINECA Italy. MS acknowledges useful discussions and fruitful exchange of ideas with Prof. H. Liu during his visit in June 2014.

%%%%%%%%%%%%%%%%%%%%%%%%%%%%%%%%%%%%%%%%%%%%%%FIG 14%%%%%%%%%%%%%%%%%%%%%%%%%%%%%%%%%%%%%%%%%%%%%%%%%%%%%%%%%%%%%%%%%%%%%%%%%%%%%%%%%%%%%%%%%%%%%%%%%%%%%%%%%%%%%%%%%%

%%%%%%%%%%%%%%%%%%%%%%%%%%%%%%%%%%%%%%%%%%%%%%%%%%%%%%%%%%%%%%%%%%%%%%%%%%%%%%%%%%%%%%%%%%%%%%%%%%%%%%%%%%%%%%%%%%%%%%%%%%%%%%%%%%%%%%%%%%%%%%%%%%%%%%%%%%%%%%%%

%%%%%%%%%%%%%%%%%%%%%%%%%%%%%%%%%%%%%%%%%%%%%%%%%%%%%%%%%%%%%%%%%%%%%%%%%%%%%%%%%%%%%%%%%%%%%%%%%%%%%%%%%%%%%%%%%%%%%%FIG 15%%%%%%%%%%%%%%%%%%%%%%%%%%%%%%%%%%%%%%%%%%%%%%%%%%%%%%%%%%%%%%%%%%%%%%%%%%%%%%%%%%%%%%%%%%%%%%%%%%%%%%%%%%%%%%%%%%%%

%%%%%%%%%%%%%%%%%%%%%%%%%%%%%%%%%%%%%%%%%%%%%%%%%%%%%%%%%%%%%%%%%%%%%%%%%%%%%%%%%%%%%%%%%%%%%%%%%%%%%%%%%%%%%%%%%%%%%%%%%%%%%%%%%%%%%%%%%%%%%%%%%%%%%%%%%%%%%%%%

%\appendix\label{sec:appendix}

\section{Hybrid Lattice Boltzmann (LB) Models - Finite Difference (FD) Schemes}\label{sec:appendix}

In this appendix we report some details of the numerical scheme used. The interested reader can find more extensive technical details in a dedicated paper~\cite{SbragagliaGupta}. We use a hybrid algorithm combining a multicomponent Lattice-Boltzmann (LB) model with Finite Differences (FD) schemes. LB is used to model the macroscopic hydrodynamic equations, while FD is used to model the dynamics of the polymers. The LB time dynamics considers the discretized probability density function $f_{\alpha i}({\bm{r}},t)$ to find at position ${\bm{r}}$ and time $t$ a fluid particle of component $\alpha=A,B$ with velocity  ${\bm{c}}_{i}$. The dispersed (d) and the continuous (c) Newtonian phases in Eqs.~(\ref{NSc}) and (\ref{NSd}) are characterized by a majority of one of the two components, i.e. majority of $A$ ($B$) in the dispersed (continuous) phase. The LB evolution scheme with a unitary time-step reads as follows~\cite{Succi01}:
\begin{equation}\label{EQ:LBapp}
f_{\alpha i} ({\bm{r}} + {\bm{c}}_{i} , t + 1)-f_{\alpha i} ({\bm{r}}, t) = \sum_{j} {\cal L}_{i j}(f_{\alpha j}-f^{(eq)}_{\alpha j}) + \Delta^{g}_{\alpha i}.
\end{equation}
The first term in the rhs of Eq.~\eqref{EQ:LBapp} is the (linear) collisional operator, expressing the relaxation of $f_{\alpha i}$ towards the local equilibrium $f^{(eq)}_{\alpha i}$. All the numerical simulations make use of the D3Q19 model (3d with 19 velocities) whose discrete velocity set is given by
\begin{equation}\label{velo}
{\bm{c}}_{i}=
\begin{cases}
(0,0,0) & i=0\\
(\pm 1,0,0), (0,\pm 1,0), (0,0,\pm 1) & i=1\ldots6\\
(\pm 1,\pm 1,0), (\pm 1,0,\pm 1), (0,\pm 1,\pm 1)  & i=7\ldots18
\end{cases}.
\end{equation}
The expression for the equilibrium distribution is~\cite{Dunweg07,DHumieres02} 
\be\label{feq}
f_{\alpha i}^{(eq)}=w_{i} \rho_{\alpha} \left[1+\frac{{\bm{u}} \cdot {\bm{c}}_{i}}{c_s^2}+\frac{{\bm{u}}{\bm{u}}:({\bm{c}}_{i}{\bm{c}}_{i}-{\Id})}{2 c_s^4} \right]
\ee
and the weights $w_{i}$ are
\begin{equation}\label{weights}
w_{i}=
\begin{cases}
1/3 & i=0\\
1/18 & i=1\ldots6\\
1/36 & i=7\ldots18
\end{cases},
\end{equation}
where $c_s$ is the isothermal speed of sound and ${\bm{u}}$ is the fluid velocity. The operator ${\cal L}_{i j}$ in Eq.~\eqref{EQ:LBapp} is the same for both components and has a diagonal representation in the {\it mode space}: the basis vectors ${\bm{H}}_{k}$ ($k=0,...,18$) of the mode space are used to calculate a complete set of moments, the so-called ``modes'' $m_{\alpha k}=\sum_{i} {\bm{H}}_{k i} f_{\alpha i}$ ($k=0,...,18$)~\cite{Dunweg07,DHumieres02,Premnath,SegaSbragaglia13}. Hydrodynamic variables are obtained from the lowest order modes: the density of both components and the total density are $\rho_{\alpha}=m_{\alpha 0}=\sum_{i} f_{\alpha i}$,  $\rho=\sum_{\alpha}m_{\alpha 0} =\sum_{\alpha}\rho_{\alpha}$, while the next three moments $\tilde{\bm{m}}_{\alpha}=(m_{\alpha 1}, m_{\alpha 2}, m_{\alpha 3})$, define the velocity of the mixture 
\be\label{totmom}
{\bm{u}} \equiv \frac{1}{\rho}\sum_{\alpha} \tilde{\bm{m}}_{\alpha}   +\frac{\bm{g}}{2 \rho} = \frac{1}{\rho}\sum_{\alpha} \sum_{i} f_{\alpha i} {\bm{c}}_{i}+\frac{\bm{g}}{2 \rho}.
\ee

%%%%%%%%%%%%%%%%%%%%%%%%%%%%%%%%%%%%%%%%%%%%%%%%%%%%%%%%%%%%%%%%%%%%%%%%%%%%%%%%%%%%%%%%%%%%%%%%%%%%%%%%%%%%%%%%%%%%%%FIG 16%%%%%%%%%%%%%%%%%%%%%%%%%%%%%%%%%%%%%%%%%%%%%%%%%%%%%%%%%%%%%%%%%%%%%%%%%%%%%%%%%%%%%%%%%%%%%%%%%%%%%%%%%%%%%%%%%%%%%%%%%%%%%%%%%%%%%%%%%%%%%%%%%%%%%%%%%%%%%%%%%%%%%%%%%%%%%%%%%%%%%%%%%%%%%%%%%

%%%%%%%%%%%%%%%%%%%%%%%%%%%%%%%%%%%%%%%%%%%%%%%%%%%%%%%%%%%%%%%%%%%%%%%%%%%%%%%%%%%%%%%%%%%%%%%%%%%%%%%
%%%%%%%%%%%%%%%%%%%%%%%%%%%%%%%%%%%%%%%%%%%%%%%%%%%%%%%%%%%%%%%%%%%%%%%%%%%%%%%%%%%%%%%%%%%%%%%%%%%%%%%

The higher order modes refer to the viscous stress tensor, and also other modes (the so called ``ghost modes'') which do not appear in the hydrodynamic equations. Since the operator ${\cal L}_{i j}$ possesses a diagonal representation in mode space, the collisional term describes a linear relaxation of the modes, $m^{post}_{\alpha k}=(1+\lambda_k)m_{\alpha k}+m_{\alpha k}^{g}$, where the ``post" indicates the post-collisional mode and where the relaxation frequencies $-\lambda_k$ are related to the transport coefficients of the modes (see also below). The term $m_{\alpha k}^{g}$ is the $k$-th moment of the forcing source $\Delta_{\alpha i}^{g}$: such term embeds the effects of a forcing term with density ${\bm{g}}_{\alpha}$~\cite{Dunweg07,Premnath}. The term ${\bm{g}}=\sum_{\alpha} {\bm g}_{\alpha}$ in Eq.~\eqref{totmom} relates to the total force. The forcing source is tuned in such a way that the correct hydrodynamic equations are obtained~\cite{SegaSbragaglia13}: given the relaxation frequencies of the momentum ($-\lambda_M$), bulk ($-\lambda_b$) and shear ($-\lambda_s$) modes, the forcing source reads
\begin{equation}
\Delta_{\alpha i}^{g}=\frac{w_{i}}{c_s^2} \left(\frac{2+\lambda_M}{2}\right) {\bm{g}}_{\alpha} \cdot {\bm{c}}_{i} +\frac{w_{i}}{c_s^2} \left[\frac{1}{2c_s^2} {\bm{G}} : ({\bm{c}}_{i} {\bm{c}}_{i}-c_s^2 {\Id} ) \right],
\end{equation}
\begin{equation}
{\bm G}=\frac{2+\lambda_s}{2}\left({\bm u} {\bm g} + ({\bm u}  {\bm g})^T-\frac{2}{3} {\Id} ({\bm u} \cdot {\bm g}) \right)+\frac{2+\lambda_b}{3} {\Id} ({\bm u} \cdot {\bm g}).
\end{equation}
LB reproduces the continuity equations and the NS equations for the total momentum~\cite{Dunweg07,Premnath,SegaSbragaglia13} with shear viscosity $\eta_s$ and bulk viscosity $\eta_b$:
\be\label{eq:2}
\partial_t \rho_{\alpha}+ {\bm \nabla} \cdot (\rho_{\alpha} {\bm u}) = {\bm \nabla} \cdot {\bm D}_{\alpha},
\ee
\be\label{eq:3}
\begin{split}
\rho & \left[ \partial_t \bm u + ({\bm u} \cdot {\bm \nabla}) \bm u \right]= -{\bm \nabla}p \\ 
&+ {\bm \nabla} \left[ \eta_{s} \left( {\bm \nabla} {\bm u}+({\bm \nabla} {\bm u})^{T}-\frac{2}{3} {\Id} ({\bm \nabla} \cdot {\bm u}) \right) +\eta_{b}{\Id} ({\bm \nabla} \cdot {\bm u}) \right] + {\bm g}.
\end{split}
\ee
The relaxation frequencies of the bulk and shear modes in (\ref{EQ:LBapp}) are related to the viscosity coefficients as
\begin{equation}\label{TRANSPORTCOEFF}
\eta_s=-\rho c_s^2 \left(\frac{1}{\lambda_s}+\frac{1}{2} \right); \hspace{.1in} \eta_b=-\frac{2}{3}\rho c_s^2  \left(\frac{1}{\lambda_b}+\frac{1}{2} \right).
\end{equation}
In the above equations, $p=\sum_{\alpha} p_{\alpha}=\sum_{\alpha} c_s^2 \rho_{\alpha}$ represents the internal (ideal) pressure of the mixture. The quantity ${\bm D}_{\alpha}$ is the diffusion flux 
\begin{equation}\label{eq:comp_Pi}
{\bm D}_{\alpha}=\mu \left[\left({\bm \nabla} p_{\alpha}-\frac{\rho_{\alpha}}{\rho}{\bm \nabla} p\right)-\left({\bm g}_{\alpha}-\frac{\rho_{\alpha}}{\rho} {\bm g} \right) \right]
\end{equation}
with $\mu$ the mobility coefficient:
\begin{equation}
\mu=-\left(\frac{1}{\lambda_M}+\frac{1}{2} \right).
\end{equation}
As for the internal forces, we will use the ``Shan-Chen'' model~\cite{SC93,CHEM09,sbragaglia12} for multicomponent fluids
\begin{equation}\label{eq:SCforce}
{\bm g}_{\alpha}({\bm{r}}) =  - {g}_{AB} \rho_{\alpha}({\bm{r}}) \sum_{i} \sum_{\alpha'\neq \alpha} w_{i} \rho_{\alpha^{\prime}} ({\bm{r}}+\bm{c}_{i}) {\bm c}_{i} \hspace{.2in} \alpha,\alpha^{\prime}=A,B
\end{equation}
where ${g}_{AB}$ is a coupling parameter that regulates the interactions between the two components. When ${g}_{AB}$ is chosen above a critical value, phase segregation occurs with the formation of stable interfaces with a positive surface tension. Due to the effect of interaction forces, the internal pressure is modified by the ``interaction'' pressure tensor ${\bm P}^{(\mbox{\tiny{int}})}$~\cite{SbragagliaBelardinelli}, i.e. ${\bm P} = p \, {\Id}+{\bm P}^{(\mbox{\tiny{int}})}$, with
\be\label{PT}
\begin{split}
{\bm P}^{(\mbox{\tiny{int}})}({\bm r}) & =  \frac{1}{2} {g}_{AB} \rho_{A}({\bm r})\sum_{i} w_{i} \rho_{B}({\bm r}+{\bm c}_{i}) {\bm c}_{i} {\bm c}_{i} \\
& +\frac{1}{2} {g}_{AB} \rho_{B}({\bm r})\sum_{i} w_{i} \rho_{A}({\bm r}+{\bm c}_{i}){\bm c}_{i} {\bm c}_{i}.
\end{split}
\ee
Wetting properties are introduced with a tuning of the density in contact with the wall~\cite{Benzi06,sbragaglia08}. The numerical simulations presented are carried out with ${g}_{AB}=1.5$ lbu in (\ref{eq:SCforce}), corresponding to a surface tension $\sigma=0.1$ lbu and associated bulk densities $\rho_A=2.0$ lbu and $\rho_B=0.1$ lbu in the $A$-rich phase. The relaxation frequencies in (\ref{TRANSPORTCOEFF}) are set to $\lambda_M=-1.0$ lbu and $\lambda_s=\lambda_b$, thus reproducing the viscous stress tensor given in Eqs.~(\ref{NSc}) and (\ref{NSd})). The viscosity ratio of the LB fluid is changed by allowing $\lambda_s$ to depend on space 
\be
-\rho c_s^2 \left(\frac{1}{\lambda_s}+\frac{1}{2} \right)=\eta_s=\eta_d (f_+(\phi))+\eta_c (f_{-}(\phi))
\ee
where $\phi=\phi({\bm r})=\frac{(\rho_A({\bm r})-\rho_B({\bm r}))}{(\rho_A({\bm r})+\rho_B({\bm r}))}$. The functions $f_{\pm}(\phi)$ are chosen as
\be
f_{\pm}(\phi)=\left(\frac{1 \pm \tanh(\phi/\xi)}{2}\right)
\ee
which allows to recover the Newtonian part of the NS equations reported in Eqs.~(\ref{NSc}) and (\ref{NSd}) with shear viscosities $\eta_d$ and $\eta_c$. The smoothing parameter $\xi$ is chosen sufficiently small so as to match analytical predictions on droplet deformation in presence of viscoelastic stresses (see~\cite{SbragagliaGupta} for all details).\\
As for the polymer evolution given in Eq.~\eqref{FENEP}, we follow the two References~\cite{perlekar06,vaithianathan2003numerical} to solve the FENE-P equation. The polymer feedback stress is computed from the FENE-P evolution equation and used to change the shear modes of the LB~\cite{SbragagliaGupta,Dunweg07,DHumieres02}. The feedback of the polymers is modulated~\cite{Yue04} in space with the functions $f_{\pm}(\phi)$ 
\be
\begin{split}
\rho & \left[ \partial_t \bm u  + ({\bm u} \cdot {\bm \nabla}) \bm u \right] = -{\bm \nabla} {\bm P} \\
&+ {\bm \nabla} \left[(\eta_d f_+(\phi)+\eta_c f_{-}(\phi)) ({\bm \nabla} {\bm u}+({\bm \nabla} {\bm u})^{T} ) \right]\\ 
&+\frac{\eta_P}{\tau_P}{\bm \nabla} [f(r_P){\bm {\mathcal C}} f_{\pm}(\phi) ].
\end{split}
\ee
By using $f_{-}(\phi)$, we recover a case where the viscoelastic properties are confined in the continuous (c) phase, while the use of the function $f_{+}(\phi)$ allows to recover a case where the viscoelastic properties are confined in the dispersed (d) phase.

%%%%%%%%%%%%%%%%%%%%%%%%%%%%%%%%%%%%%%%%%%%%%%%%%%%%%%%%%%%%%%%%%%%%%%%%%%%%%%%%%%%%%%%%%%%%%%%%%%%%%%%%%%%%%%%%%%%%%%%%%%%%%%%%%%%%%%%%%%%%%%%%%%%%%%%%%%%%%%%%

%%%%%%%%%%%%%%%%%%%%%%%%%%%%%%%%%%%%%%%%%%%%%%%%

\bibliographystyle{spphys}       % APS-like style for physics
%-- Rhodes Style --

%%%%%%%%%%%%%%%%%%%%%%%%%%%%%%%%%%%%%%%%%%%
\bibliography{sample}
%%%%%%%%%%%%%%%%%%%%%%%%%%%%%%%%%%%%%%%%%%%

\end{document}